\documentstyle [12pt,twoside]{article}

\newskip\humongous \humongous=0pt plus 1000pt minus 1000pt
\def\caja{\mathsurround=0pt}
\def\eqalign#1{\,\vcenter{\openup2\jot \caja
        \ialign{\strut \hfil$\displaystyle{##}$&$
        \displaystyle{{}##}$\hfil\crcr#1\crcr}}\,}
\newif\ifdtup
\def\panorama{\global\dtuptrue \openup2\jot \caja
        \everycr{\noalign{\ifdtup \global\dtupfalse
        \vskip-\lineskiplimit \vskip\normallineskiplimit
        \else \penalty\interdisplaylinepenalty \fi}}}

\def\eqalignnotwo#1{\panorama \tabskip=\humongous
        \halign to\displaywidth{\hfil$\displaystyle{##}$
        \tabskip=0pt&$\displaystyle{{}##}$
        \tabskip=\humongous&\llap{$##$}\tabskip=0pt
        \tabskip=0pt&$\displaystyle{{}##}$\hfil
        \crcr#1\crcr}}

\def\begintitle#1#2#3#4
        {\begin{titlepage}
         \centerline{#1 \hfill UMDGR-93-#2}
         \begin{center}\vglue .4in
         {\large\bf #3}\\[.4in]
         {\bf #4}\\[.2in]
         {\it Department of Physics}\\
         {\it University of Maryland, College Park, MD 20742}\\[.4in]
         {\bf ABSTRACT}\\
         \end{center}
         \begin{quotation}}
\def\endtitle
         {\end{quotation}
          \end{titlepage}
          \newpage}

\def\bk{{\bf k}}

\def\bv{{\bf v}}\def\bw{{\bf w}}

\def\bA{{\bf A}}\def\bB{{\bf B}}\def\bD{{\bf D}}
\def\bE{{\bf E}}\def\bF{{\bf F}}\def\bG{{\bf G}}
\def\bK{{\bf K}}
\def\bN{{\bf N}}

\def\cA{{\cal A}}\def\cC{{\cal C}}\def\cD{{\cal D}}

\def\cI{{\cal I}}\def\cL{{\cal L}}

\def\cR{{\cal R}}

\def\a{\alpha}\def\b{\beta}\def\g{\gamma}\def\d{\delta}\def\e{\epsilon}
\def\n{\eta}\def\l{\lambda}
\def\u{\mu}\def\s{\sigma}\def\t{\tau}
\def\x{\chi}
\def\G{\Gamma}\def\L{\Lambda}\def\S{\Sigma}\def\W{\Omega}

\def\pd{\partial}\def\grad{\nabla\!}
\def\id{{\rm d\! I}}
\def\iwedge{\wedge\!\!\!\!\!\!\wedge{}}
\def\ovr{\overline}\def\und{\underline}
\def\otw{\widetilde}
\def\utw#1{\rlap{\lower1ex\hbox{$\sim$}}#1{}}
\def\pb#1{\rlap{\lower1ex\hbox{$\leftarrow$}}#1{}}
\def\pf#1{\rlap{\lower1ex\hbox{$\rightarrow$}}#1{}}
\def\X{\times}
\def\dsum{\oplus}\def\dprod{\otimes}
\def\semis{\bigcirc\!\!\!\!\!\!\s\ }
\def\semit{\bigcirc\!\!\!\!\!\!\t\ }

\def\to{\rightarrow}\def\Tr{{\rm Tr}}
\def\3#1{{}^3\!#1}\def\4#1{{}^4\!#1}\def\+#1{{}^+\!#1}\def\-#1{{}^-\!#1}
\def\*#1{{}^*\!#1}
\def\const{{\rm const}}\def\dim{{\rm dim}}\def\det{{\rm det}}

\def\gr{general relativity }
\def\EH{Einstein-Hilbert }\def\CS{Chern-Simons }\def\P{Palatini }
\def\YM{Yang-Mills }\def\Pb{Poisson bracket }\def\Pbs{Poisson brackets }
\def\sd{self-dual } \def\Sd{Self-dual }\def\asd{anti self-dual }
 \def\inf{infinitesimal }\def\wrt{with respect to }
\def\st{spacetime } \def\iff{if and only if }\def\EL{Euler-Lagrange }
\def\diffeos{diffeomorphisms }
\def\eom{equation of motion }\def\eoms{equations of motion }
\def\cc{cosmological constant }\def\KG{Klein-Gordon }
\def\RHS{right hand side }

\begin{document}
\begintitle{February 1993}{129}{\hfil GEOMETRODYNAMICS\hfil\break
\centerline{VS.}\hfil\break CONNECTION DYNAMICS}{Joseph
D. Romano\footnote{romano@umdhep.umd.edu} }
The purpose of this review is to describe in some detail the mathematical
relationship between geometrodynamics and connection dynamics in the context of
the classical theories of 2+1 and 3+1 gravity. I analyze the standard \EH
theory (in any \st dimension), the \P and \CS theories in 2+1 dimensions, and
the \P and self-dual theories in 3+1 dimensions.  I also couple various matter
fields to these theories and briefly describe a pure spin-connection
formulation of 3+1 gravity. I derive the \EL equations of motion from an action
principle and perform a Legendre transform to obtain a Hamiltonian formulation
of each theory.  Since constraints are present in all these theories, I
construct constraint functions and analyze their Poisson bracket algebra. I
demonstrate, whenever possible, equivalences between the theories.

\vspace{.7cm}


\endtitle

\baselineskip=12 true pt
\noindent Contents:
\medskip
\halign to \hsize{\hfil#\quad&#\hfill\tabskip 0pt plus 3em\hskip 2 true in
&\hfill#\cr
{1.}&{Overview}                                               &\omit\cr
\noalign{\smallskip}
{2.}&{Einstein-Hilbert theory}                                &\omit\cr
\noalign{\smallskip}
\omit&2.1\quad \EL equations of motion                        &\omit\cr
\omit&2.2\quad Legendre transform                             &\omit\cr
\omit&2.3\quad Constraint algebra                             &\omit\cr
\noalign{\smallskip}
{3.}&{2+1 Palatini theory}                                  &\omit\cr
\noalign{\smallskip}
\omit&3.1\quad \EL equations of motion                        &\omit\cr
\omit&3.2\quad Legendre transform                             &\omit\cr
\omit&3.3\quad Constraint algebra                             &\omit\cr
\noalign{\smallskip}
{4.}&{Chern-Simons theory}                                    &\omit\cr
\noalign{\smallskip}
\omit&4.1\quad \EL equations of motion                        &\omit\cr
\omit&4.2\quad Legendre transform                             &\omit\cr
\omit&4.3\quad Constraint algebra                             &\omit\cr
\omit&4.4\quad Relationship to the 2+1 \P theory            &\omit\cr
\noalign{\smallskip}
{5.}&{2+1 matter couplings}                                 &\omit\cr
\noalign{\smallskip}
\omit&5.1\quad 2+1 \P theory coupled to a cosmological constant &\omit\cr
\omit&5.2\quad Relationship to \CS theory                   &\omit\cr
\omit&5.3\quad 2+1 \P theory coupled to a massless scalar field &\omit\cr
\noalign{\smallskip}
{6.}&{3+1 Palatini theory}                                  &\omit\cr
\noalign{\smallskip}
\omit&6.1\quad \EL equations of motion                        &\omit\cr
\omit&6.2\quad Legendre transform                             &\omit\cr
\omit&6.3\quad Relationship to the \EH theory                 &\omit\cr
\noalign{\smallskip}
{7.}&{Self-dual theory}                                       &\omit\cr
\noalign{\smallskip}
\omit&7.1\quad \EL equations of motion                        &\omit\cr
\omit&7.2\quad Legendre transform                             &\omit\cr
\omit&7.3\quad Constraint algebra                             &\omit\cr
\noalign{\smallskip}
{8.}&{3+1 matter couplings}                                 &\omit\cr
\noalign{\smallskip}
\omit&8.1\quad \Sd theory coupled to a cosmological constant &\omit\cr
\omit&8.2\quad \Sd theory coupled to a \YM field            &\omit\cr
\noalign{\smallskip}
{9.}&{General relativity without-the-metric}                &\omit\cr
\noalign{\smallskip}
\omit&9.1\quad A pure spin-connection formulation of 3+1 gravity &\omit\cr
\omit&9.2\quad Solution of the diffeomorphism constraints        &\omit\cr
\noalign{\smallskip}
{10.}&{Discussion}                                          &\omit\cr
\noalign{\smallskip}
{}&{References}                                             &\omit\cr}
\vfill\eject

\baselineskip=17 true pt
\noindent{\bf 1. Overview}
\vskip .5cm

Einstein's theory of \gr is by far the most attractive classical theory of
gravity today. By describing the gravitational field in terms of the structure
of spacetime, Einstein effectively equated the study of gravity with the study
of geometry.  In general relativity, \st is a 4-dimensional manifold $M$ with a
Lorentz metric $g_{ab}$ whose curvature measures the strength of the
gravitational field. Given a matter distribution described by a stress-energy
tensor $T_{ab}$, the curvature of the metric is determined by Einstein's
equation $G_{ab}=8\pi \ T_{ab}$. This equation completely describes the
classical theory.

As written, Einstein's equation is \st covariant. There is no preferred time
variable, and, as such, no evolution. However, as we shall see in Section 2,
\gr admits a Hamiltonian formulation.  The canonically conjugate variables
consist of a positive-definite metric $q_{ab}$ and a density-weighted,
symmetric, second-rank tensor field $\otw p^{ab}$---both defined on a
3-manifold $\S$. These fields are not free, but satisfy certain constraint
equations.  Evolution is defined by a Hamiltonian, which (if we ignore boundary
terms) is simply a sum of the constraints.

Now it turns out that the time evolved data defines a solution, $(M,g_{ab})$,
of the full field equations which is unique up to \st diffeomorphisms. In a
solution, $\S$ can be interpreted as a spacelike submanifold of $M$
corresponding to an initial instant of time, while $q_{ab}$ and $\otw p^{ab}$
are related to the induced metric and extrinsic curvature of $\S$ in
$M$.\footnote{More precisely, $q_{ab}$ is the induced metric on $\S$, while
$\otw p^{ab}$ is related to the extrinsic curvature $K_{ab}$ via $\otw
p^{ab}=\sqrt q(K^{ab}-K q^{ab})$.} Thus, the Hamiltonian formulation of \gr can
be thought of as describing the dynamics of 3-geometries. Following Wheeler, I
will use the phrase ``geometrodynamics'' when discussing general relativity in
this form.

On the other hand, all of the other basic interactions in physics---the strong,
weak, and electromagnetic interactions---are described in terms of connection
1-forms. For example, the Hamiltonian formulation of \YM theory has a
connection 1-form $\bA_a$ (which takes values in the Lie algebra of some gauge
group $\bG$) as its basic configuration variable. The canonically conjugate
momentum (or ``electric field'') $\otw\bE^a$ is a density-weighted vector field
which takes values in the dual to the Lie algebra of $\bG$. As in general
relativity, these variables are not free, but satisfy constraint equations: \
The Gauss constraint $\bD_a\otw\bE^a=0$ (where $\bD_a$ is the generalized
derivative operator associated with $\bA_a$) tells us to restrict attention to
divergence-free electric fields.  Thus, just as we can think of the Hamiltonian
formulation of \gr as describing the dynamics of 3-geometries, we can think of
the Hamiltonian formulation of \YM theory as describing the dynamics of
connection 1-forms. I will often use the phrase ``connection dynamics'' when
discussing \YM theory in this form.

Despite the apparent differences between geometrodynamics and connection
dynamics, many researchers have tried to recast the theory of \gr in terms of a
connection 1-form. Afterall, if the strong, weak, and electromagnetic
interactions admit a connection dynamic description, why shouldn't gravity?
Early attempts in this direction used \YM type actions, but these actions gave
rise, however, to {\it new} theories of gravity. A connection dynamic theory
was gained, but Einstein's theory of \gr was lost in the process. Later
attempts (like the ones I will concentrate on in this review) left \gr alone,
but tried to reinterpret Einstein's equation in terms of the dynamics of a
connection 1-form. The most familiar of these approaches is due to Palatini who
rewrote the standard \EH action (which is a functional of just the \st metric
$g_{ab}$) in such a way that the \st metric and an arbitrary Lorentz connection
1-form are independent basic variables. However, as we shall see in Section 6,
the 3+1 \P theory does not succeed in recasting \gr as a connection dynamical
theory. The 3+1 \P theory collapses back to the standard geometrical
description of \gr when one writes it in Hamiltonian form.

More recently, Ashtekar \cite{hf,npcg,poona} has proposed a reformulation of
\gr in which a real (densitized) triad $\otw E_i^a$ and a connection 1-form
$A_a^i$ (which takes values in the complexified Lie algebra of $SO(3)$) are the
basic canonical variables. He obtained these new variables for the real theory
by performing a canonical transformation on the standard phase space of real
general relativity. For the complex theory, Jacobson and Smolin \cite{JScqg}
and Samuel \cite{Sam} independently found a covariant action that yields
Ashtekar's new variables when one performs a 3+1 decomposition. This action is
the Palatini action for complex \gr viewed as a functional of a complex
co-tetrad and a {\it self-dual} connection 1-form.\footnote{To recover the
phase space variables for the real theory, one must impose {\it reality
conditions} to select a real section of the complex phase space.} In one sense,
it is somewhat surprising that these new variables could capture the full
content of Einstein's equation since they involve only half (i.e., the \sd
part) of a Lorentz connection 1-form. On the other hand, the special role that
\sd fields play in the theory of \gr was already evident in the work of Newman,
Penrose, and Plebanski on \sd solutions to Einstein's equation. In fact, much
of this earlier work provided the motivation for Ashtekar's search for the new
variables.

Not only did the new variables give \gr a connection dynamic description; they
also simplified the field equations of the theory---particularly the
constraints. In terms of the standard geometrodynamical variables $(q_{ab},\otw
p^{ab})$, the constraint equations are non-polynomial. However, in terms of the
new variables, the constraint equations become {\it polynomial}. This result
has rekindled interest in the canonical quantization program for 3+1 gravity.
Due to the simplicity of the constraint equations in terms of these new
variables, Jacobson, Rovelli, and Smolin \cite{JSnp,RS} and a number of other
researchers have been able to solve the quantum constraints exactly. Although
the quantization program has not yet been completed, the above results
constitute promising first steps in that direction.

The \P and \sd theories described above were attempts to give \gr in 3+1
dimensions a connection dynamic description. A few years later, Witten
\cite{Witten} considered the 2+1 theory of gravity. He was able to show that
this theory simplifies considerably when expressed in \P form. In fact, Witten
demonstrated that the 2+1 \P theory for vacuum 2+1 gravity was equivalent to
\CS theory based on the inhomogeneous Lie group $ISO(2,1)$.\footnote{\CS
theory, like \YM theory, is a theory of a connection 1-form. However, unlike
\YM theory, it is defined only in odd dimensions and does not require the
introduction of a \st metric.} He then used this fact to quantize the theory.
This result startled both relativists and field theorists alike: relativists,
since the Wheeler-DeWitt equation in geometrodynamics is as hard to solve in
2+1 dimensions as it is in 3+1 dimensions; field theorists, since a simple
power counting argument had shown that perturbation theory for 2+1 gravity
around a flat background metric is non-renormalizable---just as it is for the
3+1 theory. The success of canonical quantization and failure of perturbation
theory in 2+1 dimensions came as a welcome surprise. Despite key differences
between 2+1 and 3+1 gravity (in particular, the lack of local degrees of
freedom for 2+1 vacuum solutions), Witten's result has proven to be useful to
non-perturbative approaches to 3+1 quantum gravity. In particular, since the
overall structure of 2+1 and 3+1 gravity are the same (e.g., they are both
diffeomorphism invariant theories, there is no background time, and the
dynamics is generated in both cases by 1st class constraints), researchers have
been able to use 2+1 gravity as a ``toy model" for the 3+1 theory \cite{toy}.

Finally, the most recent developments relating geometrodynamics and connection
dynamics involve formulations of \gr that are {\it independent} of any metric
variable. This idea for 3+1 gravity dates back to Plebanski \cite{Plebanski},
and was recently developed fully by Capovilla, Dell, and Jacobson (CDJ)
\cite{grwm,sdtf,scfg,gcgt}. Shortly thereafter, Peld\'an \cite{Peldan} provided
a similar formulation for 2+1 gravity. These pure spin-connection formulations
of \gr are defined by actions that do not involve the \st metric $g_{ab}$ in
any way whatsoever---the action for the complex 3+1 theory depends only on a
connection 1-form (which takes values in the complexified Lie algebra of
$SO(3)$) and a scalar density of weight $-$1. Moreover, the Hamiltonian
formulation of this theory is the same as that of the \sd theory, and by using
their approach, CDJ have been able to write down the most general solution to
the 4 diffeomorphism constraint equations. Whether or not these results will
lead to new insights for the quantization of the 3+1 theory remains to be seen.

With this brief history of geometrodynamics and connection dynamics as
background, the purpose of this review can be stated as follows: It is to
describe in detail the theories mentioned above, and, in the process, clarify
the mathematical relationship between geometrodynamics and connection dynamics
in the context of the classical theories of 2+1 and 3+1 gravity. While
preparing the text, I made a conscious effort to make the presentation as
self-contained and internally consistent as possible. The calculations are
somewhat technical and rather detailed, but I have included many footnotes,
parenthetical remarks, and mathematical digressions to fill various gaps. I
felt that this style of presentation (as opposed to relegating the necessary
mathematics to appendices at the end of the paper) was more in keeping with the
natural interplay between mathematics and physics that occurs when one works on
an actual research problem. Also, I felt that the added details would be of
value to anyone interested in working in this area.

In Section 2, I recall the standard \EH theory and take some time to introduce
the notation and mathematical techniques that I will use repeatedly throughout
the text. Although this section is a review of fairly standard material,
readers are encouraged to at least skim through the pages to acquaint
themselves with my style of presentation. In Sections 3 and 4, I restrict
attention to 2+1 dimensions and describe the 2+1 \P and \CS theories and
demonstrate the relationship between them.  In Section 5, I couple a \cc and a
massless scalar field to the 2+1 \P theory.  2+1 \P theory coupled to a \cc
$\L$ is of interest since we shall see that the equivalence between the 2+1 \P
and \CS theories continues to hold even if $\L\not=0$; 2+1 \P theory coupled to
a massless scalar field is of interest since it is the dimensional reduction of
3+1 vacuum \gr with a spacelike, hypersurface-orthogonal Killing vector field
(see, e.g., Chapter 16 of \cite{solns}). In fact, recent work in progress (by
Ashtekar and Varadarajan) in the hamiltonian formulation of this reduced theory
indicates that its non-perturbative quantization is likely to be successful. In
Sections 6 and 7, I turn my attention to 3+1 dimensions and describe the 3+1
Palatini and \sd theories. In Section 8, I couple a \cc and a \YM field to 3+1
gravity.  Section 9 describes a pure spin-connection formulation of 3+1
gravity, and Section 10 concludes with a brief summary and discussion of the
results. All of the above theories are specified by an action. I obtain the \EL
\eoms by varying the action and perform a Legendre transform to put each theory
in Hamiltonian form.  I emphasize the similarities, differences, and
equivalences of the various theories whenever possible. While this paper is
primarily a review, some of the material is in fact new, or at least has not
appeared in the literature in the form given here.  Much of Sections 3, 4, and
5 on the 2+1 theory fall in this category.

I should also list a few of the topics that are not covered in this review.
First, I have restricted attention to the more ``standard" theories of 2+1 and
3+1 gravity.  I have made no attempt to treat higher-derivative theories of
gravity, supersymmetric theories, or any of their equivalents. Second, I have
chosen to omit any discussion of quantum theory, although it is here, in
quantum theory, that the change in emphasis from geometrodynamics to connection
dynamics has had the most success. All of the theories described in this paper
are treated at a purely classical level; issues related, for instance, to
quantum cosmology and the non-perturbative canonical quantization program for
3+1 gravity are not dealt with.  This review serves, instead, as a
thorough pre-requisite for addressing the above issues. Moreover, many books
and review articles already exist which discuss the quantum theory in great
detail. Interested readers should see, in addition to the text books
\cite{npcg,poona}, review articles \cite{abhay,lee,carlo,kodama} and references
mentioned therein. Third, in 2+1 dimensions, I have chosen to concentrate on
the relationship between the 2+1 \P and \CS theories, and have all but ignored
the equally interesting relationships between these formulations and the
standard 2+1 dimensional \EH theory. Fortunately, other researchers have
already addressed these issues, so interested readers can find details in
\cite{Moncrief,HN1,HN2}.  Also, since \CS theory is not available in 3+1
dimensions, the equivalence of the 2+1 \P and \CS theories does not have a
direct 3+1 dimensional analog.  However, recent work by Carlip
\cite{Carlip1,Carlip2} and Anderson \cite{Anderson} on the problem of time in
2+1 quantum gravity may shed some light on the corresponding issue facing the
3+1 theory. Finally, Section 9 on \gr without-the metric deals exclusively with
3+1 gravity. Readers interested in a pure spin-connection formulation of 2+1
gravity should see \cite{Peldan}.

Penrose's abstract index notation will be used throughout. Spacetime and
spatial tensor indices are denoted by latin letters from the beginning of the
alphabet $a, b, c,\cdots$ , while internal indices are denoted by latin letters
from the middle of the alphabet $i, j, k,\cdots$ or $I, J, K,\cdots$ . The
signature of the \st metric $g_{ab}$ is taken to be $(-++)$ or $(-+++)$,
depending on whether we are working in 2+1  or 3+1 dimensions. If $\grad_a$
denotes the unique, torsion-free \st derivative operator compatible with the
\st metric $g_{ab}$, then $R_{abc}{}^d k_d:=2\grad_{[a}\grad_{b]}k_c$,
$R_{ab}:=R_{acb}{}^c$, and $R:=R_{ab}g^{ab}$ define the {\it Riemann tensor},
{\it Ricci tensor}, and {\it scalar curvature} of $\grad_a$.

Finally, since I eventually want to obtain a Hamiltonian formulation for each
theory, I will assume from the beginning that the \st manifold $M$ is
topologically $\S\X R$. If the theory depends on a \st metric, I assume $\S$
to be spacelike; \ if the theory does not depend on a \st metric, I assume $\S$
to be any (co-dimension 1) submanifold of $M$. In either case, I ignore all
surface integrals and avoid any discussion of boundary conditions. In this
sense, the results I obtain are rigorous only for the case when $\S$ is
compact. Readers interested in a detailed discussion of the technically more
difficult asymptotically flat case (in the context of the standard \EH or \sd
theories) should see Chapters II.2 and III.2 of \cite{npcg}.
\vskip 1cm

\noindent{\bf 2. Einstein-Hilbert theory}
\vskip .5cm

In this section, we will describe the standard \EH theory. We obtain the
vacuum Einstein's equation starting from an action principle and perform a
Legendre transform to put the theory in Hamiltonian form. We shall see that the
phase space variables consist of a positive-definite metric $q_{ab}$ and a
density-weighted, symmetric, second-rank tensor field $\otw p^{ab}$. These are
the standard {\it geometrodynamical} variables of general relativity. We will
also analyze the motions on phase space generated by the constraint functions
and evaluate their Poisson bracket algebra. This section is basically a review
of standard material.  Our treatment will follow that given, for example, in
Appendix E of \cite{Wald} or Chapter II.2 of \cite{npcg}.

The standard \EH theory is, of course, valid in $n$+1 dimensions. Everything we
do in this section will be independent of the dimension of the \st manifold
$M$. This is an important feature which will allow us to compare the standard
\EH theory with the \CS and \sd theories. Unlike the standard \EH theory, \CS
theory is defined only in odd dimensions, while the \sd theory is defined only
in 3+1 dimensions. \vskip .5cm

\noindent{\sl 2.1 \EL equations of motion}
\medskip

Let us begin with the well-known {\it \EH action}
$$S_{EH}(g^{ab}):=\int_M\sqrt{-g}R.\eqno(2.1)$$
Here $g$ denotes the determinant of the covariant metric $g_{ab}$, and $R$
denotes the scalar curvature of the unique, torsion-free \st derivative
operator $\grad_a$ compatible with $g_{ab}$. I have taken the basic variable to
be the contravariant spacetime metric $g^{ab}$ for convenience when performing
variations of the action. The \EH action is {\it second-order} since $R$
contains second derivatives of $g_{ab}$.

To obtain the \EL equations of motion, we vary the action \wrt the field
variable $g^{ab}$. If we write the integrand as $\sqrt{-g}R_{ab} g^{ab}$ and
use the fact that $\d g=-g\ g_{ab}\d g^{ab}$, we get
$$\d S_{EH}=\int_M \sqrt{-g}(R_{ab} - {1\over 2} R g_{ab})\d g^{ab} + \int_M
\sqrt{-g}\d R_{ab} g^{ab}.\eqno(2.2)$$
The first integral is of the desired form, while the second integral requires
us to evaluate the variation of the Ricci tensor $R_{ab}$. Since one can show
that\footnote{To obtain this result, consider a 1-parameter family of \st
metrics $g_{ab}(\l)$ and their associated \st derivative operators
${}^\l\grad_a$. Define $C_{ab}{}^c$ by ${}^\l\grad_a k_b=:\grad_a k_b+\l
C_{ab}{}^c k_c$ and differentiate ${}^\l\grad_a g_{bc}(\l)=0$ \wrt $\l$.
Evaluating this expression at $\l=0$ gives $C_{ab}{}^c=-{1\over
2}g^{cd}(\grad_a\d g_{bd} + \grad_b\d g_{ad} - \grad_d\d g_{ab})$, where
$g_{ab}:=g_{ab}(0)$ and $\d g_{ab}:={d\over d\l} \big|_{\l=0} g_{ab}(\l)$.
Since $R_{abc}{}^d(\l)=R_{abc}{}^d + \l\ 2\grad_{[a} C_{b]c}{}^d + \l^2\
[C_a,C_b]_c{}^d$, it follows that $\d R_{ac}:={d\over d\l}\big|_{\l =0}
R_{abc}{}^b(\l)=2\grad_{[a}C_{b]c}{}^b$. Contracting with $g^{ac}$ (using $\d
g^{ab}=-g^{ac}g^{bd}\d g_{cd}$) yields the above result.}
$$\d R_{ab} g^{ab}=\grad_a v^a\eqno(2.3)$$
(where $v^a=\grad^a(g_{bc}\d g^{bc})-\grad_b\d g^{ab}$), we see that modulo a
surface integral, $\d S_{EH}=0$ \iff
$$G_{ab}:=R_{ab} - {1\over 2}Rg_{ab}=0.\eqno(2.4)$$
This is the desired result:  \ The vacuum Einstein's equation can be obtained
starting from an action principle.

I should note that, strictly speaking, the variation of (2.1) \wrt $g^{ab}$
does not yield the vacuum Einstein's equation $G_{ab}=0$.  The surface integral
does not vanish since $v^a$ involves derivatives of the variation $\d g^{ab}$.
Even though $\d g^{ab}$ is required to vanish on the boundary, these
derivatives need not vanish. This seems to pose a potential problem, but it can
handled by simply adding to (2.1) a boundary term which will (upon variation)
exactly cancel the surface integral. As shown in Appendix E of \cite{Wald},
this boundary term involves the trace of the extrinsic curvature of the
boundary of $M$.  For the sake of simplicity, however, we will continue to use
the unmodified \EH action (2.1) and ignore all surface integrals as mentioned
at the end of Section 1.
\vskip .5cm

\noindent{\sl 2.2 Legendre transform}
\medskip

To put the standard \EH theory in Hamiltonian form, we will follow the usual
procedure: \ We assume that $M=\S \times R$ for some spacelike submanifold
$\S$ and assume that there exists a time function $t$ (with nowhere vanishing
gradient $(dt)_a$) such that each $t=\const$ surface $\S_t$ is diffeomorphic
to $\S$. To talk about evolution from one $t=\const$ surface to the next, we
introduce a future-pointing timelike vector field $t^a$ satisfying $t^a(dt)_a
=1$. \ $t^a$ is the ``time flow'' vector field that defines the same point in
space at different instants of time.  We will treat $t^a$ and the foliation of
$M$ by the $t=\const$ surfaces as kinematical (i.e., non-dynamical) structure.
Evolution will be given by the Lie derivative \wrt $t^a$.

Since we have a \st metric $g_{ab}$ as one of our field variables, we can also
introduce a {\it unit} covariant normal $n_a$ and its associated
future-pointing timelike vector field $n^a=g^{ab}n_b$. Note that since $n^a
n_a=-1$, \ $q_b^a:=\d_b^a+n^a n_b$ is a projection operator into the $t=\const$
surfaces. We will construct the configuration variables associated with the
field variable $g^{ab}$ by contracting with $n^a$ and $q_b^a$. We define the
{\it induced metric} $q_{ab}$, the {\it lapse} $N$, and {\it shift} $N^a$ via
$$\eqalign{&q_{ab}:=q_a^m q_b^n\ g_{mn}\ (=g_{ab}+n_a n_b),\cr
&N:=-n^a t^b\ g_{ab},\quad{\rm and}\cr
&N^a:=q_b^a\ t^b.\cr}\eqno(2.5)$$
Note that in terms of $N$ and $N^a$, we can write $t^a=N n^a + N^a$.
Furthermore, since $N^a n_a=0$ and $q_{ab}n^a=0$, \ $N^a$ and $q_{ab}$ are (in
1-1 correspondence with) tensor fields defined intrinsically on $\S$.

The next step in constructing a Hamiltonian formulation of the \EH theory is to
decompose the \EH action and write it in the form
$$S_{EH}(g^{ab})=\int dt\ L_{EH}(q,\dot q).\eqno(2.6)$$
$L_{EH}$ will be the \EH Lagrangian provided it depends only on
$(q_{ab},N,N^a)$ and their first time derivatives. But, as written, (2.1) is
not convenient for such a decomposition. The integrand $\sqrt{-g}R$ contains
second time derivatives of the configuration variable $q_{ab}$. However, as we
will now show, these terms can be removed from the integrand by subtracting off
a total divergence.

To see this, let us write the scalar curvature $R$ as $R=2(G_{ab} - R_{ab})n^a
n^b$. Then the differential geometric identities
$$\eqalign{&G_{ab}n^a n^b={1\over 2}(\cR-K_{ab}K^{ab}+K^2)\quad{\rm and}\cr
&R_{ab}n^a n^b=-K_{ab}K^{ab}+K^2+\grad_b(n^a\grad_a n^b-n^b\grad_a n^a)\cr}
\eqno(2.7)$$
(where $K_{ab}:=q_a^m q_b^n\ \grad_m n_n$ is the {\it extrinsic curvature} of
the $t=\const$ surfaces and $\cR$ is the scalar curvature of the unique,
torsion-free spatial derivative operator $D_a$ compatible with the induced
metric $q_{ab}$) imply
$$R=(\cR+K_{ab}K^{ab}-K^2)+({\rm total\ divergence\ term}).\eqno(2.8)$$
Using the fact that $\sqrt{-g}=N\sqrt q\ dt$ (where $q$ denotes the
determinant of $q_{ab}$), the \EH action (2.1) becomes
$$S_{EH}(g^{ab})=\int dt\int_\S\sqrt q N(\cR + K_{ab} K^{ab} - K^2) +
({\rm surface\ integral}).\eqno(2.9)$$
If we ignore the surface integral, we get
$$L_{EH}=\int_\S\sqrt q N(\cR + K_{ab} K^{ab} - K^2).\eqno(2.10)$$
This is the desired \EH Lagrangian first proposed by Arnowitt, Deser, and
Misner (ADM) \cite{ADM}. The identity $K_{ab}={1\over 2N}(\cL_{\vec t} \
q_{ab}-2D_{(a}N_{b)})$ allows us to express $L_{EH}$ in terms of only
$(q_{ab},N,N^a)$ and their first time derivatives.

Given the \EH Lagrangian, we are now ready to perform the Legendre transform.
But before we do this, it is probably worthwhile to make a detour and first
review the standard Dirac constraint analysis for a theory with constraints
and recall some basic ideas of symplectic geometry.  I propose to examine, in
detail, a simple finite-dimensional system described by a Lagrangian
$$L(q,\dot q):={1\over 2}\dot q_1{}^2+q_3\dot q_2-q_4 f(q_2,q_3).\eqno(2.11)$$
Here $(q_1,\cdots,q_4)\in\cC_0$ are the configuration variables and $(\dot
q_1,\cdots, \dot q_4)$ are their associated time derivatives (or velocities). \
$f(q_2,q_3)$ can be any (smooth) real-valued function of $(q_2,q_3)$. The
techniques that arise when analyzing this simple system will apply not only to
the standard \EH theory but to many other constrained theories as well. Readers
interested in a more detailed description of the general Dirac constraint
analysis and symplectic geometry should see \cite{Dirac} and Appendix B of
\cite{poona}, respectively.  Readers already familiar with the standard Dirac
constraint analysis may skip to the paragraph immediately following equation
(2.25).

To perform the Legendre transform for our simple system, we first define the
{\it momentum variables} $(p_1,\cdots,p_4)$ via
$$p_{\und i}:={\d L\over\d\dot q_{\und i}}\qquad({\und i}=1,\cdots,4).
\eqno(2.12)$$
For the special form of the Lagrangian given above, they become
$$p_1=\dot q_1,\quad p_2=q_3,\quad p_3=0,\quad{\rm and}\quad p_4=0.\eqno(2.13)
$$
Since only the first equation can be inverted to give $\dot q_1$ as a function
of $(q,p)$, there are constraints:  Not all points in the phase space
$\G_0=T^*\cC_0=\{(q_{\und i},p_{\und i})| \ {\und i}=1,\cdots,4\}$ are
accessible to the system. Only those $(q,p)\in\G_0$ which satisfy
$$\phi_1:=p_2-q_3=0,\quad \phi_2:=p_3=0,\quad{\rm and}\quad\phi_3:=p_4=0
\eqno(2.14)$$
are physically allowed. The $\phi_{\und i}$'s are called {\it primary
constraints} and the vanishing of these functions define a {\it constraint
surface} in $\G_0$. It is the presence of these constraints that complicates
the standard Legendre transform.

Following the Dirac constraint analysis, we now must now write down a
Hamiltonian for the theory. But due to (2.14), the Hamiltonian will not
be unique. The usual definition $H_0(q,p):=\sum_{\und i=1}^4 p_{\und i}\dot
q_{\und i}-L(q, \dot q)$ does not work, since there exist $\dot q_{\und i}$'s
which cannot be written as functions of $q$ and $p$. If, however, we restrict
ourselves to the constraint surface defined by (2.14), we have
$$H_0(q,p)={1\over 2} p_1{}^2+q_4 f(q_2,q_3).\eqno(2.15)$$
Since the right hand side of (2.15) makes sense on all of $\G_0$, $H_0(q,p)$
actually defines one possible choice of Hamiltonian. However, as we will show
below, this Hamiltonian is definitely not the only one.

For suppose $\l_1$, $\l_2$, and $\l_3$ are three arbitrary functions on $\G_0$.
Then
$$\eqalign{H_T(q,p):&=H_0(q,p)+\l_1\phi_1+\l_2\phi_2+\l_3\phi_3\cr
&={1\over 2}p_1{}^2+q_4 f(q_2,q_3)+\l_1(p_2-q_3)+\l_2 p_3+\l_3 p_4\cr}
\eqno(2.16)$$
is another function (defined on all of $\G_0$) that agrees with $H_0(q,p)$ on
the constraint surface.\ $H_T(q,p)$ is called the {\it total Hamiltonian}, and
it differs from $H_0(q,p)$ by terms that vanish on the constraint surface. This
non-uniqueness of the total Hamiltonian exists for any theory that has
constraints.

Given $H_T(q,p)$, the next step in the Dirac constraint analysis is to require
that the primary constraints (2.14) be preserved under time evolution---i.e.,
that
$$\dot\phi_{\und i}:=\{\phi_{\und i},H_T\}_0\approx 0\qquad({\und i}=1,2,3).
\eqno(2.17)$$
Here $\approx$ means equality on the constraint surface defined by (2.14) and
$\{\ ,\ \}_0$ denotes the {\it \Pb} defined by the natural {\it symplectic
structure}\footnote{A {\it symplectic manifold} (or {\it phase space}) consists
of a pair $(\G_0,\W_0)$, where $\G_0$ is an even dimensional manifold and
$\W_0$ is a closed and non-degenerate 2-form. (i.e., $d\W_0=0$ and
$\W_0(v,w)=0$ for all $w$ implies $v=0$.) $\W_0$ is called the {\it symplectic
structure} and it allows us to define Hamiltonian vector fields and Poisson
brackets: Given any real-valued function $f:\G_0\to R$, the {\it Hamiltonian
vector field} $X_f$ is defined by $-i_{X_f}\W_0:=df$. Given any two real-valued
functions $f,g:\G_0\to  R$, the {\it Poisson bracket} $\{f,g\}_0$ is defined
by $\{f,g\}_0:=-\W(X_f,X_g)= -X_f(g)$. As a special case, if $\G_0=T^*\cC_0$ is
the cotangent bundle over some $n$-dimensional configuration space $\cC_0$,
then $\W_0=dp_1\wedge dq_1+\cdots+dp_n\wedge dq_n$ is the natural symplectic
structure on $\G_0$ associated with the chart $(q,p)$. It follows that
$\{f,g\}_0=\sum_{{\und i}=1}^n ({\pd f\over \pd q_{\und i}}{\pd g \over \pd
p_{\und i}}-{\pd f\over \pd p_{\und i}}{\pd g\over \pd q_{\und i}})$, which is
the standard textbook expression for the Poisson bracket of $f$ and $g$.}
$$\W_0=dp_1\wedge dq_1+dp_2\wedge dq_2+dp_3\wedge dq_3+dp_4\wedge dq_4
\eqno(2.18)$$
on $\G_0$. Equation (2.17) is equivalent to the requirement that the evolution
of the system take place on the constraint surface.

Evaluating (2.17) for the primary constraints (2.14), we find that
$\{\phi_3,H_T \}_0\approx 0$ implies
$$\phi_4:=f(q_2,q_3)\approx 0.\eqno(2.19)$$
The other Poisson brackets yield conditions on $\l_1$ and $\l_2$. \ $\phi_4$ is
called a {\it secondary constraint}, and for consistency we must also require
that
$$\dot\phi_4:=\{\phi_4,H_T\}_0\approx 0.\eqno(2.20)$$
Here $\approx$ now means equality on the constraint surface defined by (2.14)
and (2.19). Since one can show that (2.20) follows from the earlier conditions
on $\l_1$ and $\l_2$, \ (2.14) and (2.19) constitute all the constraints of the
theory.

The final step in the Dirac constraint analysis is to take all the constraints
$\phi_1,\cdots,\phi_4$ and evaluate their Poisson brackets. If a constraint
$\phi_{\und i}$ satisfies $\{\phi_{\und i},\phi_{\und j}\}_0\approx 0$ for all
$\phi_{\und j}$, then $\phi_{\und i}$ is said to be {\it 1st class}. If,
however, $\{\phi_ {\und i}, \phi_{\und j}\}_0\not\approx 0$ for some
$\phi_{\und j}$, then $\phi_{\und i}$ and $\phi_{\und j}$ are said to form a
{\it 2nd class} pair. (In terms of symplectic structures and Hamiltonian vector
fields, a constraint $\phi_{\und i}$ is 1st class \wrt the symplectic structure
$\W_0$ \iff the Hamiltonian vector field $X_{\phi_{\und i}}$ defined by $\W_0$
is tangent to the constraint surface defined by the vanishing of all the
constraints.) The goal is to solve all the 2nd class constraints (and possibly
some 1st class constraints) and obtain a new phase space $(\G,\W)$ where the
remaining constraints (pulled-back to $\G$) are all 1st class \wrt the \Pb
defined by $\W$. Evaluating $\{\phi_{\und i},\phi_{\und j}\}_0$ for our simple
system, we find that $\phi_3$ is the only 1st class constraint \wrt $\W_0$. By
solving the second class pair $\phi_1=0$ and $\phi_2=0$, we get
$$\W_0\Big|_{\phi_1=0,\ \phi_2=0}=dp_1\wedge dq_1+dq_3\wedge dq_2+dp_4\wedge
dq_4\eqno(2.21)$$
and
$$H_T\Big|_{\phi_1=0,\ \phi_2=0}={1\over 2}p_1{}^2+q_4 f(q_2,q_3)+\l_3
p_4.\eqno(2.22)$$
The remaining constraints $\phi_3$ and $\phi_4$ are now both 1st class \wrt
this new symplectic structure.

Although we have successively eliminated all the 2nd class constraints, we can
go one step further. We can solve the 1st class constraint $\phi_3:=p_4=0$
by gauge fixing the configuration variable $q_4$.  Even though this step is not
required by the Dirac constraint analysis, it simplifies the final phase space
structure somewhat. Solving $\phi_3=0$ and pulling-back (2.21) and (2.22) to
this new constraint surface $\G$ (coordinatized by $(q_1,q_2,q_3,p_1)$), we
obtain
$$\W:=dp_1\wedge dq_1+dq_3\wedge dq_2\eqno(2.23)$$
and
$$H(q_1,q_2,q_3,p_1):={1\over 2}p_1{}^2+q_4 f(q_2,q_3).\eqno(2.24)$$
Here $q_4$ is no longer thought of as a dynamical variable---it is a {\it
Lagrange multiplier} of the theory associated with the 1st class constraint
$f(q_2,q_3)=0$.

To summarize: Given a Lagrangian of the form
$$L(q,\dot q):={1\over 2}\dot q_1{}^2+q_3\dot q_2-q_4 f(q_2,q_3),\eqno(2.25)$$
the Dirac constraint analysis says that the momentum $p_1$ is unconstrained,
while $p_2=q_3$ and $p_3=p_4=0$.  Demanding that the constraints be preserved
under evolution, we obtain a secondary constraint $f(q_2,q_3) =0$. The
constraints $p_2-q_3=0$ and $p_3=0$ form a 2nd class pair and are easily
solved; the remaining constraints $p_4=0$ and $f(q_2,q_3)=0$ now form a 1st
class set. By gauge fixing $q_4$ we can solve $p_4=0$, and thus obtain a new
phase space $(\G,\W)$ coordinatized by $(q_1,q_2,q_3,p_1)$ with symplectic
structure (2.23) and Hamiltonian (2.24). We are left with a single 1st class
constraint, $f(q_2,q_3)=0$.

Let us now return to our analysis of the standard \EH theory. Given $L_{EH}$,
we find that the momentum $\otw p^{ab}$ canonically conjugate to $q_{ab}$ is
given by
$$\otw p^{ab}:={\d L_{EH}\over\d\cL_{\vec t} \ q_{ab}}=\sqrt q(K^{ab}-K
q^{ab}),\eqno(2.26)$$
while the momenta canonically conjugate to $N$ and $N^a$ are zero. Since
equation (2.26) can be inverted to give
$$\cL_{\vec t} \ q_{ab}=2N q^{-1/2}(\otw p_{ab}-{1\over 2}\otw p q_{ab})
+2D_{(a}N_{b)},\eqno(2.27)$$
it does not define a constraint. However, $N$ and $N^a$ play the role of
Lagrange multipliers.

Thus, by following the Dirac constraint analysis we find that the phase space
$(\G_{EH},\W_{EH})$ of the standard \EH theory is coordinatized by the pair
$(q_{ab},\otw p^{ab})$ and has symplectic structure\footnote{I use \ $\id$ \
and $\ \iwedge\ $ to denote the infinite-dimensional exterior derivative and
infinite-dimensional wedge product of forms on $\G_{EH}$. They are to be
distinguished from $d$ and $\wedge$ which are the finite-dimensional exterior
derivative and finite-dimensional wedge product of forms on $\S$. Note that in
terms of the Poisson bracket $\{\ ,\ \}$ defined by $\W_{EH}$, we have
$\{q_{ab}(x),\otw p^{cd}(y)\}= \d_{(a}^c\d_{b)}^d\d(x,y)$.}
$$\W_{EH}=\int_\S\id\otw p^{ab}\ \iwedge\ \id q_{ab}.\eqno(2.28)$$
The Hamiltonian is given by
$$H_{EH}(q,\otw p)=\int_\S N\Big(-q^{1/2}\cR + q^{-1/2}(\otw p^{ab}\otw p_{ab}
-{1\over 2}\otw p^2)\Big)-2N^a q_{ab}D_c \otw p^{bc}.\eqno(2.29)$$
As we shall see in the next subsection, this is just a sum of 1st class
constraint functions associated with
$$\eqalignnotwo{&\otw C(q,\otw p):=-q^{1/2}\cR + q^{-1/2}(\otw p^{ab}\otw
p_{ab}-{1\over 2}\otw p^2)\approx 0\quad{\rm and}&(2.30a)\cr
&\otw C_a(q,\otw p):=-2q_{ab}D_c \otw p^{bc}\approx 0.&(2.30b)
\cr}$$
Note that constraint equation ($2.30a$) is {\it non-polynomial} in the
canonically conjugate variables due to the dependence of $\cR$ on the inverse
of $q_{ab}$. This is a major stumbling block for the canonical quantization
program in terms of $(q_{ab},\otw p^{ab})$. To date, there exist no exact
solutions to the quantum version of this constraint in full (i.e.,
non-truncated) general relativity.
\vskip .5cm

\noindent{\sl 2.3 Constraint algebra}
\medskip

To evaluate the Poisson brackets of the constraints and to determine the
motions they generate on phase space, we must first construct {\it constraint
functions} (i.e., mappings $\G_{EH}\rightarrow R$) associated with the
constraint equations ($2.30a$) and ($2.30b$). To do this, we smear $\otw
C(q,\otw p)$ and $\otw C_a(q,\otw p)$ with test fields $N$ and $N^a$ on
$\S$---i.e., we define
$$\eqalignnotwo{&C(N):=\int_\S N\Big(-q^{1/2}\cR + q^{-1/2}(\otw p^{ab}\otw
p_{ab}-{1\over 2}\otw p^2)\Big)\quad{\rm and}&(2.31a)\cr
&C(\vec N):=\int_\S-2N^a q_{ab}D_c\otw p^{bc}. &(2.31b)\cr}$$
They are called the {\it scalar} and {\it vector constraint functions} of the
standard \EH theory.

The next step is to evaluate the functional derivatives of $C(N)$ and $C(\vec
N)$. For recall that if $f,g:\G_{EH}\rightarrow R$ are any two real-valued
functions on phase space, the Hamiltonian vector field $X_f$ (defined by the
symplectic structure (2.28)) is given by
$$X_f=\int_\S{\d f\over\d\otw p^{ab}}{\d\over\d q_{ab}} - {\d f\over\d q_{ab}}
{\d\over\d\otw p^{ab}}\eqno(2.32)$$
and the Poisson bracket $\{f,g\}$ \ (defined by $\{f,g\}:=-X_f(g)$) is given by
$$\{f,g\}=\int_\S{\d f\over\d q_{ab}}{\d g\over\d\otw p^{ab}}-{\d f\over\d
\otw p^{ab}}{\d g\over\d q_{ab}}.\eqno(2.33)$$
Note that under the 1-parameter family of diffeomorphisms on $\G_{EH}$
associated with $X_f$,
$$\eqalignnotwo{&q_{ab}\mapsto q_{ab}+\e{\d f\over\d\otw p^{ab}}+O(\e^2)\quad
{\rm and}&(2.34a)\cr
&\otw p^{ab}\mapsto\otw p^{ab}-\e{\d f\over\d q_{ab}}+O(\e^2).
&(2.34b)\cr}$$
We will use (2.33) to determine the various Poisson brackets between $C(N)$ and
$C(\vec N)$; we will use ($2.34a$) and ($2.34b$) to determine the motions that
they generate on phase space.

Let us begin with the vector constraint $C(\vec N)$. Integrating ($2.31b$) by
parts and noting that $2D_{(a}N_{b)}=\cL_{\vec N}q_{ab}$, we get
$$C(\vec N)=\int_\S(\cL_{\vec N}q_{ab})\otw p^{ab}\ \ \ \left(=-\int_\S q_{ab}
(\cL_{\vec N}\otw p^{ab})\right).\eqno(2.35)$$
By inspection,
$${\d C(\vec N)\over\d q_{ab}}=-\cL_{\vec N}\otw p^{ab}\quad{\rm and}\quad
{\d C(\vec N)\over\d\otw p^{ab}}=\cL_{\vec N}q_{ab}.\eqno(2.36)$$
Thus, we see that
$$\eqalignnotwo{&q_{ab}\mapsto q_{ab} + \e\cL_{\vec N} q_{ab} + O(\e^2)\quad
{\rm and}&(2.37a)\cr
&\otw p^{ab}\mapsto \otw p^{ab} + \e\cL_{\vec N}\otw p^{ab} +
O(\e^2)&(2.37b) \cr}$$
is the motion on $\G_{EH}$ generated by $C(\vec N)$. Note that ($2.37a$) and
($2.37b$) are the maps on the tensor fields $q_{ab}$ and $\otw p^{ab}$ induced
by the 1-parameter family of diffeomorphisms on $\S$ associated with the vector
field $N^a$. In other words, the Hamiltonian vector field $X_{C(\vec N)}$ on
$\G_{EH}$ is the {\it lift} of the vector field $N^a$ on $\S$.

Let us now consider the scalar constraint $C(N)$. Due to the non-polynomial
dependence of $\cR$ on $q_{ab}$, the functional derivative $\d C(N)/\d q_{ab}$
is much harder to evaluate. After a fairly long calculation, one finds
that\footnote{To obtain this result we used the facts that $\d q=q\
q^{ab}\d q_{ab}$ and $\d\cR_{ab} q^{ab}=D_a v^a$ for $v^a=-D^a(q^{bc}\d
q_{bc})+D^b(q^{ac} \d q_{bc})$. These are just the spatial analogs of the
results used in subsection 2.1 when we varied the \EH action \wrt $g^{ab}$.}
$$\eqalign{{\d C(N)\over \d q_{ab}}=-{1\over 2}N\otw C(q,\otw p)&q^{ab}+2N
q^{-1/2}(\otw p^{ac}\otw p^b{}_c - {1\over 2}\otw p\otw p^{ab})\cr
&+N q^{1/2}(\cR^{ab}-\cR q^{ab}) - q^{1/2}(D^aD^b N - q^{ab} D^c D_c N).\cr}
\eqno(2.38)$$
A much simpler calculation gives
$${\d C(N)\over\d\otw p^{ab}}=2N q^{-1/2}(\otw p_{ab}-{1\over 2}\otw p q_{ab}).
\eqno(2.39)$$
Recall that for the vector constraint function $C(\vec N)$, the motion on
$\G_{EH}$ along $X_{C(\vec N)}$ corresponded to the Lie derivative of $q_{ab}$
and $\otw p^{ab}$ \wrt $N^a$. Thus, one might expect the motion on $\G_{EH}$
along $X_{C(N)}$ to correspond to the Lie derivative \wrt $t^a:=N n^a$. We will
now show that if we restrict ourselves to the constraint surface
$\ovr\G_{EH}\subset\G_{EH}$ (defined by ($2.30a$) and ($2.30b$)), then this is
actually the case.

Comparing (2.39) with equation (2.27) (setting $N^a=0$), we see that $\d
C(N)/\d\otw p^{ab}=\cL_{\vec t} \  q_{ab}$, so
$$q_{ab}\mapsto q_{ab}+\e\cL_{\vec t} \  q_{ab}+O(\e^2)\eqno(2.40)$$
as conjectured. Similarly, writing $\otw p^{ab}=\sqrt q(K^{ab}-K q^{ab})$ and
using the differential geometric identity
$$\cL_{N\vec n}K_{ab}=-N\cR_{ab}+2NK_a{}^c K_{bc}-NK K_{ab}+D_aD_b N
\eqno(2.41)$$
(which holds in this form when $R_{ab}=0$), we see that
$${\d C(N)\over\d q_{ab}}=-\left({1\over 2}N\otw C(q,\otw p) q^{ab}
+\cL_{\vec t} \ \otw p^{ab}\right).\eqno(2.42)$$
Thus,
$$\otw p^{ab}\mapsto \otw p^{ab} + \e\left({1\over 2}N\otw C(q,\otw p)
q^{ab} +\cL_{\vec t} \ \otw p^{ab}\right) + O(\e^2).\eqno(2.43)$$
If we now restrict ourselves to $\ovr\G_{EH}$ (so that $\otw C(q,\otw p)=0$),
we get
$$\otw p^{ab}\mapsto \otw p^{ab} + \e\cL_{\vec t} \ \otw p^{ab} + O(\e^2).
\eqno(2.44)$$
This is the desired result: \ When restricted to $\ovr\G_{EH}\subset \G_{EH}$,
the Hamiltonian vector field $X_{C(N)}$ on $\ovr\G_{EH}$ is the lift of the
vector field $t^a:=Nn^a$ on $\S$.

We are now ready to evaluate the Poisson brackets between the constraint
functions. But first note that if $f(M):\G_{EH}\rightarrow R$ is any
real-valued function on phase space of the form
$$f(M):=\int_\S  M^{a\cdots b}{}_{c\cdots d}\ \otw f_{a\cdots b}{}^{c\cdots d}
(q,\otw p)\eqno(2.45)$$
(were $M^{a\cdots b}{}_{c\cdots d}$ is any tensor field on $\S$ which is {\it
independent} of $q_{ab}$ and $\otw p^{ab}$), then
$$\eqalign{\{C(\vec N),f(M)\}&=\int_\S-\cL_{\vec N}\otw p^{ab}\Big({\d f(M)
\over\d\otw p^{ab}}\Big)-\cL_{\vec N}q_{ab}\Big({\d f(M)\over\d q_{ab}}\Big)\cr
&=\int_\S-M^{a\cdots b}{}_{c\cdots d}\ \cL_{\vec N}\otw f_{a\cdots b}{}^{c
\cdots d}(q,\otw p).\cr}\eqno(2.46)$$
Integrating the last line of (2.46) by parts and throwing away the surface
integral, we get
$$\{C(\vec N),f(M)\}=f(\cL_{\vec N} M).\eqno(2.47)$$
Thus, the Poisson bracket of $C(\vec N)$ with any other constraint function is
easy to evaluate. We have
$$\eqalignnotwo{&\{C(\vec N),C(\vec M)\}=C([\vec N,\vec M])\quad{\rm and}
&(2.48a)\cr
&\{C(\vec N),C(M)\}=C(\cL_{\vec N} M),&(2.48b)\cr}$$
where $[\vec N,\vec M]:=\cL_{\vec N} M^a$ is the commutator of the vector
fields
$N^a$ and $M^a$ on $\S$. Note that ($2.48a$) tells us that the subset of vector
constraint functions is closed under Poisson brackets. In fact, $N^a\mapsto
C(\vec N)$ is a representation of the Lie algebra of vector fields on $\S$. The
commutator of vector fields on $\S$ is mapped to the Poisson bracket of the
corresponding vector constraint functions.

We are left with only the Poisson bracket $\{C(N),C(M)\}$ of two scalar
constraint functions to evaluate. Using (2.38) and (2.39) (and eliminating all
terms symmetric in $M$ and $N$), we get
$$\eqalign{\{C(N),C(M)\}&=\int_\S -2M(D^aD^b N-q^{ab} D^c D_c N)(\otw p_{ab}
-{1\over 2}\otw p q_{ab})-(N\leftrightarrow M)\cr
&=\int_\S -2(N\pd^aM-M\pd^aN) q_{ab} D_c\otw p^{bc}\cr
&=C(\vec K),\cr}\eqno(2.49)$$
where $K^a:=(N\pd^a M-M\pd^a N)=q^{ab}(N\pd_b M-M\pd_b N)$. Thus, the Poisson
bracket of two scalar constraints is a vector constraint. Although this implies
that the subset of scalar constraint functions is not closed under Poisson
bracket, the totality of constraint functions (scalar and vector) is---i.e.,
the constraint functions form a 1st class set as claimed in subsection 2.2.
Note, however, that since the vector field $K^a$ depends on the phase space
variable $q_{ab}$ (through its inverse), the Poisson bracket (2.49) involves
{\it structure functions}. The constraint functions do not form a Lie algebra.
\vskip 1cm

\noindent{\bf 3. 2+1 Palatini theory}
\vskip .5cm

In this section, we will describe the 2+1 \P theory which, as we shall see at
the end of subsection 3.2, is defined for {\it any} Lie group $G$. We will
discuss the relationship between the \P and \EH actions, and show how the
2+1 \P theory based on $SO(2,1)$ reproduces the standard results of
2+1 gravity. After performing a Legendre transform to put this theory in
Hamiltonian form, we shall see that the phase space variables consist of a
connection 1-form $A_a^I$ (which takes values in the Lie algebra of $G$) and
its canonically conjugate momentum (or ``electric field'') $\otw E_I^a$. Thus,
for $G=SO(2,1)$, the 2+1 \P theory gives us a connection dynamic description
of 2+1 gravity. The constraint equations are {\it polynomial} in the basic
variables and the constraint functions form a Lie algebra \wrt Poisson bracket.

Once we write the 2+1 \P action in its generalized form, we will let $G$ be
an arbitrary Lie group. To reproduce the results of 2+1 gravity, we simply take
$G$ to be $SO(2,1)$.  Note that much of the material in subsections 3.2 and 3.3
can also be found in \cite{CSP}.
\vskip .5cm

\noindent{\sl 3.1 \EL equations of motion}
\medskip

Recall the standard \EH action of Section 2,
$$S_{EH}(g^{ab})=\int_\S\sqrt{-g} R.\eqno(3.1)$$
To define the 2+1 \P action, it is convenient to first rewrite the integrand
$\sqrt{-g} R$ in {\it triad notation}. But in order to do this, we will have
to make a short mathematical digression. Readers interested in a more detailed
discussion of what follows should see \cite{AHM}. Readers already familiar
the method of orthonormal bases may skip to the paragraph immediately following
equation (3.9).

Consider an $n$-dimensional manifold $M$, and let $V$ be a fixed
$n$-dimensional
vector space with Minkowski metric $\n_{IJ}$ having signature $(-+\cdots +)$.
A {\it soldering form} at $p\in M$ is an isomorphism $e_a^I(p):T_pM\to V.$
(Here $T_pM$ denotes the tangent space to $M$ at $p$.) Although an $n$-manifold
does not in general admit a globally defined soldering form $e_a^I$, we
can use $e_a^I$ to define tensor fields locally on $M$.  For instance,
$$g_{ab}:=e_a^I e_b^J \n_{IJ}\eqno(3.2a)$$
is a (locally defined) \st metric having the same signature as $\n_{IJ}$. The
inverse of $e_a^I$ will be denoted by $e_I^a$; it satisfies
$$g_{ab} e_I^a e_J^b=\n_{IJ}.\eqno(3.2b)$$
Spacetime tensor fields with additional {\it internal} indices $I,J,K,\cdots$
will be called {\it generalized tensor fields} on $M$. Spacetime indices are
raised and lowered with the spacetime metric $g_{ab}$; internal indices are
raised and lowered with the Minkowski metric $\n_{IJ}$.

If one introduces a standard basis $\{b_{\und I}^I\ |\und I=1,\cdots,n\}$ in
$V$, then the vector fields $e_{\und I}^a:=e_I^a b_{\und I}^I$ form an {\it
orthonormal basis} of $g_{ab}$. These $n$-vector fields will be called a {\it
triad} when $n=3$ and a {\it tetrad} when $n=4$. The dual co-vector fields,
$e_a^{\und I} :=g_{ab}\n^{\und I\und J} e_{\und J}^b$, will be called a {\it
co-triad} and a {\it co-tetrad} when $n=3$ and 4, respectively. I should note,
however, that from now on I will ignore the distinction between a soldering
form $e_a^I$ and the co-vector fields $e_a^{\und I}$. I will call a
3-dimensional soldering form $\3 e_a^I$ a co-triad and a 4-dimensional
soldering form $\4 e_a^I$ a co-tetrad in what follows.

To do calculus with these generalized tensor fields, it is necessary to extend
the definition of spacetime derivative operators so that they also ``act''
on internal indices.  We require (in addition to the usual properties that a
spacetime derivative operator satisfies) that a {\it generalized derivative
operator} obey the linearity, Leibnitz, and commutativity with contraction
rules with respect to the internal indices. Furthermore, we require that all
generalized derivative operators be compatible with $\n_{IJ}$. Given these
properties, it is straightforward to show that the set of all generalized
derivative operators has the structure of an affine space. In other words, if
$\pd_a$ is some fiducial generalized derivative operator (which we treat as an
origin in the space of generalized derivative operators), then any other
generalized derivative operator $\cD_a$ is completely characterized by a pair
of generalized tensor fields $A_{ab}{}^c$ and $A_{aI}{}^J$ defined by
$$\cD_a k_{bI}=:\pd_a k_{bI} + A_{ab}{}^c k_{cI} + A_{aI}{}^J k_{bJ}.
\eqno(3.3)$$
We will call $A_{ab}{}^c$ and $A_{aI}{}^J$ the {\it \st connection 1-form} and
{\it internal connection 1-form} of $\cD_a$. It is easy to show that
$$A_{aIJ}=A_{a[IJ]}\quad{\rm and}\quad A_{ab}{}^c=A_{(ab)}{}^c.\eqno(3.4)$$
These conditions follow from the requirements that all generalized derivative
operators be compatible with $\n_{IJ}$ and that they be torsion-free. Later
in this section, we will consider what happens if we allow derivative operators
to have non-zero torsion---i.e., if $A_{[ab]}{}^c\not=0$. Finally, note that
$A_{ab}{}^c$ need not equal $A_{aI}{}^J e_b^I e_J^c$, in general.

As usual, given a generalized derivative operator $\cD_a$, we can construct
curvature tensors by commuting derivatives. The {\it internal curvature tensor}
$F_{abI}{}^J$ and the {\it spacetime curvature tensor} $F_{abc}{}^d$ are
defined by
$$\eqalignnotwo{&2\cD_{[a}\cD_{b]}k_I=:F_{abI}{}^J k_J\quad{\rm and}&(3.5a)\cr
&2\cD_{[a}\cD_{b]}k_c=:F_{abc}{}^d k_d.&(3.5b)\cr}$$
If our fiducial generalized derivative operator is chosen to be flat on both
spacetime and internal indices, then
$$\eqalignnotwo{&F_{abI}{}^J=2\pd_{[a}A_{b]I}{}^J + [A_a,A_b]_I{}^J\quad
{\rm and}&(3.6a)\cr
&F_{abc}{}^d=2\pd_{[a}A_{b]c}{}^d + [A_a,A_b]_c{}^d.&(3.6b)\cr}$$
Here $[A_a,A_b]_I{}^J:=(A_{aI}{}^K A_{bK}{}^J - A_{bI}{}^K A_{aK}{}^J)$ and
$[A_a,A_b]_c{}^d:=(A_{ac}{}^e A_{be}{}^d - A_{bc}{}^e A_{ae}{}^d)$ are the
commutators of linear operators.

Just as a compatibility with a \st metric $g_{ab}$ defines a unique,
torsion-free {\it spacetime} derivative operator $\grad_a$, compatibility with
an orthonormal basis $e_I^a$ (and thus with $g_{ab}$) defines a unique
torsion-free {\it generalized} derivative operator, which we also denote by
$\grad_a$. The {\it Christoffel symbols} $\G_{aI}{}^J$ and $\G_{ab}{}^c$ are
defined by
$$\grad_a k_{bI}=:\pd_a k_{bI} + \G_{ab}{}^c k_{cI} + \G_{aI}{}^J k_{bJ},
\eqno(3.7)$$
and satisfy
$$\eqalignnotwo{&\G_{aI}{}^J=-e^{bJ}(\pd_a e_{bI} + \G_{ab}{}^c e_{cI})\quad
{\rm and}&(3.8a)\cr
&\G_{ab}{}^c=-{1\over 2}g^{cd}(\pd_a g_{bd}+\pd_b g_{ad}-\pd_d
g_{ab}).&(3.8b)\cr}$$
It also follows that internal and \st curvature tensors $R_{abI}{}^J$ and
$R_{abc}{}^d$ of $\grad_a$ are related by
$$R_{abI}{}^J=R_{abc}{}^d e_I^c e_d^J.\eqno(3.9)$$
We will need the above result in this and later sections to show that the \P
and \sd actions reproduce Einstein's equation.

Now let us return to our discussion of the 2+1 \P theory. Recall that we
wanted to write the integrand $\sqrt{-g} R$ in triad notation. Using
$$R_{abI}{}^J=R_{abc}{}^d\ \3 e_I^c\ \3 e_d^J\eqno(3.10)$$
(which is equation (3.9) written in terms of a triad $\3 e_I^a$) and
$$\e_{abc}=\ \3 e_a^I\ \3 e_b^J\ \3 e_c^K\ \e_{IJK}\eqno(3.11)$$
(which relates the volume element $\e_{abc}$ of $g_{ab}=\3 e_a^I\ \3 e_b^J
\n_{IJ}$ to the volume element $\e_{IJK}$ of $\n_{IJ}$), we find that
$$\eqalign{\sqrt{-g} R&=\sqrt{-g}\ \d_{[d}^b\d_{e]}^c\ R_{bc}{}^{de}\cr
&={1\over 2}\otw\n^{abc}\e_{ade}R_{bc}{}^{de}\cr
&={1\over 2}\otw\n^{abc}\ \3 e_a^I\ \3 e_d^J\ \3 e_e^K\
\e_{IJK}R_{bc}{}^{de}\cr
&={1\over 2}\otw\n^{abc}\e_{IJK}\ \3 e_a^I\ R_{bc}{}^{JK}.\cr}\eqno(3.12)$$
Thus, viewed as a functional of a co-triad $\3 e_a^I$, the standard \EH
action is given by
$$S_{EH}(\3 e)={1\over 2}\int_\S\otw\n^{abc}\e_{IJK}\ \3 e_a^I\ R_{bc}{}^{JK}.
\eqno(3.13)$$

To obtain the 2+1 \P action, we simply replace $R_{abI}{}^J$ in (3.13) with
the internal curvature tensor $\3 F_{abI}{}^J$ of an {\it arbitrary}
generalized derivative operator $\3\cD_a$ defined by
$$\3\cD_a k_I:=\pd_a k_I + \3 A_{aI}{}^J k_J.\eqno(3.14)$$
We define the 2+1 {\it \P action based on} $SO(2,1)$ to be
$$S_P(\3 e,\3 A):={1\over 4}\int_M\otw\n^{abc}\e_{IJK}\ \3 e_a^I\ \3
F_{bc}{}^{JK},\eqno(3.15)$$
where $\3 F_{abI}{}^J=2\pd_{[a}\3 A_{b]I}{}^J + [\3 A_a,\3 A_b]_I{}^J$.
Note that I have included an additional factor of $1/2$ in definition (3.15).
This overall factor will  not affect the \EL \eoms in any way, but it
will change the canonically conjugate variables. I have chosen to use this
action so that the expressions for our canonically conjugate variables agree
with those used in the literature (see, e.g., \cite{CSP}).

As defined above, $S_P(\3 e,\3 A)$ is a functional of both a co-triad $\3
e_a^I$ and a connection 1-form $\3 A_{aI}{}^J$ which takes values in the
defining representation of the Lie algebra of $SO(2,1)$. Note also that
$\3\cD_a$ as defined by (3.14) knows how to act only on internal indices. We
do not require that $\3\cD_a$ know how to act on spacetime indices.
However, when performing calculations, we will find that it is often convenient
to consider a {\it torsion-free} extension of $\3\cD_a$ to spacetime tensor
fields. It turns out that all calculations and all results will be independent
of our choice of torsion-free extension. In fact, we will see that these
results hold for extensions of $\3\cD_a$ that have non-zero torsion as well.

Since the 2+1 \P action is a functional of both a co-triad and a connection
1-form, we will obtain two Euler-Lagrange equations of motion. When we vary $\3
e_a^I$, we get
$$\otw\n^{abc}\e_{IJK}\ \3 F_{bc}{}^{JK}=0.\eqno(3.16)$$
When we vary $\3 A_a{}^{IJ}$, we get
$$\3\cD_b(\otw\n^{abc}\e_{IJK}\ \3 e_c^K)=0.\eqno(3.17)$$
To arrive at (3.17), we considered a torsion-free extension of $\3\cD_a$ to
spacetime tensor fields (so that $\d\3 F_{bc}{}^{JK}=2\ \3\cD_{[b}\d\3 A_{c]}
{}^{JK}$) and then integrated by parts. The surface integral vanished since
$\d\3 A_c{}^{JK}=0$ on the boundary, while the remaining term gave (3.17). Note
that since the left hand side of (3.17) is the divergence of a skew spacetime
tensor density of weight +1 on $M$, it is independent of the choice of
torsion-free extension of $\3\cD_a$. Since $\otw\n^{abc}\e_{IJK}\ \3 e_c^K=
2(\3 e)\ \3 e_I^{[a}\ \3 e_J^{b]}$ (where $(\3 e):=\sqrt{-g}$), we can rewrite
(3.17) as
$$\3\cD_b\left((\3 e)\ \3 e_I^{[a}\ \3 e_J^{b]}\right)=0.\eqno(3.18)$$
We shall see in Section 6 that the form of equation (3.18) holds for the 3+1 \P
theory as well.

To determine the content of equation (3.18), let us express $\3\cD_a$ in terms
of the unique, torsion-free generalized derivative operator $\grad_a$
compatible with $\3 e_a^I$, and $\3 C_{aI}{}^J$ defined by
$$\3\cD_a k_I=:\grad_a k_I+\3 C_{aI}{}^J k_J.\eqno(3.19)$$
Since (3.18) is the divergence of a skew spacetime tensor density of weight
+1 on $M$, and since $\grad_a$ is compatible with $\3 e_a^I$, we get
$$\3 C_{bI}{}^K\ \3 e_K^{[a}\ \3 e_J^{b]}+\3 C_{bJ}{}^K\ \3 e_I^{[a}
\ \3 e_K^{b]}=0.\eqno(3.20)$$
This is equivalent to the statement that the (internal) commutator of $\3
C_{bIJ}$ and $\3 e_I^{[a}\ \3 e_J^{b]}$ vanishes.  We will now show that (3.20)
implies that $\3 C_{aI}{}^J=0$.

To see this, define a spacetime tensor field $\3 S_{abc}$ via
$$\3 S_{abc}:=\3 C_{aIJ}\ \3 e_b^I\ \3 e_c^J.\eqno(3.21)$$
(Note, incidently, that $\3 S_{abc}$ is not the spacetime connection of
$\3\cD_a$ relative to $\grad_a$.) Then the condition $\3 C_{aIJ}=\3 C_{a[IJ]}$
is equivalent to $\3 S_{abc}=\3 S_{a[bc]}$. Now contract equation (3.20) with
$\3 e_a^I\ \3 e_c^J$. This yields $\3 S_{bc}{}^b=0$, so $\3 S_{abc}$  is
trace-free on its first and last indices. Using this result, (3.20) reduces to
$$\3 C_{bI}{}^K\ \3 e_K^a\ \3 e_J^b-\3 C_{bJ}{}^K\ \3 e_I^b\ \3 e_K^a=0.
\eqno(3.22)$$
If we now contract (3.22) with $\3 e_c^I\ \3 e_d^J$, we get
$$\3 S_{cd}{}^a=\3 S_{(cd)}{}^a.\eqno(3.23)$$
Thus, $\3 S_{abc}$ is symmetric in its first two indices. Since $\3 S_{abc}=\3
S_{a[bc]}$ and $\3 S_{abc}=\3 S_{(ab) c}$, we can successively interchange the
first two and last two indices (with the appropriate sign changes) to show $\3
S_{abc}=0$. Futhermore, since $e_a^I$ are invertible, we get $\3 C_{aI}{}^J=0$.
This is the desired result.\footnote{This method of proving $\3
C_{aI}{}^J=0$---which  generalizes to the 3+1 \P and \sd actions---was shown to
me by J. Samuel and A. Ashtekar.}

Since $\3 C_{aI}{}^J=0$, we can conclude that the generalized derivative
operator $\3\cD_a$ must agree with $\grad_a$ when acting on internal
indices. Thus, although the \P action started as a functional of a co-triad and
an arbitrary generalized derivative operator $\3\cD_a$, we find that one
equation of motion implies that $\3\cD_a=\grad_a$.  In terms of connection
1-forms, $\3 C_{aI}{}^J=0$ implies that $\3 A_{aI}{}^J=\G_{aI}{}^J$, where
$\G_{aI}{}^J$ is the internal Christoffel symbol of $\grad_a$. Using this
result, the remaining Euler-Lagrange equation of motion (3.16) becomes
$$\otw\n^{abc}\e_{IJK}R_{bc}{}^{JK}=0.\eqno(3.24)$$
When (3.24) is contracted with $\3 e^{dI}$, we get $G^{ad}=0$. Thus, the \P
action based on $SO(2,1)$ reproduces the standard 2+1 vacuum Einstein's
equation.

It is interesting to note that to show that the \P action reduces to the
standard \EH action in 2+1 dimensions, we need only vary the connection
1-form $\3 A_{aI}{}^J$. Since we found that (3.17) could be solved uniquely for
$\3 A_{aI}{}^J$ in terms of the remaining basic variables $\3 e_a^I$, we can
pull-back $S_P(\3 e,\3 A)$ to the solution space $\3 A_{aI}{}^J= \G_{aI}{}^J$
and obtain a new action $\und S_P(\3 e)$, which depends only on a co-triad.
This pulled-back action is just $1/2$ times the standard \EH action $S_{EH}(\3
e)$ given by (3.13). But what about the boundary term that one should strictly
include in the standard \EH action? It looks as if $\und S_P(\3 e)$ is missing
this needed term.

The answer to this question is the following: Whereas the standard \EH action
is a second-order action,  the 2+1 \P action is {\it first-order}. As
mentioned at the beginning of Section 2, varying the standard \EH action (3.1)
with respect to $g^{ab}$ produces a surface integral involving derivatives of
the variation $\d g^{ab}$. Since we are allowed only to keep $g^{ab}$  fixed on
the boundary, this surface integral is non-vanishing and must be compensated
for by adding a boundary term to (3.1). This is also the case if we vary
$S_{EH}(\3 e)$ given by (3.13) with respect to $\3 e_a^I$. On the other hand,
when we vary the Palatini action (3.15) \wrt $\3 A_{aI}{}^J$, we hold $\3
A_{aI}{}^J$ fixed on the boundary and $\3 e_a^I$ fixed throughout. Then by
solving (3.17) uniquely for $\3 A_{aI}{}^J$, we can pull-back $S_P(\3 e,\3 A)$
to the solution space $\3 A_{aI}{}^J=\G_{aI}{}^J$. But now when we vary $\und
S_P(\3 e)$ with respect to $\3 e_a^I$ which lie entirely in the solution
space, fixing $\3 e_a^I$ on the boundary also fixes certain derivatives of $\3
e_a^I$ on the boundary. This is a reflection of the fact that the reduction
procedure comes with a prescription on how to do variations. It is precisely
the vanishing of these derivatives of $\d\3 e_a^I$ which eliminates the need
of a boundary term for $\und S_P(\3 e)$.

It is also interesting to note that we could obtain the same result ($\3
A_{aI}{}^J=\G_{aI}{}^J$) by considering an extension of $\3\cD_a$ to \st tensor
fields with non-zero torsion $\3 T_{ab}{}^c$. (Recall that if $\3 A_{ab}{}^c$
denotes the \st connection 1-form of the extension of $\3\cD_a$, then the {\it
torsion tensor} $\3 T_{ab}{}^c$ is defined by $2 \3\cD_{[a}\ \3\cD_{b]}f=:\3
T_{ab}{}^c \ \3\cD_c f$ and satisfies $\3 T_{ab}{}^c=2\ \3 A_{[ab]}{}^c$.) By
varying the 2+1 \P action (3.15) \wrt $\3 A_{aI}{}^J$, we would find
$$2\ \3\cD_{[a}\3 e_{b]}^I - \3 T_{ab}{}^c\ \3 e_c^I=0.\eqno(3.25)$$
This is the field equation for $\3 A_{aI}{}^J$ which holds for any
extension of $\3\cD_a$ to \st tensor fields. If we restrict ourselves to
torsion-free extensions, we get back equation (3.17). Then by following the
argument given there, we would find $\3 A_{aI}{}^J=\G_{aI}{}^J$ as before.

However, there exists an alternative approach to solving equation (3.25) which
is often used by particle physicists. Namely, instead of considering a
torsion-free extension of $\3\cD_a$ to \st tensor fields, one considers an
extension of $\3\cD_a$ to \st tensor fields which is compatible with the
co-triad $\3 e_a^I$. This can always be done, but the price of such an
extension is in general a non-zero torsion tensor $\3 T_{ab}{}^c$. \ But since
we now have $\3\cD_a\ \3 e_b^I=0$, equation (3.25) implies
$$\3 T_{ab}{}^c\ \3 e_c^I=0.\eqno(3.26)$$
Invertibility of $\3 e_c^I$ then implies that $\3 T_{ab}{}^c=0$. Since there
exists only one torsion-free derivative  operator compatible with $\3 e_a^I$,
we can conclude that $\3\cD_a=\grad_a$ \ (or equivalently, $\3 A_{aI}{}^J
=\G_{aI}{}^J$). This is the desired result.

Finally, to conclude this section, let us write the 2+1 \P action (3.15)
in a form which is valid for {\it any} Lie group $G$. Recall that the
connection 1-form $\3 A_{aI}{}^J$---being a linear operator on the internal
3-dimensional vector space (equipped with the Minkowski metric $\n_{IJ}$)
and satisfying $\3 A_{aIJ}=\3 A_{a[IJ]}$---takes values in the defining
representation of the Lie algebra of $SO(2,1)$. Since $\dim(SO(2,1))= 3$ (which
is the same as the dimension of the internal vector space), we can define an
$SO(2,1)$ {\it Lie algebra-valued} connection 1-form, $\3 A_a^I$, via
$$\3 A_{aI}{}^J=:\3 A_a^K\e^J{}_{IK}.\eqno(3.27)$$
This is just the {\it adjoint representation} of the Lie algebra of $SO(2,1)$
with respect to the structure constants
$\e^I{}_{JK}:=\n^{IM}\e_{MJK}$.\footnote{Given a Lie algebra $\cL$ with
structure constants $C^I{}_{JK}$, the {\it adjoint representation} of $\cL$ by
linear operators on $\cL$ is defined by the mapping $v^I\in\cL \mapsto
(ad_v)_I{}^J:=v^K C^J{}_{IK}$. Under $ad$, the Lie bracket $[v,w]^I:=
C^I{}_{JK} v^J w^K\ \in\cL$ maps to the commutator of linear operators
$[ad_v,ad_w]_I{}^J:=(ad_v)_I{}^K(ad_w)_K{}^J -(ad_w)_I{}^K(ad_v)_K{}^J$. I
should note that since $(ad_v)_I{}^J w^I=-[v,w]^J$, the above definition of the
adjoint representation differs in sign from that given in most math and physics
textbooks. The sign difference can be traced to my definition of the commutator
of linear operators, which also differs in sign from the standard definition.}
That the defining representation and adjoint representation agree is a property
that holds  only in 2+1 dimensions since $\dim(SO(n,1))=n+1$ if and only if
$n=2$. In terms of $\3 A_a^I$, the generalized derivative operator $\3\cD_a$
satisfies
$$\3\cD_a v^I=\pd_a v^I+[\3 A_a,v]^I,\eqno(3.28)$$
where $[\3 A_a,v]^I:=\e^I{}_{JK}\ \3 A_a^J v^K$. From (3.27), it also follows
that the Lie algebra valued-curvature tensor $\3 F_{ab}^I$ (which is related to
$\3 F_{abI} {}^J$ via $\3 F_{abI}{}^J=\3 F_{ab}^K\e^J{}_{IK}$) can be written
as
$$\3 F_{ab}^I=2\pd_{[a}\3 A_{b]}^I+[\3 A_a,\3 A_b]^I.\eqno(3.29)$$
Thus, in terms $\3 A_a^I$ and $\3 F_{ab}^I$, the \P action becomes
$$\eqalign{S_P(\3 e,\3 A)&={1\over 4}\int_M\otw\n^{abc}\e_{IJK}\ \3 e_a^I\
\3 F_{bc}{}^{JK}\cr
&={1\over 4}\int_M\otw\n^{abc}\e_{IJK}\ \3 e_a^I\ \3 F_{bc}^L\ \e^{KJ}{}_L\cr
&={1\over 2}\int_M\otw\n^{abc}\ \3 e_{aI}\ \3 F_{bc}^I.\cr}\eqno(3.30)$$

But now note that the last line above suggests a natural generalization.
Namely, let $G$ be any Lie group with Lie algebra $\cL_G$, and let $\3
A_a^I$ and $\3 e_{aI}$ be $\cL_G$- and $\cL_G^*$-valued 1-forms, respectively.
Although the action given by (3.30) was originally defined for the Lie group
$SO(2,1)$, it is well-defined in the above sense for any Lie group $G$. \ $\3
F_{ab}^I$ is still the curvature tensor of $\3 A_a^I$, but $\3 e_{aI}$ can no
longer be thought of as a co-triad. In fact, since $G$ is now arbitrary, the
index $I$ can take any value $1,2, \cdots,\dim(G)$. Nonetheless, we can still
define the {\it \P action based on} $G$ via
$${}^G\!S_P(\3 e,\3 A):={1\over 2}\int_M\otw\n^{abc}\ \3 e_{aI}\ \3
F_{bc}^I,\eqno(3.31)$$
which we treat it as a functional of an $\cL_G$-valued connection 1-form $\3
A_a^I$ and an $\cL_G^*$-valued covector field $\3 e_{aI}$. The equations of
motion we obtain by varying $\3 e_{aI}$ and $\3 A_a^I$ are
$$\otw\n^{abc}\ \3 F_{bc}^I=0\quad{\rm and}\quad\3\cD_{b}(\otw\n^{abc}
\ \3 e_{cI})=0,\eqno(3.32)$$
which are the analogs of equations (3.16) and (3.17). As before, the second
equation requires a torsion-free extension of $\3\cD_a$ to spacetime tensor
fields, but again, all results will be independent of this choice.
\vskip .5cm

\noindent{\sl 3.2 Legendre transform}
\medskip

Given the action (3.31), it is a straightforward exercise to put the 2+1 \P
theory based on $G$ in Hamiltonian form. We will assume that $M$ is
topologically $\S\times R$ and that there exists a function $t$ (with nowhere
vanishing gradient $(dt)_a$) such that each $t=\const$ surface $\S_t$ is
diffeomorphic to $\S$. As usual, $t^a$  will denote the flow vector field
satisfying $t^a(dt)_a=1$. Since the Lie group $G$ is arbitrary, the 2+1 \P
theory based on $G$ is not a theory of a \st metric; it does not involve a \st
metric in any way whatsoever. Thus, in particular, $t$ does not necessarily
have the interpretation of time. Nonetheless, we can still define ``evolution''
from one $t=\const$ surface to the next using the Lie derivative \wrt $t^a$.

To write (3.31) in 2+1 form, we decompose $\otw\n^{abc}$ in terms of $t^a$
and $\otw\n^{ab}$ (the Levi-Civita tensor density of weight +1 on $\S$). Using
$\otw\n^{abc}=3t^{[a}\otw\n^{bc]}dt$, we get
$$\eqalign{{}^G\!S_P(\3 e,\3 A)&={1\over 2}\int_M\otw\n^{abc}\ \3 e_{aI}
\ \3 F_{bc}^I\cr
&={1\over 2}\int dt\int_\S(t^a\otw\n^{bc}+t^b\otw\n^{ca}+t^c\otw\n^{ab})
\ \3 e_{aI}\ \3 F_{bc}^I\cr
&=\int dt\int_\S{1\over 2}(\3 e\cdot t)_I\otw\n^{bc}F_{bc}^I + \otw E_I^c
\cL_{\vec t} \ A_c^I - \otw E_I^c\cD_c(\3 A\cdot t)^I,\cr}\eqno(3.33)$$
where $(\3 e\cdot t)_I:=t^a\ \3 e_{aI}$, \ $\otw E_I^a:=\otw\n^{ab}\ \3
e_{bI}$, \ $(\3 A\cdot t)^I:=t^a\ \3 A_a^I$, and $A_a^I:=t_a^b\ \3 A_b^I$
are the configuration variables which specify all the information contained in
the field variables $\3 e_{aI}$ and $\3A_a^I$. Note that:

\begin{enumerate}
\item Since $G$ is an arbitrary Lie group, the internal index $I$ can take any
value $I=1,2,\cdots,\dim(G)$. Thus, $\otw E_I^a$  cannot in general be
interpreted as a dyad. In fact, this is true even when $G=SO(2,1)$, since
$\dim(SO(2,1))=3$. However, for $SO(2,1)$ we have $\otw E_I^a\otw
E^{bI}=\otw{\otw q}{}^{ab}\ (=q q^{ab})$.
\item $t^a\ \3 F_{ab}^I=\cL_{\vec t} \ \3 A_b^I-\ \3\cD_b(\3 A\cdot t)^I$,
which follows from a generalization of {\it Cartan's identity} $\cL_{\vec v}\a=
i_{\vec v}d\a+d(i_{\vec v}\a)$. The Lie derivative $\cL_{\vec t}$ treats
fields with only internal indices as scalars.
\item $\cL_{\vec t} \ t_b^a=0$, where $t_b^a:=\d_b^a-t^a(dt)_b$ is the natural
projection operator into the $t=\const$ surfaces defined by $t$ and $t^a$.
\item $\cD_a:=t_a^b\ \3\cD_b$ is the generalized derivative operator
on $\S$ associated with $A_a^I$.
\item $F_{ab}^I:=t_a^c t_b^d\ \3 F_{cd}^I$ is the curvature tensor of
$\cD_a$ and satisfies $F_{ab}^I=2\pd_{[a}A_{b]}^I+[A_a,A_b]^I$.
\end{enumerate}

{}From (3.33), we see that (modulo a surface integral) the Lagrangian
${}^G\!L_P$
of the 2+1 \P theory based on $G$ is given by
$${}^G\!L_P=\int_\S{1\over 2}(\3 e\cdot t)_I\otw\n^{ab}F_{ab}^I + \otw
E_I^a\cL_{\vec t} \  A_a^I + (\cD_a\otw E_I^a)(\3 A\cdot t)^I.\eqno(3.34)$$
By inspection, we see that the momentum conjugate to $A_a^I$ is $\otw E_I^a$,
while $(\3 e\cdot t)_I$ and $(\3 A\cdot t)^I$ both play the role of Lagrange
multipliers. Thus, the Dirac constraint analysis says that the phase space
$({}^G\G_P,{}^G\W_P)$ is coordinatized by the pair $(A_a^I, \otw E_I^a)$ and
has symplectic structure\footnote{Note that in terms of the Poisson bracket
$\{\ ,\ \}$ defined by ${}^G\W_P$, we have $\{A_a^I(x),\otw E_J^b(y)\}=\d_a^b
\d_J^I\d(x,y)$.}
$${}^G\W_P=\int_\S\ \id\otw E_I^a\ \iwedge\ \id A_a^I.\eqno(3.35)
$$
The Hamiltonian is given by
$${}^G\!H_P(A,\otw E)=\int_\S-{1\over 2}(\3 e\cdot t)_I\otw\n^{ab}F_{ab}^I  -
(\cD_a \otw E_I^a)(\3 A\cdot t)^I.\eqno(3.36)$$
As we shall see in the next subsection, this is just a sum of 1st class
constraint functions associated with
$$\otw\n^{ab} F_{ab}^I\approx 0\quad{\rm and}\quad\cD_a\otw E_I^a\approx 0.
\eqno(3.37)$$
Note that these equations are the field equations (3.32) pulled-back to $\S$
with $\otw\n^{ab}$. Note also that they are {\it polynomial} in the canonically
conjugate variables $(A_a^I,\otw E_I^a)$. This is to be contrasted with the
constraint equations for the standard \EH theory. Recall that the scalar
constraint of that theory depended non-polynomially on $q_{ab}$.
\vskip .5cm

\noindent{\sl 3.3 Constraint algebra}
\medskip

As usual, to evaluate the Poisson brackets of the constraints and to determine
the motions they generate on phase space, we must first construct constraint
functions associated with (3.37).  Given test fields $v^I$ and $\a_I$, which
take values in the Lie algebra $\cL_G$ and its dual $\cL_G^*$, we define
$$F(\a):={1\over 2}\int_\S\a_I\otw\n^{ab} F_{ab}^I\quad{\rm and}\quad
G(v):=\int_\S v^I (\cD_a\otw E_I^a)\eqno(3.38)$$
We will call $G(v)$ the {\it Gauss constraint function} since it will play the
same role as the Gauss constraint of \YM theory. We will see that $G(v)$
generates the usual gauge transformations of the connection 1-form $A_a^I$ and
it conjugate momentum (or ``electric field'') $\otw E_I^a$.

We are now ready to evaluate the functional derivatives of $F(\a)$ and $G(v)$.
Since $F(\a)$ is independent of the momentum $\otw E_I^a$, and since
$\d F_{ab}^I=2\cD_{[a}\d A_{b]}^I$, we find
$${\d F(\a)\over \d\otw E_I^a}=0\quad{\rm and}\quad{\d F(\a)\over \d A_a^I}
=\otw\n^{ab}\cD_b\a_I.\eqno(3.39)$$
Similarly, if we vary $G(v)$ \wrt $\otw E_I^a$ and $A_a^I$, we find
$${\d G(v)\over \d\otw E_I^a}=-\cD_a v^I\quad{\rm and}\quad{\d G(v)\over
\d A_a^I}=\{v,\otw E^a\}_I\ \left(:=C^K{}_{JI}v^J\otw E_K^a\right).\eqno(3.40)
$$
Here $C^I{}_{JK}$ denote the structure constants of the Lie algebra $\cL_G$
and $\{\ ,\ \}:\cL_G\times\cL_G^*\rightarrow\cL_G^*$ denotes the co-adjoint
bracket. \ $\{\ ,\ \}$ is defined in terms of the Lie bracket $[\ ,\ ]:
\cL_G\times\cL_G\rightarrow\cL_G$ via $\{v,\a\}_I w^I:=\a_K[v,w]^K$.

Given (3.39) and (3.40), we can now write down the Hamiltonian vector fields
$X_{F(\a)}$ and $X_{G(v)}$ associated with $F(\a)$ and $G(v)$. They are
$$\eqalignnotwo{&X_{F(\a)}=\int_\S -\otw\n^{ab}(\cD_b\a_I){\d\over \d\otw
E_I^a}
\quad{\rm and}&(3.41a)\cr
&X_{G(v)}=\int_\S -(\cD_a v^I){\d\over\d A_a^I}-\{v,\otw E^a\}_I{\d\over
\d\otw E_I^a}.&(3.41b)\cr}$$
Thus, under the 1-parameter family of \diffeos on ${}^G\G_P$ associated with
$X_{F(\a)}$, we have
$$\eqalignnotwo{&A_a^I\mapsto A_a^I\quad{\rm and}&(3.42a)\cr
&\otw E_I^a\mapsto\otw E_I^a - \e(\otw\n^{ab}\cD_b\a_I) + O(\e^2).&(3.42b)\cr}
$$
Similarly, under the 1-parameter family of \diffeos on ${}^G\G_P$ associated
with $X_{G(v)}$, we have
$$\eqalignnotwo{&A_a^I\mapsto A_a^I-\e\cD_a v^I+O(\e^2)\quad{\rm
and}&(3.43a)\cr
&\otw E_I^a\mapsto\otw E_I^a - \e\{v,\otw E^a\}_I + O(\e^2).&(3.43b)\cr}$$
Note that ($3.43a$) and ($3.43b$) are the usual gauge transformations of the
connection 1-form $A_a^I$ and its conjugate momentum $\otw E_I^a$ that we find
in \YM theory. Equations (3.42) and (3.43) are the motions on phase space
generated by $F(\a)$ and $G(v)$.

Given (3.39) and (3.40), we can also evaluate the \Pbs of the constraint
functions. Since  the $G(v)$'s play the same role as the Gauss constraint
functions of \YM theory, we would expect their \Pb algebra to be isomorphic to
the Lie algebra $\cL_G$. This is indeed the case. We find
$$\{G(v),G(w)\}=G([v,w]),\eqno(3.44)$$
where $[v,w]^I=C^I{}_{JK} v^J w^K$ is the Lie bracket of $v^I$ and $w^I$. Thus,
$v^I\mapsto G(v)$ is a representation of the Lie algebra $\cL_G$. The Lie
bracket in $\cL_G$ is mapped to the \Pb of the corresponding Gauss constraint
functions.

Since $F(\a)$ is independent of $\otw E_I^a$, it follows trivially that
$$\{F(\a),F(\b)\}=0.\eqno(3.45)$$
With only slightly more effort, we can show that
$$\{G(v),F(\a)\}=-F(\{v,\a\}),\eqno(3.46)$$
where $\{v,\a\}_I=C^K{}_{JI}v^J\a_K$ is the co-adjoint bracket of $v^I$ and
$\a_I$. Thus, the totality of constraint functions ($G(v)$ and $F(\a)$) is
closed under \Pb---i.e., they form a 1st class set.  Furthermore, since
(3.44), (3.45), and (3.46) do  not involve any structure functions (unlike
the constraint algebra of the \EH theory discussed in subsection 2.3), the set
of constraint functions form a Lie algebra \wrt Poisson bracket. In fact,
$(\a,v)\in\cL_G^*\times\cL_G\mapsto (F(\a),G(v))$ is a representation of the
Lie algebra $\cL_{IG}$ of the {\it inhomogeneous Lie group} $IG$ associated
with $G$.\footnote{We will discuss the construction of the inhomogeneous
Lie group $IG$ and its Lie algebra $\cL_{IG}$ in subsection 4.4 where we show
the equivalence between the 2+1 \P theory based on $G$ and \CS theory based
on $IG$. When $G$ is the Lorentz group $SO(2,1)$, \ $IG$ is the corresponding
Poincar\'e group $ISO(2,1)$.} The action $\t_v(\a):=-\{v,\a\}_I$ of
$v^I\in\cL_G$ on $\a_I\in\cL_G^*$ is mapped to the \Pb $\{G(v),F(\a)\}$ of the
corresponding constraint functions. The $\ F(\a)$'s play the role of
``translations'' and the $G(v)$'s play the role of ``rotations'' in the
inhomogeneous group.
\vskip 1cm

\noindent{\bf 4. Chern-Simons theory}
\vskip .5cm

So far in this review, we have written down two different actions for 2+1
gravity: \ The standard \EH action, which gave us a description of 2+1 gravity
in terms of a spacetime metric (or equivalently, a co-triad); and the 2+1 \P
action based on $SO(2,1)$, which gave us a description in terms of a co-triad
and a connection 1-form. In this section, we shall see that 2+1 gravity can be
described by an action that depends {\it only} on a connection 1-form. We shall
see that the 2+1 \P action based on $SO(2,1)$ is equal to the \CS action based
on $ISO(2,1)$, modulo a surface integral that does not affect the equations of
motion. This result was first shown by A. Achucarro and P.K. Townsend
\cite{AT}; it was later rediscovered  and used by Witten \cite{Witten} to
quantize 2+1 gravity. In this section, we will follow the treatment of
\cite{CSP} in which the result for 2+1 gravity follows as a special case. We
will show that the 2+1 \P theory based on any Lie group $G$ is equivalent to
\CS theory based on the inhomogeneous Lie group $IG$ associated with $G$.

The above equivalence between the 2+1 \P and \CS theories is at the level of
actions. The gauge groups, $G$ and $IG$, are different, but the actions are the
same. It is interesting to note that the 2+1 \P and \CS theories based on the
{\it same} Lie group $G$ are also related, but this time at the level of their
Hamiltonian formulations. We shall see that the {\it reduced phase space} of
the \CS theory based on $G$ is the {\it reduced configuration space} of the 2+1
\P theory based on the same $G$.  Since \CS theory is not available in 3+1
dimensions, the relationships that we find in this section do not,
unfortunately, extend to 3+1 theories of gravity.
\vskip .5cm

\noindent{\sl 4.1 \EL equations of motion}
\medskip

Unlike the standard \EH and \P theories which are well-defined in $n$+1
dimensions, \CS theory is defined only in odd dimensions.  In 2+1 dimensions,
the basic variable is a connection 1-form $\3 A_a^i$ which takes values in a
Lie algebra $\cL_G$  equipped with an invariant, non-degenerate bilinear form
$k_{ij}$.\footnote{We will require that $k_{ij}$ be invariant under the {\it
adjoint action} of the Lie algebra $\cL_G$ on itself---i.e., that
$k_{ij}[x,v]^i w^j + k_{ij}v^i[x,w]^j=0$ for all $v^i,w^i,x^i\in\cL_G$. If
$C_{ijk}$ is defined in terms of the structure constants $C^i{}_{jk}$ via
$C_{ijk}:=k_{im}C^m{}_{jk}$, then invariance of $k_{ij}$ under the adjoint
action is equivalent to $C_{ijk}=C_{[ijk]}$. If the Lie group is {\it
semi-simple} (i.e., if it does not admit any non-trivial abelian normal
subgroup), then we are guaranteed that such a $k_{ij}$ exists. This is just the
{\it Cartan-Killing metric} defined by $k_{ij}:=C^m{}_{ni} C^n{}_{mj}$.
Invariance of $k_{ij}$ is equivalent to the invariance of $C^i{}_{jk}$---that
is, the {\it Jacobi identity} $C^m{}_{[ij}C^n{}_{k]m}=0$.} The {\it \CS action
based on G} is defined by
$${}^G\!S_{CS}(\3 A):={1\over 2}\int_M\otw\n^{abc}k_{ij}\left(\ \3 A_a^i\pd_b\3
A_c^j+ {1\over 3}\ \3 A_a^i[\3 A_b,\3 A_c]^j\right),\eqno(4.1)$$
where $[\3 A_b,\3 A_c]^j:=C^j{}_{mn}\ \3 A_b^m\ \3 A_c^n$ denotes the Lie
bracket of $\3 A_b^i$ and $\3 A_c^i$. It is important to note that \CS theory
is  not defined for arbitrary Lie groups---we need the additional structure
provided by the invariant, non-degenerate bilinear form $k_{ij}$.

To obtain the \EL equations of motion, we vary ${}^G\!S_{CS}(\3 A)$ \wrt
$\3 A_a^i$. Using the fact that $C_{ijk}:=k_{im} C^m{}_{jk}$ is totally
anti-symmetric, we obtain
$$\otw\n^{abc}k_{ij}\3 F_{bc}^j=0,\eqno(4.2)$$
where $\3 F_{ab}^i=2\pd_{[a}\3 A_{b]}^i+[\3 A_a,\3 A_b]^i$ is the Lie
algebra-valued curvature tensor of the generalized derivative operator
$\3\cD_a$ defined by
$$\3\cD_a v^i:=\pd_a v^i + [\3 A_a,v]^i.\eqno(4.3)$$
If we also use the fact that $k_{ij}$ is non-degenerate, we get $\3
F_{ab}^i=0$. Thus, \CS theory is a theory of a {\it flat} connection 1-form. We
will see the role that this equation plays in the next two subsections when we
put the theory in Hamiltonian form.
\vskip .5cm

\noindent{\sl 4.2 Legendre transform}
\medskip

Just like the 2+1 \P theory based on an arbitrary Lie group $G$, \CS theory is
not a theory of a \st metric. However, we can still put this theory in
Hamiltonian form if we assume that $M$ is topologically $\S\times R$ for
some submanifold $\S$ and assume that there exists a function $t$ (with nowhere
vanishing gradient $(dt)_a$) such that each $t=\const$ surface $\S_t$ is
diffeomorphic to $\S$. As usual, we let $t^a$ denote the flow vector field
satisfying $t^a(dt)_a=1$.

Given $t$ and $t^a$, we are now ready to write the \CS action (4.1) in 2+1
form. Using the decomposition $\otw\n^{abc}=3t^{[a}\otw\n^{bc]}dt$, we get
$$\eqalign{{}^G\!S_{CS}(\3 A)&={1\over 2}\int_M\otw\n^{abc}k_{ij}(\ \3
A_a^i\pd_b \ \3 A_c^j+{1\over 3}\ \3 A_a^i[\3 A_b,\3 A_c]^j)\cr
&={1\over 2}\int dt\int_\S(\3 A\cdot t)^i k_{ij}\otw\n^{bc} F_{bc}^j +
\otw\n^{ca} k_{ij} A_a^i\cL_{\vec t} \  A_c^j,\cr}\eqno(4.4)$$
where the last equality holds modulo a surface integral. Here $(\3 A\cdot
t)^i:= t^a\ \3 A_a^i$ and $ A_a^i:=t_a^b\ \3 A_b^i\ (=(\d_a^b-t^b(dt)_a)\ \3
A_b^i)$ are the configuration variables which specify all the information
contained in the field variable $\3 A_a^i$. The Lie derivative treats fields
with only internal indices as scalars, and $F_{ab}^i=2\pd_{[a} A_{b]}^i +[ A_a,
A_b]^i$ is the curvature tensor associated with $ A_a^i$.

{}From (4.4), it follows that the Lagrangian ${}^G\!L_{CS}$ of the \CS theory
based on $G$ is given by
$${}^G\!L_{CS}={1\over 2}\int_\S(\3 A\cdot t)^i k_{ij}\otw\n^{ab} F_{ab}^j
+ \otw\n^{ab}k_{ij} A_b^j\cL_{\vec t} \ A_a^i.\eqno(4.5)$$
The momentum canonically conjugate to $ A_a^i$ is ${1\over 2}\otw\n^{ab} k_{ij}
A_b^j$, while $(\3 A\cdot t)^i$ plays the role of a Lagrange multiplier. Thus,
the Dirac constraint analysis says that the phase space
$({}^G\G_{CS},{}^G\W_{CS})$ is coordinatized by $( A_a^i)$ and has symplectic
structure\footnote{Note that in terms of the \Pb $\{\ ,\ \}$ defined by
${}^G\W_{CS}$, we have $\{A_a^i(x), A_b^j(y)\}=\utw\n{}_{ab}k^{ij}\d(x,y)$,
where $\utw\n {}_{ab}$ and $k^{ij}$ denote the inverses of $\otw\n^{ab}$ and
$k_{ij}$. This result follows from the fact that for any $f:\
{}^G\G_{CS}\rightarrow R$, the Hamiltonian vector field $X_f$ is given by
$X_f=\int_\S \utw\n{}_{ab}k^{ij}\ {\d f\over\d A_b^j}\ {\d\over\d A_a^i}$.
Hence the \Pb of any two functions $f,g:\ {}^G\G_{CS}\to  R$ is
$\{f,g\}=\int_\S \utw\n{}_{ab}k^{ij}\ {\d f\over\d A_a^i}\ {\d g\over\d
A_b^j}$.}
$${}^G\W_{CS}=-{1\over 2}\int_\S\otw\n^{ab} k_{ij}\id A_a^i\ \iwedge\
\id A_b^j. \eqno(4.6)$$
The Hamiltonian is given by
$${}^G\!H_{CS}( A)=-{1\over 2}\int_\S(\3 A\cdot t)^i k_{ij}\otw\n^{ab}
F_{ab}^j.\eqno(4.7)$$
As we shall see in the next subsection, this is just a 1st class constraint
function associated with
$$k_{ij}\otw\n^{ab} F_{ab}^j=0.\eqno(4.8)$$
Note that constraint equation (4.8) is the field equation (4.2)
pulled-back to $\S$ with $\otw\n^{ab}$. Just as in the 2+1 \P theory, the
constraint equation is polynomial in the basic variable $A_a^i$. Note also that
although (4.8) may not look like the standard Gauss constraint of \YM theory,
we shall see that its associated constraint function generates the same motion
of $A_a^i$.
\vskip .5cm

\noindent{\sl 4.3 Constraint algebra}
\medskip

Following the same procedure that we used in Sections 2 and 3, we first
construct a constraint function associated with (4.8). Given a test field
$v^i$ (which takes values in the Lie algebra $\cL_G$), we define
$$G(v):={1\over 2}\int_\S v^i k_{ij}\otw\n^{ab} F_{ab}^j.\eqno(4.9)$$
Since the phase space is coordinatized by the single field $ A_a^i$, we need
to evaluate only one functional derivative. Varying $G(v)$ \wrt $ A_a^i$, we
get
$${\d G(v)\over\d A_a^i}=k_{ij}\otw\n^{ab}\cD_b v^j,\eqno(4.10)$$
where $\cD_a$ is any torsion-free extension of the generalized derivative
operator associated with $ A_a^i$. From (4.10) it then follows that the
Hamiltonian vector field $X_{G(v)}$ is given by
$$X_{G(v)}=\int_\S-(\cD_a v^i){\d\over\d A_a^i},\eqno(4.11)$$
so that
$$ A_a^i\mapsto A_a^i-\e\cD_a v^i+ O(\e^2)\eqno(4.12)$$
under the 1-parameter family of \diffeos on ${}^G\G_{CS}$ associated with
$X_{G(v)}$. This is the usual gauge transformation of the connection 1-form
that we find in \YM theory. Thus, $G(v)$ can be appropriately called a Gauss
constraint function.

Given (4.10), it is also straightforward to evaluate the \Pbs of the
constraints. We find that
$$\{G(v),G(w)\}=G([v,w]),\eqno(4.13)$$
which is the expected \Pb algebra of the Gauss constraint functions. The map
$v^i\mapsto G(v)$ is a representation of the Lie algebra $\cL_G$. The Lie
bracket in $\cL_G$ is mapped to the \Pb of the corresponding Gauss constraint
functions.
\vskip .5cm

\noindent{\sl 4.4 Relationship to the 2+1 \P theory}
\medskip

Before we can show the relationship between the \CS and 2+1 \P theories, we
will first have to recall the construction of the {\it inhomogeneous Lie group}
$IG$ associated with any Lie group $G$. This will allow us to generalize the
equivalence of the 2+1 \P and \CS theories (as shown in \cite{Witten,AT}) to
arbitrary Lie groups $G$. We will be able to show that the 2+1 \P theory based
on any $G$ is equivalent to \CS theory based on $IG$.

Consider any Lie group $G$ with Lie algebra $\cL_G$, and let $\cL_G^*$ denote
the vector space dual of $\cL_G$. If $v^I$, $w^I$ denote typical elements of
$\cL_G$ and $\a_I$, $\b_I$ denote typical elements of $\cL_G^*$, then
$(\a,v)^i:=(\a_I,v^I)$ and $(\b,w)^i:=(\b_I,w^I)$ are typical elements of the
direct sum vector space $\cL_G^*\dsum\cL_G$. We can define a bracket on
$\cL_G^*\dsum\cL_G$ via
$$[(\a,v),(\b,w)]^i:=(-\{v,\b\}+\{w,\a\},[v,w])^i,\eqno(4.14)$$
where $[v,w]^I:=C^I{}_{JK}v^J w^K$ and $\{v,\b\}_I:=C^K{}_{JI}v^J\b_K$ are
the Lie bracket and co-adjoint bracket associated with $\cL_G$. By inspection,
we see that (4.14) is linear and anti-symmetric. If we use
$$\{[v,w],\a\}_I=-\{v,\{w,\a\}\}_I+\{w,\{v,\a\}\}_I\eqno(4.15)$$
(which follows as a consequence of the Jacobi identity for $\cL_G$), we can
show that (4.14) satisfies the Jacobi identity as well. Thus, the vector space
$\cL_{IG}:=\cL_G^*\dsum\cL_G$ together with (4.14) is actually a Lie algebra.
We call $\cL_{IG}$ the {\it inhomogeneous Lie algebra} associated with $G$; the
{\it inhomogeneous Lie group} $IG$ is obtained by exponentiating the Lie
algebra $\cL_{IG}$. As we shall see later in this subsection, $IG$ is simply
the cotangent bundle over $G$.

The terminology inhomogeneous is due to the fact that $\cL_{IG}$ admits
an abelian Lie ideal isomorphic to $\cL_G^*$, and that the quotient of
$\cL_{IG}$ by this ideal is isomorphic to $\cL_G$.\footnote{A {\it Lie ideal}
$\cI$ of a Lie algebra $\cL$ is a vector subspace $\cI\subset\cL$ such that
$[i,x]\in\cI$ for any $i\in\cI, \ x\in\cL$.} Thus, elements of $\cL_G^*$ are
analogous to \inf ``translations," while elements of $\cL_G$ are analogous to
\inf ``rotations."  Note, however, that the space of translations and rotations
have the same dimension. As a special case, if one chooses $G$ to be the 2+1
dimensional Lorentz group $SO(2,1)$, then the above construction yields for
$IG$ the 2+1 dimensional Poincar\'e group $ISO(2,1)$.

In addition to the above Lie algebra structure, $\cL_G^*\dsum\cL_G$ is equipped
with a (natural) invariant, non-degenerate bilinear form $k_{ij}$ defined by
$$k_{ij}(\a,v)^i(\b,w)^j:=\a_I w^I+\b_I v^I.\eqno(4.16)$$
Since $\cL_{IG}$ is not semi-simple (because it admits a non-trivial abelian
Lie ideal), $\ k_{ij}$ is  not the (degenerate) Cartan-Killing metric of
$\cL_{IG}$. Nevertheless, the existence of $k_{ij}$ will allow us to construct
\CS theory for $IG$. Recall that without an invariant, non-degenerate bilinear
form, \CS theory could  not be defined. Note also that for $G= SO(2,1)$, the
above construction of $k_{ij}$ reduces to that used by Witten \cite{Witten}.

Given these remarks, we can now show that the 2+1 \P theory based on any Lie
group $G$ is equivalent to \CS theory based on $IG$. To do this, recall that
for any Lie group $G$, the 2+1 \P action based on $G$ is given by
$${}^G\!S_P(\3 e,\3 A)={1\over 2}\int_M\otw\n^{abc}\ \3 e_{aI}\ \3 F_{bc}^I,
\eqno(4.17)$$
where $\3 A_a^I$ and $\3 e_{aI}$ are $\cL_G$- and $\cL_G^*$-valued 1-forms.
We now construct the inhomogeneous Lie algebra $\cL_{IG}$ associated with $G$
and define an $\cL_{IG}$-valued connection 1-form $\3 A_a^i$ via
$$\3 A_a^i:=(\3 e_{aI},\ \3 A_a^I).\eqno(4.18)$$
By simply substituting this expression for $\3 A_a^i$ into the \CS action
${}^{IG}\!S_{CS}(\3 A)$, we find that
$$\eqalign{{}^{IG}\!S_{CS}(\3 A)&={1\over 2}\int_M\otw\n^{abc}k_{ij}\Big(\
\3 A_a^i\pd_b\3 A_c^j+ {1\over 3}\ \3 A_a^i[\3 A_b,\3 A_c]^j\Big)\cr
&={1\over 2}\int_M\otw\n^{abc}\Big(\ \3 e_{aI}\pd_b\3 A_c^I+(\pd_b
\3 e_{cI})\3 A_a^I\cr
&\qquad\qquad+{1\over 3}(\ \3 e_{aI}[\3 A_b,\3 A_c]^I-\{\3 A_b,\3
e_c\}_I\ \3 A_a^I +\{\3 A_c,\3 e_b\}_I\ \3 A_a^I)\Big)\cr
&={1\over 2}\int_M\otw\n^{abc}\Big(\ \3 e_{aI}\pd_b\3 A_c^I+\pd_b
(\3 e_{cI}\3 A_a^I)-\3 e_{cI}\pd_b\3 A_a^I\cr
&\qquad\qquad+{1\over 3}(\ \3 e_{aI}[\3 A_b,\3 A_c]^I-\3 e_{cI}[\3 A_b,\3
A_a]^I +\3 e_{bI}[\3 A_c,\3 A_a]^I)\Big)\cr
&={1\over 2}\int_M\otw\n^{abc}\Big(\ \3 e_{aI}(2\pd_b\3 A_c^I)+\3 e_{aI}
[\3 A_b,\3 A_c]^I+\pd_b(\3 e_{cI}\3 A_a^I)\Big)\cr
&={1\over 2}\int_M\otw\n^{abc}\ \3 e_{aI}\ \3 F_{bc}^I+(\hbox{\rm surface \
integral})\cr
&={}^G\!S_P(\3 e,\3 A)+(\hbox{\rm surface \ integral}),\cr}\eqno(4.19)$$
where we have used definitions (4.14), (4.16), and (4.18) repeatedly.  Since
the surface term does not affect the Euler-Lagrange equations of motion, we can
conclude that the 2+1 \P theory based on $G$ is equivalent to \CS theory based
on $IG$. This is the desired result. Note that as a special case, we can
conclude that 2+1 gravity as described by the 2+1 \P action based on $SO(2,1)$
is equivalent to \CS theory based on $ISO(2,1)$.

Up to now, we have only described the inhomogeneous Lie group $IG$ in terms of
its associated Lie algebra $\cL_{IG}$. We just exponentiated the Lie algebra
$\cL_{IG}$ to obtain $IG$.  However, it is also instructive to give an explicit
construction of $IG$ at the level of groups and manifolds.  But to do this, we
will need to make another short digression, this time on {\it semi-direct
products} and {\it semi-direct sums}. Readers already familiar with these
definitions may skip to the paragraph immediately following equations $(4.24)$.

Let $G$ and $H$ be any two groups. To define the semi-direct product $H\semis
G$, we need a homomorphism $\s$ from the group $G$ into the group of
automorphisms of $H$---i.e., for each $g,g'\in G$ and $h',h'\in H$, the map
$\s_g:H\rightarrow H$ must be 1-1, onto, and satisfy
$$\s_g(hh')=\s_g(h)\s_g(h')\quad{\rm and}\quad\s_{gg'}(h)=\s_g(\s_{g'}(h))
\eqno(4.20)$$
for every $g,g'\in G$ and $h',h'\in H$. Given this structure, one can check
that
$$(h,g)(h',g')=(h\s_g(h'),gg')\eqno(4.21)$$
defines a group multiplication law on the set $H\times G$. The identity element
is $(e_H,e_G)$, where $e_H,\ e_G$ are the identities in $H$ and $G$, and the
inverse $(h,g)^{-1}$ of $(h,g)$ is $(\s_{g^{-1}}(h^{-1}), g^{-1})$.  The set
$H\times G$ together with this group multiplication law defines the {\it
semi-direct product} $H\semis G$. Note that $H$ is homomorphic to a (not
necessarily abelian) normal subgroup of $H\semis G$, and the quotient of
$H\semis G$ by this normal subgroup is homomorphic to $G$. As a trivial
example, if $\s_g(h)=h$ for all $g\in G,\ h\in H$, then $H\semis G$ is the
usual direct product $H\dprod G$ of groups.

Now assume that $G$ and $H$ are Lie groups with Lie algebras $\cL_G$ and
$\cL_H$. If $H\semis G$ is the semi-direct product of $G$ and $H$ \wrt some
action $\s$ of $G$ on $H$ satisfying (4.20), we would now like to determine the
relationship between the Lie algebras $\cL_{H\semis G}$, $\cL_G$, and $\cL_H$.
To do this, we differentiate the action $\s$ of $G$ on $H$ to obtain an action
$\t$ of $\cL_G$ on $\cL_H$. More precisely, if $g(\e)$ is a 1-parameter curve
in $G$ with $g(0)=e_G$ and tangent vector $v:={d\over d\e}\big|_{\e=0}g(\e)$
and $h(\l)$ is a 1-parameter curve in $H$ with $h(0)=e_{H}$ and tangent vector
$\a:={d\over d\l}\big|_{\l=0}h(\l)$, then we define
$$\t_v(\a):={d\over d\e}\Big|_{\e=0}\ {d\over d\l}\Big|_{\l=0}\ \s_{g(\e)}
(h(\l)).\eqno(4.22)$$
In terms of $\t$, the Lie bracket of $(\a,v)$ and $(\b,w)$ in $\cL_{H\semis G}$
becomes\footnote{To obtain this result, consider 1-parameter curves
$(h(\e),g(\e))$ and $(h'(\e'),g'(\e'))$ in $H\semis G$ with
$(h(0),g(0))=(h'(0),g'(0))=(e_H,e_G)$ and tangent vectors $(\a,v):={d\over
d\e}|_{\e=0}(h(\e),g(\e))$ and $(\b,w):={d\over d\e'}|_{\e'=0}(h'(\e'),
g'(\e'))$.  Then use the definition of the Lie bracket in terms
of the group multiplication law (4.21), $[(\a,v),(\b,w)]:={d\over d\e}|_{\e=0}
{d\over d\e'}|_{\e'=0}(h(\e),g(\e))(h'(\e'),g'(\e'))(h(\e),g(\e))^{-1}(h'(\e'),
g'(\e'))^{-1}$. This leads to (4.23).}
$$[(\a,v),(\b,w)]=(\t_v(\b)-\t_w(\a)+[\a,\b],[v,w]),\eqno(4.23)$$
where $[v,w]$ and $[\a,\b]$ are the Lie brackets of $v,w\in\cL_G$ and $\a,\b
\in\cL_H$. Note that (4.23) satisfies the Jacobi identity as a consequence of
$$\eqalignnotwo{&\t_v([\a,\b])=[\t_v(\a),\b]+[\a,\t_v(\b)]\quad{\rm and}
&(4.24a)\cr
&\t_{[v,w]}(\a)=\t_v(\t_w(\a))-\t_w(\t_v(\a)),&(4.24b)\cr}$$
which follow from the definition of $\t$ and the properties (4.20) satisfied by
$\s$. Thus, if $\cL_G$ and $\cL_H$ are two Lie algebras and $\t$ is an action
of $\cL_G$ on $\cL_H$ satisfying (4.24), then the direct sum vector space
$\cL_H\dsum\cL_G$ together with the bracket defined by (4.23) is a Lie algebra.
This Lie algebra, denoted $\cL_H\semit\cL_G$, is called the {\it semi-direct
sum} of $\cL_G$ and $\cL_H$. Note that $\cL_H$ is isomorphic to a (not
necessarily abelian) Lie ideal of $\cL_H\semit\cL_G$, and the quotient of
$\cL_H\semit\cL_G$ by this ideal is isomorphic to $\cL_G$. If $\s_g(h)=h$ for
all $g\in G,\ h\in H$ (so that $H\semis G=H\dprod G$), then $\cL_H\semit\cL_G$
is the usual direct sum $\cL_H\dsum \cL_G$ of Lie algebras.

Given these general remarks, let us now return to our discussion of the
inhomogeneous Lie group $IG$ and its associated Lie algebra $\cL_{IG}$. From
the above definitions, we see that $\cL_{IG}$ is simply the semi-direct sum
$\cL_G^*\semit\cL_G$. \ $\cL_G^*$ is to be thought of as a Lie algebra with the
trivial Lie bracket $[\a,\b]=0$ for all $\a_I,\ \b_I \in\cL_G^*$; \ the action
$\t$ of $\cL_G$ on $\cL_G^*$ is given by $\t_v(\b)=-\{v,\b\}_I$. Equations
(4.24) hold for this action as a consequence of the Jacobi identity in $\cL_G$:
\ Equation ($4.24a$) is satisfied since $[\a,\b]=0$ for all $\a_I,\
\b_I\in\cL_G^*$, while equation ($4.24b$) is equivalent to equation (4.15).
Furthermore, the inhomogenized Lie group $IG$ is simply the semi-direct product
$\cL_G^*\semis G$. \ $\cL_G^*$ is to be thought of as an abelian group \wrt
vector addition, and the action $\s$ of $G$ on $\cL_G^*$ is induced by the
adjoint action of $G$ on itself.\footnote{The {\it adjoint action} of $G$ on
itself is defined by $A_g(g')=gg'g^{-1}$ for all $g,g'\in G$. By
differentiating $A_g$ at the identity $e$, we obtain a map $Ad_g:\cL_G
\rightarrow\cL_G$ via $Ad_g(v):=A_g'(e)\cdot v$.  \ $Ad$ defines the {\it
adjoint representation} of the Lie group $G$ by linear operators on the Lie
algebra $\cL_G$.  The action $\s$ of $G$ on $\cL_G^*$ is then given by
$(\s_g(\a))(v):=\a(Ad_g(v))$ for any $\a\in\cL_G^*$ and $v\in\cL_G$.} This
implies that as a manifold $IG$ is the cotangent bundle $T^*G$. At each point
$g\in G$, the cotangent space $T_g^*G$ is isomorphic to $\cL_G^*$.

Moreover, the above relationship between $G$ and $IG$ allows us to prove an
interesting mathematical result involving the space of connection 1-forms on a
2-dimensional manifold. We can show that for any Lie group $G$
$$T^*({}^G\!\cA)=\ {}^{IG}\!\cA,\eqno(4.25)$$
where ${}^G\!\cA$ and ${}^{IG}\!\cA$ denote the space of $\cL_G$- and
$\cL_{IG}$-valued connection 1-forms on a 2-dimensional manifold $\S$. The map
$$(A_a^I,\otw E_I^a)\in T^*({}^G\!\cA)\mapsto A_a^i:=(e_{aI},A_a^I)\in{}^{IG}
\!\cA\eqno(4.26)$$
(where $e_{aI}:=-\utw\n{}_{ab}\otw E_I^b$) is a diffeomorphism from the
manifold $T^*({}^G\!\cA)$ to the manifold ${}^{IG}\!\cA$ that sends the natural
symplectic structure
$${}^G\W:=\int_\S\id\otw E_I^a\ \iwedge\ \id A_a^I\eqno(4.27)$$
on $T^*({}^G\!\cA)$ to the natural symplectic structure
$${}^{IG}\W:=-{1\over 2}\int_\S\otw\n^{ab}k_{ij}\id A_a^i\ \iwedge\ \id
A_b^j\eqno(4.28)$$
on ${}^{IG}\!\cA$.  (Here $k_{ij}$ denotes the (natural) invariant,
non-degenerate bilinear form on $\cL_{IG}$ defined by (4.16).) Note that (4.27)
and (4.28) are the symplectic structures of the 2+1 \P theory based on $G$
and the \CS theory based on $IG$. However, the above result (4.25) does not
require any knowledge of the 2+1 \P or \CS actions.

Finally, to conclude this subsection, I would like to verify the claim made at
the start of Section 4 that the reduced phase space of \CS theory based on any
Lie group $G$ is the reduced configuration space of the 2+1 \P theory based on
the {\it same} Lie group $G$. This result will be simpler to prove than the
previous two results since most of the preliminary work has already been done.

Let $G$ be any Lie group whose Lie algebra $\cL_G$ admits an invariant,
non-degenerate bilinear form $k_{IJ}$. Then \CS theory based on $G$ is
well-defined, and, as we saw in subsection 4.2, the phase space ${}^G\G_{CS}$
is coordinatized by $\cL_G$-valued connection 1-forms $ A_a^I$ on $\S$. The
symplectic structure is
$${}^G\W_{CS}=-{1\over 2}\int_\S\otw\n^{ab}k_{IJ}\id A_a^I\ \iwedge\ \id A_b^J.
\eqno(4.29)$$
In subsection 4.3, we then verified that the constraint functions $G(v)$
associated with the the constraint equation
$$k_{IJ}\otw\n^{ab} F_{ab}^J=0\eqno(4.30)$$
formed a 1st class set and generated the usual gauge transformations
$$ A_a^I\mapsto A_a^I-\e\cD_a v^I+O(\e^2).\eqno(4.31)$$
To pass to the reduced phase space, we must factor-out the constraint surface
(defined by (4.30)) by the orbits of the Hamiltonian vector fields
$X_{G(v)}$.\footnote{Recall that given a symplectic manifold $(\G,\W)$, a set
of constraints $\phi_{\und i}$ form a 1st class set \iff each Hamiltonian
vector field $X_{\phi_{\und i}}$ is tangent to the constraint surface
$\ovr\G\subset\G$ defined by the vanishing of all the constraints. The
pull-back, $\ovr\W$, of $\W$ to $\ovr\G$ is {\it degenerate} with the
degenerate directions given by the $X_{\phi_{\und i}}$. Thus, $(\ovr\G,\ovr\W)$
is not a symplectic manifold. However, by factoring-out the constraint surface
by the orbits of the $X_{\phi_{\und i}}$, we obtain a {\it reduced phase space}
$(\hat\G, \hat\W)$ whose coordinates are precisely the true degrees of freedom
of the theory. $\hat\W$ is non-degenerate; it is the projection of $\ovr\W$ to
$\hat\G$.} From (4.30) and (4.31) we see that the {\it reduced phase space}
${}^G\hat\G_{CS}$ of the \CS theory based on $G$ is coordinatized by
equivalence classes of flat $\cL_G$-valued connection 1-forms on $\S$, where
two such connection 1-forms are said to be equivalent \iff they are related by
(4.31). This space is called the {\it moduli space} of flat $\cL_G$-valued
connection 1-forms on $\S$.

Now recall the Hamiltonian formulation of the 2+1 \P theory based on the {\it
same} Lie group $G$. In subsection 3.3, we saw that the phase space ${}^G\G_P$
was coordinatized by pairs $(A_a^I,\otw E_I^a)$ consisting of $\cL_G$-valued
connection 1-forms $A_a^I$ on $\S$ and their canonically conjugate momentum
$\otw E_I^a$. \ ${}^G\G_P$ was the cotangent bundle $T^*({}^G\!\cC_P)$ over the
configuration space ${}^G\!\cC_P$ of $\cL_G$-valued connection 1-forms $A_a^I$
on $\S$ with symplectic structure
$${}^G\W_P=\int_\S\id\otw E_I^a\ \iwedge\ \id A_a^I.\eqno(4.32)$$
We also saw that the constraint equations of the 2+1 \P theory were
$$\otw\n^{ab} F_{ab}^I=0\quad{\rm and}\quad \cD_a\otw E_I^a=0,\eqno(4.33)$$
and verified that their associated constraint functions $F(\a)$ and $G(v)$
formed a 1st class set. They generated the motions
$$\eqalignnotwo{&A_a^I\mapsto A_a^I\quad{\rm and}&(4.34a)\cr
&\otw E_I^a\mapsto\otw E_I^a - \e(\otw\n^{ab}\cD_b\a_I) + O(\e^2)&(4.34b)\cr}$$
and
$$\eqalignnotwo{&A_a^I\mapsto A_a^I-\e\cD_a v^I+O(\e^2)\quad{\rm
and}&(4.35a)\cr
&\otw E_I^a\mapsto\otw E_I^a - \e\{v,\otw E^a\}_I + O(\e^2),&(4.35b)\cr}$$
respectively. Thus, the reduced phase space ${}^G\hat\G_P$ of the 2+1 \P
theory based on $G$ is coordinatized by equivalence classes of pairs consisting
of flat $\cL_G$-valued connection 1-forms on $\S$ and divergence-free
$\cL_G^*$-valued vector densities of weight +1 on $\S$, where two such pairs
are said to be equivalent \iff they are related by (4.34) and (4.35). Since
$F(\a)$ and $G(v)$ are independent and linear in the momentum $\otw E_I^a$, it
follows that \ ${}^G \hat\G_P$ is naturally the cotangent bundle
$T^*({}^G\!\hat\cC_P)$ over the {\it reduced configuration space}
${}^G\!\hat\cC_P$ of the 2+1 \P theory based on $G$. From ($4.34a$) and
($4.35a$) we see that ${}^G\!\hat\cC_P$ is again the moduli space of flat
$\cL_G$-valued connection 1-forms on $\S$. Thus, ${}^G\hat\G_{CS}=
{}^G\!\hat\cC_P$ as desired. In particular, the reduced configuration space of
the 2+1 \P theory based on $G$ has the structure of a symplectic manifold.

This last result has interesting consequences. It can be used, for example, to
show the relationship between the $\hat T^0[\g]$ and $\hat T^1[\g]$ observables
for 2+1 gravity. These are the 2+1 dimensional analogs of the classical
$T$-observables constructed by Rovelli and Smolin \cite{RS} for the 3+1 theory.
As shown in \cite{CSP},  \ $\hat T^0[\g]$ is the trace of the holonomy of
the connection around a closed loop $\g$ in $\S$, while $\hat T^1[\g]$ is the
function on the reduced phase space of the 2+1 \P theory defined by the
Hamiltonian vector field associated with $\hat T^0 [\g]$. Thus, many properties
satisfied by the $\hat T^1[\g]$'s can be derived from similar properties
satisfied by the $\hat T^0[\g]$'s.
\vskip 1cm

\noindent{\bf 5. 2+1 matter couplings}
\vskip .5cm

In this section, we will couple various matter fields to 2+1 gravity via the
2+1 \P action. We will consider the inclusion of a \cc and a massless scalar
field. One can couple other fundamental matter fields (e.g., \YM and Dirac
fields) to 2+1 gravity in a similar fashion---I have chosen to consider a
massless scalar field in detail since 2+1 gravity coupled to a massless scalar
field is the dimensional reduction of 3+1 vacuum \gr with a spacelike,
hypersurface-orthogonal Killing vector field \cite{solns}. As noted in Section
1, this is an interesting case since it appears likely that the
non-perturbative canonical quantization program for 3+1 gravity can be carried
through to completion for this reduced theory.

In subsection 5.1, we define the 2+1 \P theory based on a Lie group $G$ with
\cc $\L$. We derive the \EL \eoms and perform a Legendre transform to obtain a
Hamiltonian formulation of the theory. Just as in Section 3 (when $\L$ was
equal to zero), we shall find that the constraint equations of the theory are
polynomial in the canonically conjugate variables. We shall also find that
their associated constraint functions still form a Lie algebra \wrt Poisson
bracket.

In subsection 5.2, we will show that the 2+1 \P theory based on $G$ with \cc
$\L$ is equivalent to \CS theory based on the $\L$-deformation, $\L G$, of $G$.
This is a generalization of Witten's result \cite{Witten} for $G=SO(2,1)$ (and
$\L G=SO(3,1)$ or $SO(2,2)$ depending on the sign of $\L$) which holds for any
Lie group $G$ that admits an invariant, totally anti-symmetric tensor
$\e^{IJK}$. \ This result also generalizes the $\L=0$ equivalence of the 2+1 \P
and \CS theories given in subsection 4.4.

Finally, in subsection 5.3, we define the action for a massless scalar field
and couple this field to 2+1 gravity by adding the action to the 2+1 \P action
based on $G=SO(2,1)$. It is when we wish to couple matter with local degrees of
freedom to 2+1 gravity (as it is for the case of a massless scalar field) that
we are forced to take the Lie group $G$ to be such that the fields $\3 e_a^I$
have the interpretation of a co-triad. We obtain the \EL \eoms for the coupled
theory and then perform a Legendre transform to obtain the Hamiltonian
formulation.  We shall find that the constraint equations remain polynomial in
the canonically conjugate variables and the associated constraint functions
form a 1st class set, but they no longer form a Lie algebra \wrt Poisson
bracket.

The basis for much of the material in this section can be found in
\cite{Witten,CSP,matter}.
\vskip .5cm

\noindent{\sl 5.1 2+1 \P theory coupled to a cosmological constant}
\medskip

Recall that the equation of motion for gravity coupled to the cosmological
constant $\L$ is
$$G_{ab}+\L g_{ab}=0,\eqno(5.1)$$
where $G_{ab}:=R_{ab}-{1\over 2}R g_{ab}$ is the Einstein tensor of $g_{ab}$.
We can obtain (5.1) via an action principle if we modify the standard \EH
action by a term proportional to the volume of the spacetime. Defining
$$S_\L(g^{ab}):=\int_\S\sqrt{-g}(R-2\L),\eqno(5.2)$$
we find that the variation of (5.2) \wrt $g^{ab}$ yields (5.1) (modulo the
usual boundary term associated with the standard \EH action). These results are
valid in $n$+1 dimensions.

To write this action in 2+1 \P form, we proceed as in Section 3. We replace
the \st metric $g_{ab}$ with a co-triad $\3 e_{aI}$ and replace the unique,
torsion-free \st derivative operator $\grad_a$ (compatible with $g_{ab}$) with
an arbitrary generalized derivative operator $\3\cD_a$. Recalling that
$\sqrt{-g}={1\over 3!}\otw\n^{abc}\e^{IJK}\ \3 e_{aI}\ \3 e_{bJ} \ \3 e_{cK}$,
we define
$$\eqalign{S_\L(\3 e,\3 A):&={1\over 2}\int_M\otw\n^{abc}\ \3 e_{aI}\ \3
F_{bc}^I -{\L\over 3!}\int_M\otw\n^{abc}\e^{IJK}\ \3 e_{aI}\ \3 e_{bJ}\ \3
e_{cK}\cr
&={1\over 2}\int_M\otw\n^{abc}\ \3 e_{aI}(\ \3 F_{bc}^I-{\L\over 3}\e^{IJK}\ \3
e_{bJ} \ \3 e_{cK}),\cr}\eqno(5.3)$$
where $\3 F_{ab}^I=2\pd_{[a}\3 A_{b]}^I+[\3 A_a,\3 A_b]^I$ is the internal
curvature tensor of the generalized derivative operator $\3\cD_a$ defined
by
$$\3\cD_a v^I:=\pd_a v^I+[\3 A_a,v]^I.\eqno(5.4)$$
Note that $[\3 A_a,v]^I:=\e^I{}_{JK}\ \3 A_a^J v^K$ where $\e^I{}_{JK}:=
\e^{IMN}\n_{MJ}\n_{NK}$. Just as we did for the vacuum 2+1 \P theory, we have
included an additional overall factor of $1/2$ in definition (5.3).

Although the action (5.3) was originally defined for $G=SO(2,1)$, it is
well-defined for any Lie group $G$ that admits an invariant, totally
anti-symmetric tensor $\e^{IJK}$.\footnote{We will require that $\e^{IJK}$ be
invariant under the adjoint action of the Lie algebra $\cL_G$ on its dual
$\cL_G^*$---i.e., that $\e^{IJK}\{v,\a\}_I\b_J\g_K+\e^{IJK}\a_I\{v,\b\}_J\g_K
+\e^{IJK}\a_I\b_J\{v,\g\}_K=0$ for all $v^I\in\cL_G$ and $\a_I,\ \b_I,\ \g_I\in
\cL_G^*$. ($\{v,\a\}_I$ is the co-adjoint bracket of $v^I$ and $\a_I$ defined
in terms of the structure constants $C^I{}_{JK}$ via $\{v,\a\}_I:=C^K{}_{JI}v^J
\a_K$.) Invariance of $\e^{IJK}$ is equivalent to $\e^{M[IJ}C^{K]}{}_{MN} =0$.}
This is additional structure that does not naturally exist for an arbitrary Lie
group $G$, so unlike the 2+1 \P theory with $\L=0$, the 2+1 \P theory with
non-zero cosmological constant $\L$ is not defined for arbitrary $G$. If the
Lie algebra $\cL_G$ admits an invariant, non-degenerate bilinear form $k_{IJ}$,
then we are guaranteed that such an $\e^{IJK}$ exists---we can take
$\e^{IJK}:=k^{JM}k^{KN}C^I{}_{MN}$. Thus, in particular, 2+1 \P theory with a
non-zero cosmological constant is well-defined for semi-simple Lie groups.  We
should emphasize, however, that it is not necessary to restrict ourselves to
semi-simple Lie groups.  In what follows, we will only assume that $\e^{IJK}$
exists. Given such a Lie group $G$, the action
$${}^G\!S_\L(\3 e,\3 A):={1\over 2}\int_M\otw\n^{abc}\ \3 e_{aI}(\ \3
F_{bc}^I-{\L\over 3}\e^{IJK}\ \3 e_{bJ}\ \3 e_{cK})\eqno(5.5)$$
will be called the 2+1 {\it \P action based on $G$ with \cc $\L$}. Note that
$$\3 F_{ab}^I=2\pd_{[a}\3 A_{b]}^I+[\3 A_a,\3 A_b]^I,\eqno(5.6)$$
where $[\3 A_a,\3 A_b]^I:=C^I{}_{JK}\ \3 A_a^J\ \3 A_b^K$ is the Lie bracket in
$\cL_G$.  It is only for $G=SO(2,1)$ that $C^I{}_{JK}=\e^I{}_{JK}=
\e^{IMN}\n_{MJ}\n_{NK}$.

To obtain the \EL equations of motion, we vary ${}^G\!S_\L(\3 e,\3 A)$ \wrt
both $\3 e_{aI}$ and $\3 A_a^I$. We find
$$\otw\n^{abc}(\ \3 F_{bc}^I-\L\e^{IJK}\ \3 e_{bJ}\ \3 e_{cK})=0\quad{\rm
and}\quad\3\cD_b(\otw\n^{abc}\ \3 e_{cI})=0,\eqno(5.7)$$
where the second equation, as usual, requires a torsion-free extension of the
generalized derivative operator $\3\cD_a$ to \st tensor fields, but is
independent of this choice. Note further that if $\L\not=0$, $\3\cD_a$ is not
flat. In fact, for the special case $G=SO(2,1)$, equations (5.7) imply that the
\st $(M,g_{ab}:=\3 e_{aI}\3 e_{bJ}\n^{IJ})$ has constant curvature equal to
$6\L$.

To show that the above two equations reproduce (5.1), let us restrict ourselves
to $G=SO(2,1)$ (with $\e^{IJK}$ being the volume element of $\n_{IJ}$) so that
$\3 e_{aI}$ is, in fact, a co-triad. Then by following the argument given in
Section 3, we find that the second equation implies $\3 A_a^I=\G_a^I$, where
$\G_a^I$ is the (internal) Christoffel symbol of the unique, torsion-free
generalized derivative operator $\grad_a$ compatible with the co-triad $\3
e_{aI}$. Thus, $\3\cD_a$ is not arbitrary, but equals $\grad_a$ when acting on
internal indices. Substituting this solution back into the first equation, we
find
$$\otw\n^{abc}(R_{bc}^I-\L\e^{IJK}\ \3 e_{bJ}\ \3 e_{cK})=0, \eqno(5.8)$$
where $R_{ab}^I$ is the (internal) curvature tensor of $\grad_a$. Contracting
(5.8) with $\3 e_I^d$ gives
$$G^{ad}+\L g^{ad}=0.\eqno(5.9)$$
This is the desired result.

To put this theory in Hamiltonian form, we will assume that $M$ is
topologically $\S\times R$, and assume that there exists a function $t$
(with nowhere vanishing gradient $(dt)_a$) such that each $t=\const$ surface
$\S_t$ is diffeomorphic to $\S$. By $t^a$  we will denote the flow vector field
satisfying $t^a(dt)_a=1$. Using $\otw\n^{abc}=3t^{[a}\otw\n^{bc]}dt$ (and our
decomposition of ${}^G\!S_P(\3 e,\3 A)$ from Section 3), we obtain
$$\eqalign{{}^G\!S_\L(\3 e,\3 A)=\int dt\int_\S{1\over 2}(\3 e\cdot
t)_I(\otw\n^{ab} F_{ab}^I &- \L\e^{IJK}\utw\n{}_{ab}\otw E_J^a\otw E_K^b)\cr
&+\otw E_I^a\cL_{\vec t} \ A_a^I - \otw E_I^a\cD_a(\3 A\cdot t)^I.\cr}
\eqno(5.10)$$
The configuration variables are $(\3 e\cdot t)_I:=t^a\ \3 e_{aI}$, \ $\otw
E_I^a:=\otw\n^{ab}\ \3 e_{bI}$, \ $(\3 A\cdot t)^I:=t^a\ \3 A_a^I$, and
$A_a^I:=t_a^b\ \3 A_b^I$. Thus, (modulo a surface integral) the Lagrangian
${}^G\!L_\L$ of the 2+1 \P theory based on $G$ with cosmological constant
$\L$ is given by
$$\eqalign{{}^G\!L_\L=\int_\S{1\over 2}(\3 e\cdot t)_I(\otw\n^{ab}F_{ab}^I &-
\L\e^{IJK} \utw\n{}_{ab}\otw E_J^a\otw E_K^b)\cr
&+\otw E_I^a\cL_{\vec t} \ A_a^I + (\cD_a\otw E_I^a)(\3 A\cdot t)^I.\cr}
\eqno(5.11)$$
${}^G\!L_\L$ is to be viewed as a functional of the configuration variables and
their first derivatives.

Following the standard Dirac constraint analysis, we find that the momentum
canonically conjugate to $A_a^I$ is $\otw E_I^a$, while $(\3 e\cdot t)_I$ and
$(\3 A\cdot t)^I$ both play the role of Lagrange multipliers. Thus, the phase
space and symplectic structure are the same as those found for the 2+1 \P
theory with $\L=0$, and the Hamiltonian is given by
$${}^G\!H_\L(A,\otw E)=\int_\S-{1\over 2}(\3 e\cdot t)_I(\otw\n^{ab}F_{ab}^I -
\L\e^{IJK} \utw\n{}_{ab}\otw E_J^a\otw E_K^b)-(\cD_a\otw E_I^a)(\3 A\cdot
t)^I.\eqno(5.12)
$$
We will see that this is just a sum of 1st class constraint functions
associated with
$$\otw\n^{ab} F_{ab}^I - \L\e^{IJK}\utw\n{}_{ab}\otw E_J^a\otw
E_K^b\approx 0\quad{\rm and}\quad\cD_a\otw E_I^a\approx 0.\eqno(5.13)$$
By inspection, constraint equations (5.13) are polynomial in the canonically
conjugate variables $(A_a^I,\otw E_I^a)$. They are the field equations (5.7)
pulled-back to $\S$ with $\otw\n^{ab}$.

As usual, given test fields $\a_I$ and $v^I$, which take values in $\cL_G^*$
and $\cL_G$, we can define constraint functions
$$F(\a):={1\over 2}\int_\S\a_I(\otw\n^{ab} F_{ab}^I-\L\e^{IJK}
\utw\n{}_{ab}\otw E_J^a \otw E_K^b)\quad{\rm and}\quad G(v):=\int_\S
v^I(\cD_a\otw E_I^a).\eqno(5.14)$$
Note that $G(v)$ is unchanged from the 2+1 \P theory with $\L=0$, while $F(\a)$
has an additional term {\it quadratic} in the momentum $\otw E_I^a$. There is
only one new functional derivative,
$${\d F(\a)\over \d\otw E_I^a}=-\L\e^{IJK}\utw\n{}_{ab}\otw E_J^b\a_K.\eqno
(5.15)$$
All the others are the same as before.

Under the 1-parameter family of \diffeos associated with the Hamiltonian
vector field $X_{F(\a)}$, we have
$$\eqalignnotwo{&A_a^I\mapsto A_a^I - \e(\L\e^{IJK}\utw\n{}_{ab}\otw E_J^b\a_K)
+O(\e^2)\quad{\rm and}&(5.16a)\cr
&\otw E_I^a\mapsto\otw E_I^a - \e(\otw\n^{ab}\cD_b\a_I) +
O(\e^2).&(5.16b)\cr}$$
Similarly, under the 1-parameter family of \diffeos associated with the
Hamiltonian vector field $X_{G(v)}$, we have
$$\eqalignnotwo{&A_a^I\mapsto A_a^I-\e\cD_a v^I+O(\e^2)\quad{\rm
and}&(5.17a)\cr
&\otw E_I^a\mapsto\otw E_I^a - \e\{v,\otw E^a\}_I + O(\e^2).&(5.17b)\cr}$$
Comparing these results with those from the 2+1 \P theory with $\L=0$, we see
that the motion of $\3 A_a^I$ generated by the constraint functions  no
longer corresponds to the usual gauge transformation of \YM theory. This is due
to the non-zero contribution from $F(\a)$. In fact, since $F(\a)$ depends
quadratically on the momentum $\otw E_I^a$, the reduced phase space of the 2+1
\P theory with non-zero \cc $\L$ is not naturally a cotangent bundle over a
reduced configuration space.  Thus, the result of Section 4 that the reduced
phase space of the \CS theory based on $G$ equals the reduced configuration
space of the 2+1 \P theory based on the same $G$ does not extend in general to
the case $\L\not=0$.

Nevertheless, we can still evaluate the Poisson brackets of the constraint
functions $F(\a)$ and $G(v)$. As in the $\L=0$ case, we find that
$$\{G(v),G(w)\}=G([v,w]),\eqno(5.18)$$
where $[v,w]^I=C^I{}_{JK} v^J w^K$ is the Lie bracket of $v^I$ and $w^I$, so
$v^I\mapsto G(v)$ is a representation of the Lie algebra $\cL_G$. Although
$F(\a)$ has changed, we again find that
$$\{G(v),F(\a)\}=-F(\{v,\a\}),\eqno(5.19)$$
where $\{v,\a\}_I=C^K{}_{JI}v^J\a_K$ is the co-adjoint bracket of $v^I$ and
$\a_I$. However, the \Pb of $F(\a)$ with $F(\b)$ is no longer zero; it equals
$$\{F(\a),F(\b)\}=-\L G(\e(\a,\b)),\eqno(5.20)$$
where $\e(\a,\b)^I:=\e^{IJK}\a_J\b_K$. Thus, the totality of constraint
functions is closed under \Pb---i.e., they form a 1st class set.  In fact,
since (5.18), (5.19), and (5.20) do not involve any structure functions,
the constraint functions form a Lie algebra \wrt Poisson bracket. The mapping
$(\a,v)\in\cL_G^*\times\cL_G\mapsto (F(\a),G(v))$ is a representation of the
Lie algebra $\cL_{\L G}$ of the $\L$-{\it deformation}, $\L G$, of the Lie
group $G$.\footnote{We will discuss the construction of the $\L$-deformation
of a Lie group $G$ in the following section where we show the equivalence
between the 2+1 \P theory based on $G$ with cosmological constant $\L$ and
\CS theory based on $\L G$. When $\L=0$, $\L G$ is just the inhomogeneous Lie
group $IG$ constructed in subsection 4.4.} The $\ F(\a)$'s play the role of
``boosts'' while the $G(v)$'s play the role of ``rotations'' in the
$\L$-deformation of $G$.
\vskip .5cm

\noindent{\sl 5.2 Relationship to \CS theory}
\medskip

In a manner similar to that used in subsection 4.4, we will now show that if
$G$ is any Lie group which admits an invariant, totally anti-symmetric tensor
$\e^{IJK}$, then the 2+1 \P theory based on $G$ with cosmological constant
$\L$ is equivalent to \CS theory based on the $\L$-deformation, $\L G$, of the
Lie group $G$. The actions for these two theories are the same modulo a surface
term that does not affect the equations of motion.

Given a Lie group $G$ with an invariant, totally anti-symmetric tensor
$\e^{IJK}$, we first construct the $\L$-{\it deformation}, $\L G$, of $G$ as
follows: \ Form the direct sum vector space $\cL_G^*\dsum\cL_G$ (having typical
elements $(\a,v)^i :=(\a_I,v^I)$ and $(\b,w)^i:=(\b_I,w^I)$) and then define a
bracket on $\cL_G^*\dsum\cL_G$ via
$$[(\a,v),(\b,w)]^i:=(-\{v,\b\}+\{w,\a\},[v,w]-\L\e(\a,\b))^i,\eqno(5.21)$$
where $[v,w]^I:=C^I{}_{JK}v^J w^K$, \ $\{v,\b\}_I:=C^K{}_{JI}v^J\b_K$, and
$\e(\a,\b)^I:=\e^{IJK}\a_J\b_K$. By inspection, (5.21) is linear and
anti-symmetric. By using the Jacobi identity $C^M{}_{[IJ}C^N{}_{K]M}=0$ on
$\cL_G$ together with the anti-symmetry and invariance of $\e^{IJK}$, one can
show that (5.21) satisfies the Jacobi identity as well. Thus, the vector space
$\cL_{\L G}:=\cL_G^*\dsum\cL_G$ together with (5.21) is actually a Lie algebra.
We call $\cL_{\L G}$ the $\L$-{\it deformed Lie algebra} associated with $G$.
The $\L$-{\it deformation}, $\L G$, of $G$ is obtained by exponentiating
$\cL_{\L G}$. We can think of $\L G$ as an extension of the inhomogeneous Lie
group $IG$ in the sense that $\L G$ reduces to $IG$ when $\L=0$. Note also that
if $G=SO(2,1)$, then the above construction for $\L G$ yields $SO(3,1)$ if
$\L<0$ and $SO(2,2)$ if $\L>0$.

In addition to the above Lie algebra structure, $\cL_G^*\dsum\cL_G$ is also
equipped with a (natural) invariant, non-degenerate bilinear form
$$k_{ij}(\a,v)^i(\b,w)^j:=\a_I w^I+\b_I v^I.\eqno(5.22)$$
This is the same $k_{ij}$ that we had when $\L=0$.  As before, the existence
of $k_{ij}$ will allow us to construct \CS theory for $\L G$.

Given these remarks, we are now ready to verify that the 2+1 \P theory based
on $G$ with cosmological constant $\L$ is equivalent to \CS theory based on $\L
G$. Recall that
$${}^G\!S_\L(\3 e,\3 A)={1\over 2}\int_M\otw\n^{abc}\ \3 e_{aI}(\ \3 F_{bc}^I -
{\L\over 3}\e^{IJK}\ \3 e_{bJ}\ \3 e_{cK}),\eqno(5.23)$$
where $\3 A_a^I$ and $\3 e_{aI}$ are $\cL_G$- and $\cL_G^*$-valued 1-forms.
If we now construct the $\L$-deformed Lie algebra $\cL_{\L G}$ associated with
$G$ and define an $\cL_{\L G}$-valued connection 1-form $\3 A_a^i$ via
$$\3 A_a^i:=(\3 e_{aI},\ \3 A_a^I),\eqno(5.24)$$
then a straightforward calculation along the lines of that used in subsection
4.4 shows that the \CS action
$${}^{\L G}\!S_{CS}(\3 A)={1\over 2}\int_M\otw\n^{abc}k_{ij}\left(\ \3
A_a^i\pd_b\3 A_c^j +{1\over 3}\ \3 A_a^i[\3 A_b,\3 A_c]^j\right)\eqno(5.25)$$
equals ${}^G\!S_\L(\3 e,\3 A)$ modulo a surface term which does not affect the
Euler-Lagrange equations of motion. Specializing to the case $G=SO(2,1)$, we
see that 2+1 gravity coupled to the cosmological constant $\L$ is equivalent to
\CS theory based on $SO(3,1)$ if $\L<0$ or $SO(2,2)$ if $\L>0$. This was the
observation of Witten \cite{Witten}.
\vskip .5cm

\noindent{\sl 5.3 2+1 \P theory coupled to a massless scalar field}
\medskip

So far, we have seen that the 2+1 \P theory (with or without a \cc $\L$) is
well-defined for a wide class of Lie groups. If $\L=0$, the Lie group $G$ can
be completely arbitrary; \ if $\L\not=0$, then $G$ has to admit an invariant,
totally anti-symmetric tensor $\e^{IJK}$.  We are not forced to restrict
ourselves to $G=SO(2,1)$. However, in order to couple fundamental matter fields
with local degrees of freedom to 2+1 gravity via the 2+1 \P action, we will
need to take $G=SO(2,1)$. The matter actions require the existence of a \st
metric $g_{ab}$, and, as such, \ $\3 e_a^I$ must have the interpretation of
a co-triad.  We will only consider coupling a massless scalar field to 2+1
gravity in this section---a similar treatment would work for \YM and Dirac
fields as well.

Let us first recall that the theory of a massless scalar field $\phi$ can be
defined in $n$+1 dimensions. If $g^{ab}$ denotes the inverse of the \st metric
$g_{ab}$, then the {\it \KG action} $S_{KG}(g^{ab},\phi)$ is defined by
$$S_{KG}(g^{ab},\phi):=-8\pi\int_M\sqrt{-g}\ g^{ab}\pd_a\phi\pd_b\phi,
\eqno(5.26)$$
where $\pd_a\phi$ denotes the gradient of $\phi$. To couple the scalar field to
gravity, we simply add the \KG action (5.26) to the standard \EH action
$$S_{EH}(g^{ab})=\int_M\sqrt{-g} R.\eqno(5.27)$$
The {\it total action} $S_T(g^{ab},\phi)$ is then given by the sum
$$S_T(g^{ab},\phi):=S_{EH}(g^{ab})+S_{KG}(g^{ab},\phi),\eqno(5.28)$$
and the \EL \eoms are obtained by varying $S_T(g^{ab},\phi)$ \wrt both $g^{ab}$
and $\phi$. The variation of $\phi$ yields
$$g^{ab}\grad_a\grad_b\phi=0,\eqno(5.29)$$
while the variation of $g^{ab}$ yields
$$G_{ab}=8\pi T_{ab}(KG).\eqno(5.30)$$
$\grad_a$ is the unique, torsion-free \st derivative operator compatible with
the metric $g_{ab}$, and $T_{ab}(KG)$ is the {\it stress-energy tensor} of the
massless scalar field. In terms of $g_{ab}$ and $\phi$, we have
$$T_{ab}(KG):=\pd_a\phi\pd_b\phi-{1\over 2} g_{ab}\pd_c\phi\pd^c\phi.
\eqno(5.31)$$

Now let us restrict ourselves to 2+1 dimensions and rewrite the above actions
and \eoms in 2+1 \P form.  As we saw in Section 3, the 2+1 \P action can be
written as
$$S_P(\3 e,\3 A):={1\over 2}\int_M\otw\n^{abc}\ \3 e_{aI}\ \3 F_{bc}^I,
\eqno(5.32)$$
where $\3 e_{aI}$ is the co-triad related to the \st metric $g_{ab}$
via $g_{ab}:=\3 e_a^I\3 e_b^J\n_{IJ}$ and $\3 F_{ab}^I=2\pd_{[a}\3 A_{b]}^I
+\e^I{}_{JK}\3 A_a^J\3 A_b^K$ is the internal curvature tensor of the
generalized derivative operator $\3\cD_a$ defined by $\3 A_a^I$.  The \KG
action, viewed as a functional of $\3 e_a^I$ and the scalar field $\phi$, is
given by
$$S_{KG}(\3 e,\phi)=-8\pi\int_M\sqrt{-g}\ g^{ab}\pd_a\phi\pd_b\phi.
\eqno(5.33)$$
Note that although the \KG action depends on the co-triad $\3 e_{aI}$ through
its dependence on $\sqrt{-g}$ and $g^{ab}$, it is independent of the connection
1-form $\3 A_a^I$.  In fact, of all the fundamental matter couplings, only the
action for the Dirac field would depend on $\3 A_a^I$.

Given (5.32) and (5.33), we define the total action as the sum
$$S_T(\3 e,\3 A,\phi):=S_P(\3 e,\3 A)+{1\over 2}S_{KG}(\3 e,\phi).\eqno(5.34)
$$
The additional factor of $1/2$ is needed in front of $S_{KG}(\3 e,\phi)$ so
that the above definition of the total action will be consistent with the
definition of $S_P(\3 e,\3 A)$. The \EL \eoms are obtained by varying $S_T(\3
e,\3 A,\phi)$ \wrt each field. Varying $\phi$ gives
$$g^{ab}\grad_a\grad_b\phi=0,\eqno(5.35)$$
while varying $\3 A_a^I$ and $\3 e_{aI}$ imply
$$\eqalignnotwo{&\3\cD_b(\otw\n^{abc}\ \3 e_{cI})=0\quad{\rm and}&(5.36)\cr
&\otw\n^{abc}\ \3 F_{bc}^I-8\pi\sqrt{-g}(\3 e^{aI}g^{bc}-2 \ \3 e^{bI}
g^{ac})\pd_b\phi\pd_c\phi=0,&(5.37)\cr}$$
respectively. Note that equation (5.35) is just the standard \eom for $\phi$,
while equation (5.36) implies that $\3 A_a^I=\G_a^I$, as in the vacuum case.
Substituting this result for $\3 A_a^I$ back into (5.37) and contracting with
$\3 e_I^d$ gives
$$G^{ad}=8\pi T^{ad}(KG).\eqno(5.38)$$
These are the desired results.

To put this theory in Hamiltonian form, we will basically proceed as we have in
the past, but use additional structure provided by the \st metric $g_{ab}$. We
will assume that $M$ is topologically $\S\X R$ for some spacelike
submanifold $\S$ and assume that there exists a time function $t$ (with nowhere
vanishing gradient $(dt)_a$) such that each $t=\const$ surface $\S_t$ is
diffeomorphic to $\S$. \ $t^a$ will denote the time flow vector field
($t^a(dt)_a=1$), while $n_a$ will denote the {\it unit} covariant normal to the
$t=\const$ surfaces. \ $n^a:=g^{ab}n_b$ will be the associated future-pointing
timelike vector field ($n^a n_a=-1$).  Given $n_a$ and $n^a$, it follows that
$q_b^a:=\d_b^a+n^a n_b$ is a projection operator into the $t=\const$ surfaces.
We can then define the induced metric $q_{ab}$, the lapse $N$, and shift $N^a$
as we did for the standard \EH theory in Section 2.  Recall that $t^a=N n^a+
N^a$, with $N^a n_a=0$.

Now let us write the total action (5.34) in 2+1 form by decomposing each
of its pieces. Using $\otw\n^{abc}=3 t^{[a}\otw\n^{bc]}dt$, it follows that
$$S_P(\3 e,\3 A)=\int dt\int_\S{1\over 2}(\3 e\cdot t)_I\otw\n^{ab} F_{ab}^I +
\otw E_I^a\cL_{\vec t} \ A_a^I - \otw E_I^a\cD_a(\3 A\cdot t)^I,\eqno(5.39)$$
where $(\3 e\cdot t)_I:=t^a\ \3 e_{aI}$, $\otw E_I^a:=\otw\n^{ab}\ \3 e_{bI}$,
$(\3 A\cdot t)^I:=t^a\ \3 A_a^I$, and $A_a^I:=q_a^b\ \3 A_b^I$. To obtain (39),
we used the fact that $\cL_{\vec t} \ q_b^a=0$. Note also that
$F_{ab}^I :=q_a^cq_b^d\ \3 F_{cd}^I$
is the curvature tensor of the generalized derivative operator
$\cD_a$ ($:=q_a^b\ \3\cD_b$) on $\S$ associated with $A_a^I$. Since
$${1\over 2}(\3 e\cdot t)_I\otw\n^{ab} F_{ab}^I=-{1\over 2}\utw N\e^{IJK}\otw
E_I^a\otw E_J^b F_{abK}- N^a\otw E_I^b F_{ab}^I\eqno(5.40)$$
(where $\utw N:=q^{-{1\over 2}}N$), we see that (modulo a surface integral)
the Lagrangian $L_P$ of the 2+1 \P theory is given by
$$L_P=\int_\S -{1\over 2}\utw N\e^{IJK}\otw E_I^a\otw E_J^b F_{abK}-N^a
\otw E_I^b F_{ab}^I+\otw E_I^a\cL_{\vec t} \ A_a^I+(\cD_a\otw E_I^a)
(\3 A\cdot t)^I.\eqno(5.41)$$
Similarly, using $g^{ab}=q^{ab}-n^a n^b$ and the decomposition $\sqrt{-g}=N
\sqrt{q}dt$, it follows that
$$S_{KG}(\3 e,\phi)=-8\pi\int dt\int_\S\Big\{\utw N \otw{\otw q}{}^{ab}
\pd_a\phi\pd_b\phi-\utw N{}^{-1}(\cL_{\vec t} \ \phi-N^a\pd_a\phi)^2\Big\},
\eqno(5.42)$$
where $\otw{\otw q}{}^{ab}:=q q^{ab}\ (=\otw E_I^a\otw E^{bI})$. Thus, the \KG
Lagrangian $L_{KG}$ is simply given by
$$L_{KG}=-8\pi\int_\S\Big\{\utw N \otw{\otw q}{}^{ab}
\pd_a\phi\pd_b\phi-\utw N{}^{-1}(\cL_{\vec t} \ \phi-N^a\pd_a\phi)^2\Big\}.
\eqno(5.43)$$
The total Lagrangian $L_T$ is the sum $L_T=L_P+{1\over 2}L_{KG}$ and is to be
viewed as a functional of the configuration variables $(\3 A\cdot t)^I$, $\utw
N$, $N^a$, $A_a^I$, $\otw E_I^a$, $\phi$ and their first time derivatives.

Following the standard Dirac constraint analysis, we find that
$$\otw\pi:={\d L_T\over\d(\cL_{\vec t} \ \phi)}=8\pi\utw N{}^{-1}
(\cL_{\vec t} \ \phi-N^a\pd_a\phi)\eqno(5.44)$$
is the momentum canonically conjugate to $\phi$. Since this equation can be
inverted to give
$$\cL_{\vec t} \ \phi ={1\over 8\pi}\utw N\otw\pi+N^a\pd_a\phi,\eqno(5.45)$$
it does not define a constraint. On the other hand, $\otw E_I^a$ is
constrained to be the momentum canonically conjugate to $A_a^I$, while
$(\3 A\cdot t) ^I$, $\utw N$, and $N^a$ play the role
of Lagrange multipliers. The resulting total phase space $(\G_T,\W_T)$
is coordinatized by the pairs of fields $(A_a^I,\otw E^a_I)$ and $(\phi,
\otw\pi)$ with symplectic structure
$$\W_T=\int_\S \id\otw E_I^a\ \iwedge\ \id A_a^I + \Tr(\id\otw\pi\ \iwedge
\ \id\phi).\eqno(5.46)$$
The Hamiltonian is given by
$$\eqalign{H_T(A,\otw E,\phi,\otw\bE)=\int_\S \utw N&\Big({1\over 2}
\e^{IJK}\otw E_I^a\otw E_J^b F_{abK}+(4\pi\otw{\otw q}{}^{ab}\pd_a\phi
\pd_b\phi+{1\over 16\pi} \ \otw\pi{}^2)\Big)\cr
&+N^a(\otw E_I^b F_{ab}^I+\otw\pi\pd_a\phi)-(\cD_a\otw E_I^a)(\3 A\cdot
t)^I,\cr}\eqno(5.47)$$
We shall see that this is just a sum of 1st class constraint functions
associated with
$$\eqalignnotwo{&{1\over 2}\e^{IJK}\otw E_I^a\otw E_J^b F_{abK}+
(4\pi\otw{\otw q}{}^{ab}\pd_a\phi\pd_b\phi+{1\over 16\pi} \ \otw\pi{}^2)
\approx 0,&(5.48)\cr
&\otw E_I^b F_{ab}^I+\otw\pi\pd_a\phi\approx 0,\quad{\rm and}&(5.49)\cr
&\cD_a\otw E_I^a\approx 0.&(5.50)
\cr}$$
These are the constraint equations associated with the Lagrange multipliers
$\utw N$, $N^a$, $(\3 A\cdot t)^I$, respectively.

Two remarks are in order: First, note that just as we found for the 2+1 \P
theory with or without a \cc $\L$, the constraint equations for the 2+1 \P
theory coupled to a \KG field are {\it polynomial} in the basic canonically
conjugate variables $(A_a^I,\otw E_I^a)$ and $(\phi,\otw\pi)$. Since the
Hamiltonian is just a sum of these constraints, it follows that the evolution
equations will be polynomial as well. Second, since the constraint equations do
not involve the inverse of $\otw E_I^a$, the above Hamiltonian formulation is
well-defined even if $\otw E_I^a$ is non-invertible. Thus, we have a slight
extension of the standard 2+1 theory of gravity coupled to a massless scalar
field. It can handle those cases where the spatial metric $\otw{\otw
q}{}^{ab}:=\otw E_I^a\otw E^{bI}$ becomes degenerate.

Our next goal is to verify the claim that the constraint functions associated
with (5.48)-(5.50) form a 1st class set. To do this, we let $v^I$ (which takes
values in the Lie algebra of $SO(2,1)$), $\utw N$, and $N^a$ be arbitrary test
fields on $\S$.  Then we define
$$\eqalignnotwo{&C(\utw N):=\int_\S\utw N\Big({1\over 2}\e^{IJK}\otw E_I^a\otw
E_J^b F_{abK}+ (4\pi\otw{\otw q}{}^{ab}\pd_a\phi\pd_b\phi+{1\over 16\pi} \
\otw\pi{}^2)\Big),&(5.51)\cr
&C'(\vec N):=\int_\S N^a(\otw E_I^b F_{ab}^I+\otw\pi\pd_a\phi),\quad{\rm and}
&(5.52)\cr
&G(v):=\int_\S v^I(\cD_a\otw E_I^a),&(5.53)\cr}$$
to be the {\it scalar}, {\it vector}, and {\it Gauss constraint functions}.

As we saw in subsection 3.3 for the 2+1 \P theory, the subset of Gauss
constraint functions form a Lie algebra \wrt Poisson bracket. Given $G(v)$ and
$G(w)$, we have
$$\{G(v),G(w)\}=G([v,w]),\eqno(5.54)$$
where $[v,w]^I$ is the Lie bracket in $\cL_{SO(2,1)}$. Thus, the mapping
$v\mapsto G(v)$ is a representation of the Lie algebra $\cL_{SO(2,1)}$. Given
its geometrical interpretation as the generator of internal rotations, it
follows that
$$\eqalignnotwo{&\{G(v),C(\utw N)\}=0\quad{\rm and}&(5.55)\cr
&\{G(v),C'(\vec N)\}=0,&(5.56)\cr}$$
as well.

Since one can show that the vector constraint function does not by itself have
any direct geometrical interpretation (see, e.g., \cite{AMT}), we will define a
new constraint function $C(\vec N)$ by taking a linear combination of the
vector and Gauss constraints. We define
$$C(\vec N):=C'(\vec N)-G(N),\eqno(5.57)$$
where $N^I:=N^a A_a^I$. We will call $C(\vec N)$ the {\it diffeomorphism
constraint function} since the motion it generates on phase space corresponds
to the 1-parameter family of diffeomorphisms on $\S$ associated with the vector
field $N^a$. To see this, we can write
$$\eqalign{C(\vec N):&=C'(\vec N)-G(N)\cr
&=\int_\S N^a(\otw E_I^b F_{ab}^I+\otw\pi\pd_a\phi)-\int_\S N^I
(\cD_a\otw E_I^a)\cr
&=\int_\S N^a(\otw E_I^b F_{ab}^I-A_a^I\cD_b\otw E_I^b) + \otw\pi N^a\pd_a
\phi\cr
&=\int_\S \otw E_I^a\cL_{\vec N} A_a^I+\otw\pi\cL_{\vec N}\phi,\cr}
\eqno(5.58)$$
where the Lie derivative \wrt $N^a$ treats fields having only internal indices
as scalars. To obtain the last line of (5.58), we ignored a surface integral
(which would vanish anyways for $N^a$ satisfying the appropriate boundary
conditions). By inspection, it follows that $A_a^I\mapsto A_a^I+\e\cL_{\vec N}
A_a^I+O(\e^2)$, etc. Using this geometric interpretation of $C(\vec N)$, it
follows that
$$\eqalignnotwo{&\{C(\vec N),G(v)\}=G(\cL_{\vec N}v),
&(5.59)\cr
&\{C(\vec N),C(\utw M)\}=C(\cL_{\vec N}\utw M),\quad{\rm and}&(5.60)\cr
&\{C(\vec N),C(\vec M)\}=C([\vec N,\vec M]).&(5.61)\cr}$$

We are left to evaluate the \Pb $\{C(\utw N),C(\utw M)\}$ of two scalar
constraints. After a fairly long but straightforward calculation, one can show
that
$$\{C(\utw N),C(\utw M)\}=C'(\vec K) \ \ \left(=C(\vec K)+G(\bK,K)\right),
\eqno(5.62)$$
where $K^a:=\otw{\otw q}{}^{ab}(\utw N\pd_b\utw M-\utw M\pd_b\utw N)$ and
$\otw{\otw q}{}^{ab}=\otw E_I^a\otw E^{bI}$. This result makes crucial use
of the fact that
$$\e^{IJK}\e_I{}^{MN}=(-1)(\n^{JM}\n^{KN}-\n^{JN}\n^{KM}).\eqno(5.63)$$
This is a property of the structure constants $\e^I{}_{JK}:=\e^{IMN}\n_{MJ}
\n_{NK}$ of the Lie algebra of $SO(2,1)$. Thus, the constraint functions are
closed under Poisson bracket---i.e., they form a 1st class set. Note, however,
that since the vector field $K^a$ depends on the phase space variable $\otw
E_I^a$, the \Pb (5.62) involves {\it structure functions}. The constraint
functions do not form a Lie algebra. This result is similar to what we
found for the standard \EH theory in subsection 2.3.

It is interesting to note that even if we did not couple matter to the 2+1 \P
theory, but performed the Legendre transform as we did above (i.e., using the
additional structure provided by the \st metric $g_{ab}:=\3 e_a^I\ \3
e_b^J\n_{IJ}$), we would still obtain the same \Pb algebra. The constraint
functions would still fail to form a Lie algebra due to the structure functions
in (5.62). At first, something seems to be wrong with this statement, since we
saw in subsection 3.3 that the constraint functions $G(v)$ and $F(\a)$ of the
2+1 \P theory form a Lie algebra \wrt Poisson bracket. One may ask why the
constraints functions obtained via one decomposition of the 2+1 \P theory form
a Lie algebra, while those obtained from another decomposition do not.

The answer to this question is actually fairly simple. Namely, it is easy to
destroy the ``Lie algebra-ness'' of a set of constraint functions. If $\phi_
{\und i}$ \ $(\und i=1,\cdots,m)$ denote $m$ constraint functions which form a
Lie algebra under \Pb (i.e., $\{\phi_{\und i},\phi_{\und j}\}= C^{\und
k}{}_{\und i \und j}\phi_{\und k}$ where $C^{\und k}{}_{\und i\und j}$ are
constants), then a linear combination of these constraints, $\x_{\und
i}=\L_{\und i}{}^{\und j}\phi_{\und j}$, will not in general form a Lie
algebra if $\L_{\und i}{}^{\und j}$ are  not constants on the phase space. In
essence, this is what happens when one passes from the $G(v)$ and $F(\a)$
constraint functions of subsection 3.3 to the $G(v)$, $C(\utw N)$, and $C(\vec
N)$ constraint functions of this subsection. The transition from $F(\a)$ to
$C(\utw N)$ and $C(\vec N)$ involve functions of the phase space variables.
\vskip 1cm

\noindent{\bf 6. 3+1 Palatini theory}
\vskip .5cm

In this section, we will describe the 3+1 \P theory. In subsection 6.1, we
define the 3+1 \P action and show that the \EL \eoms are equivalent to the
standard vacuum Einstein's equation. In subsection 6.2, we will follow the
standard Dirac constraint analysis to put the 3+1 \P theory in Hamiltonian
form. We obtain a set of constraint equations which include a 2nd class pair.
By solving this pair, we find that the remaining (1st class) constraints become
{\it non-polynomial} in the (reduced) phase space variables. In essence, we are
forced into using the standard geometrodynamical variables of general
relativity. In fact, as we shall see in subsection 6.3, the Hamiltonian
formulation of the 3+1 \P theory is just that of the standard \EH theory. Thus,
the 3+1 \P theory does not give us a connection-dynamic description of 3+1
gravity.

Much of the material in subsection 6.2 is based on an an analysis of the 3+1 \P
theory given in Chapter 4 of \cite{poona}.
\vskip .5cm

\noindent{\sl 6.1 \EL equations of motion}
\medskip

To obtain the \P action for 3+1 gravity, we will first write the standard \EH
action
$$S_{EH}(g^{ab})=\int_\S\sqrt{-g} R\eqno(6.1)$$
in {\it tetrad notation}. Using
$$R_{abI}{}^J=R_{abc}{}^d\ \4 e_I^c\ \4 e_d^J\eqno(6.2)$$
(which relates the internal and \st curvature tensors of the unique,
torsion-free generalized derivative operator $\grad_a$ compatible with the
tetrad $\4 e_I^a$) and
$$\e_{abcd}=\ \4 e_a^I\ \4 e_b^J\ \4 e_c^K\ \4 e_d^L\ \e_{IJKL}\eqno(6.3)$$
(which relates the volume element $\e_{abcd}$ of $g_{ab}=\4 e_a^I\ \4 e_b^J
\n_{IJ}$ to the volume element $\e_{IJKL}$ of $\n_{IJ}$), we find that
$$\sqrt{-g} R={1\over 4}\otw\n^{abcd}\e_{IJKL}\ \4 e_a^I\ \4 e_b^J\ R_{cd}
{}^{KL}.\eqno(6.4)$$
Thus, viewed as a functional of the co-tetrad $\4 e_a^I$, the standard \EH
action is given by
$$S_{EH}(\4 e)={1\over 4}\int_\S\otw\n^{abcd}\e_{IJKL}\ \4 e_a^I\ \4 e_b^J
\ R_{cd}{}^{KL}.\eqno(6.5)$$

To obtain the 3+1 \P action, we simply replace $R_{abI}{}^J$ in (5) with the
internal curvature tensor $\4 F_{abI}{}^J$ of an arbitrary generalized
derivative operator $\4\cD_a$ defined by
$$\4\cD_a k_I:=\pd_a k_I + \4 A_{aI}{}^J k_J.\eqno(6.6)$$
We define the 3+1 {\it \P action} to be
$$S_P(\4 e,\4 A):={1\over 8}\int_M\otw\n^{abcd}\e_{IJKL}\ \4 e_a^I\ \4
e_b^J\ \4 F_{cd}{}^{KL},\eqno(6.7)$$
where $\4 F_{abI}{}^J=2\pd_{[a}\4 A_{b]I}{}^J + [\4 A_a,\4 A_b]_I{}^J$. Just as
we did for the 2+1 \P theory, we have included an additional factor of $1/2$ in
definition (6.7) so our canonically conjugate variables will agree with those
used in the literature (see, e.g., \cite{poona}). Note also, that as defined
above, $\4\cD_a$ knows how to act only on internal indices. We do not require
that $\4\cD_a$ know how to act on \st indices. However, we will often find it
convenient to consider a torsion-free extension of $\4\cD_a$ to \st tensor
fields. All calculations and all results will be independent of this choice of
extension.

Since the 3+1 \P action is a functional of both a co-tetrad and a connection
1-form, there will be two \EL equations of motion. When we vary $S_P(\4 e,\4
A)$ \wrt $\4 e_a^I$ and $\4 A_a{}^{IJ}$, we find
$$\eqalignnotwo{&\otw\n^{abcd}\e_{IJKL}\ \4 e_b^J\ \4 F_{cd}{}^{KL}=0 \quad{\rm
and}&(6.8)\cr
&\4\cD_b(\otw\n^{abcd}\e_{IJKL}\ \4 e_c^K\ \4 e_d^L)=0,&(6.9)\cr}$$
respectively. The last equation requires a torsion-free extension of $\4\cD_a$
to spacetime tensor fields, but since the left hand side of (6.9) is the
divergence of a skew spacetime tensor density of weight +1 on $M$, it is
independent of this choice. Noting that $\otw\n^{abcd}\e_{IJKL}\ \4 e_c^K \ \4
e_d^L=4(\4 e)\ \4 e_I^{[a}\ \4 e_J^{b]}$ \ (where $(\4 e):=\sqrt{-g}$), we can
rewrite (6.9) as
$$\4\cD_b\left((\4 e)\ \4 e_I^{[a}\ \4 e_J^{b]}\right)=0.\eqno(6.10)$$
This equation is identical in form to equation (3.18) obtained in Section 3
for the 2+1 \P theory.

By following exactly the same argument used in subsection 3.1, equation (6.10)
implies that $\4 A_{aI}{}^J=\G_{aI}{}^J$, where $\G_{aI}{}^J$ is the internal
Christoffel symbol of $\grad_a$. Using this result, the remaining \EL equation
of motion (6.8) becomes
$$\otw\n^{abcd}\e_{IJKL}\ \4 e_b^J\ R_{cd}{}^{KL}=0.\eqno(6.11)$$
When (6.11) is contracted with $\4 e^{eI}$, we get $G^{ae}=0$. Thus, we
can produce the 3+1 vacuum Einstein's equation starting from the 3+1 \P
action given by (6.7). Note that just as in the 2+1 theory, the equation of
motion (6.9) for $\4 A_{aI}{}^J$ can be solved uniquely for $\4 A_{aI}{}^J$ in
terms of the remaining basic variables $\4 e_a^I$.  The pulled-back action
$\und S_P(\4 e)$ defined on the solution space $\4 A_{aI}{}^J=\G_{aI} {}^J$ is
just $1/2$ times the standard \EH action $S_{EH}(\4 e)$ given by (6.5).
\vskip .5cm

\noindent{\sl 6.2 Legendre transform}
\medskip

To put the 3+1 \P theory in Hamiltonian form, we will use the additional
structure provided by the \st metric $g_{ab}$.  We will assume that $M$ is
topologically $\S\times R$ for some spacelike submanifold $\S$ and assume
that there exists a time function $t$ (with nowhere vanishing gradient
$(dt)_a$) such that each $t=\const$ surface $\S_t$ is diffeomorphic to $\S$. \
$t^a$  will denote the time flow vector field ($t^a(dt)_a=1$), while $n_a$ will
denote the unit covariant normal to the $t=\const$ surfaces. \ $n^a:=g^{ab}n_b$
will be the associated future-pointing timelike vector field ($n^a n_a=-1$).
Given $n_a$ and $n^a$, it follows that $q_b^a:=\d_b^a+n^a n_b$ is a projection
operator into the $t=\const$ surfaces. We can then define the induced metric
$q_{ab}$, the lapse $N$, and shift $N^a$ as we did for the standard \EH theory
in Section 2. Recall that $t^a=N n^a+N^a$, with $N^a n_a=0$.

Now let us write (6.7) in 3+1 form.  Using the decomposition $\otw\n^{abcd}=4
t^{[a}\otw\n^{bcd]}dt$ (where $\otw\n^{abc}$ is the Levi-Civita tensor density
of weight +1 on $\S$), we get
$$\eqalign{S_P(\4 e,\4 A)&={1\over 8}\int_M\otw\n^{abcd}\e_{IJKL}\ \4 e_a^I
\ \4 e_b^J\ \4 F_{cd}{}^{KL}\cr
&=\int dt\int_\S{1\over 4}(\4 e\cdot t)^I\e_{IJKL}\otw\n^{bcd}e_b^J F_{cd}
{}^{KL}+{1\over 2}\otw E^a{}_{IJ}\cL_{\vec t} \ A_a{}^{IJ}-
{1\over 2}\otw E^a{}_{IJ}
\cD_a(\4 A\cdot t)^{IJ}, \cr}\eqno(6.12)$$
where $(\4 e\cdot t)_I:=t^a\ \4 e_{aI}$, $\otw E^a{}_{IJ}:={1\over 2}\e_{IJKL}
\otw\n^{abc}\ \4 e_b^K\ \4 e_c^L$, $(\4 A\cdot t)^{IJ}:=t^a\ \4 A_a{}^{IJ}$,
$A_a{}^{IJ}:=q_a^b\ \4 A_b{}^{IJ}$, and $e_a^I:=q_a^b\ \4 e_b^I$. To obtain the
last line of (6.12), we used the fact that $\cL_{\vec t} \ q_b^a=0$. Note also
that $F_{ab}{}^{IJ}:=q_a^c q_b^d\ \4 F_{cd}{}^{IJ}$ is the curvature tensor of
the generalized derivative operator $\cD_a\ (:=q_a^b\ \4\cD_b)$ on $\S$
associated with $A_a{}^{IJ}$. Since
$${1\over 4}(\4 e\cdot t)^I\e_{IJKL}\otw\n^{bcd}e_b^J F_{cd}{}^{KL}=-{1\over 2}
\utw N\Tr(\otw E^a\otw E^b F_{ab})+{1\over 2}N^a\Tr(\otw E^b F_{ab})\eqno(6.13)
$$
(where $\utw N:=q^{-{1\over 2}}N$ and $\Tr$ denotes the trace operation on
internal indices), we see that (modulo a surface integral) the Lagrangian $L_P$
of the 3+1 \P theory is given by
$$\eqalign{L_P=\int_\S -{1\over 2}\utw N\Tr&(\otw E^a\otw E^b F_{ab})+
{1\over 2}N^a \Tr(\otw E^b F_{ab})\cr
&+{1\over 2}\otw E^a{}_{IJ}\cL_{\vec t} \ A_a{}^{IJ} + {1\over 2}
(\cD_a\otw E^a{}_{IJ})(\4 A\cdot t)^{IJ}.\cr}\eqno(6.14)$$
The configuration variables of the theory are $(\4 A\cdot t)^{IJ}$, $\utw N$,
$N^a$, $A_a{}^{IJ}$, and $\otw E^a{}_{IJ}$.

But before we perform the Legendre transform, we should note that the
configuration variable $\otw E^a{}_{IJ}$ is not free to take on arbitrary
values. In fact, from its definition
$$\otw E^a{}_{IJ}:={1\over 2}\e_{IJKL}\otw\n^{abc}\ \4 e_b^K\ \4 e_c^L,
\eqno(6.15)$$
one can show that
$$\otw{\otw\phi}{}^{ab}:=\e^{IJKL}\otw E^a{}_{IJ}\otw E^b{}_{KL}=0\quad{\rm
and}\quad\Tr(\otw E^a\otw E^b)>0.\eqno(6.16)$$
The second condition follows from the fact that $\Tr(\otw E^a\otw
E^b)=2\otw{\otw q}{}^{ab} \ (=2q q^{ab})$, where $q^{ab}$ is the inverse of the
induced positive-definite metric $q_{ab}$ on $\S$. Thus, the starting point for
the Legendre transform is $L_P$ {\it together} with the primary constraint
$\otw{\otw\phi}{}^{ab}=0$. Since the inequality is a non-holonomic constraint,
it will not reduce the number of phase space degrees of freedom.

If we now follow the standard Dirac constraint analysis, we find that $(\4
A\cdot t)^{IJ}$, $\utw N$, and $N^a$ play the role of Lagrange multipliers.
Their associated constraint equations (which arise as secondary constraints in
the analysis) are
$$\eqalignnotwo{&\Tr(\otw E^a\otw E^b F_{ab})\approx 0,&(6.17)\cr
\quad&\Tr(\otw E^b F_{ab})\approx 0,\quad{\rm and}&(6.18)\cr
&\cD_a\otw E^a{}_{IJ}\approx 0.&(6.19)\cr}$$
There is also a primary constraint which says that ${1\over 2}\otw E^a{}_{IJ}$
is the momentum canonically conjugate to $A_a{}^{IJ}$. By demanding that the
\Pb of this constraint with the total Hamiltonian and the \Pb of
$\otw{\otw\phi}{}^{ab}$ with the total Hamiltonian be weakly zero, we find that
$$\x^{ab}:=\e^{IJKL}(\cD_c\otw E^a{}_{IJ})[\otw E^b,\otw E^c]_{KL}
+(a\leftrightarrow b)\approx 0.\eqno(6.20)$$
This is an additional secondary constraint which must be appended to constraint
equations (6.16)-(6.19). In virtue of $\otw{\otw\phi}{}^{ab}=0$, the expression
for $\x^{ab}$ is independent of the choice of torsion-free extension of
$\cD_a$ to \st tensor fields. If we further demand that the \Pb of $\x^{ab}$
with the total Hamiltonian be weakly zero, we find nothing new---i.e., there
are no tertiary constraints.

Let us summarize the situation so far: \ Out of the original set of
configuration variables $\{(\4 A\cdot t)^{IJ},\ \utw N,\ N^a,\ A_a{}^{IJ},\
\otw E^a{}_{IJ}\}$, the first three are non-dynamical. We also found that
${1\over 2}\otw E^a{}_{IJ}$ is the momentum canonically conjugate to
$A_a{}^{IJ}$. Thus, the phase space $(\G'_P,\W'_P)$ of the 3+1 \P theory is
coordinatized by the pair $(A_a{}^{IJ},\otw E^a{}_{IJ})$ and has the symplectic
structure
$$\W'_P={1\over 2}\int_\S\id\otw E^a{}_{IJ}\ \iwedge \ \id A_a{}^{IJ}.
\eqno(6.21)$$
The Hamiltonian is given by
$$\eqalign{H'_P(A,\otw E)=\int_\S {1\over 2}\utw N\Tr(\otw E^a\otw E^b &F_{ab})
-{1\over 2}N^a\Tr(\otw E^b F_{ab})-{1\over 2}(\cD_a\otw E^a{}_{IJ})(\4
A\cdot t)^{IJ}\cr
&+\utw\l{}_{ab}\e^{IJKL}\otw E^a{}_{IJ}\otw E^b{}_{KL},\cr}\eqno(6.22)$$
where $\utw\l{}_{ab}$ is a Lagrange multiplier. The constraints of the
theory are (6.16)-(6.20). Note also that at this stage of the Dirac constraint
analysis all constraint (and evolution) equations are {\it polynomial} in the
canonically conjugate variables.

The next step in the Dirac constraint analysis is to evaluate the Poisson
brackets of the constraints and solve all 2nd class pairs.  Since only
$$\{\otw{\otw\phi}{}^{ab}(x),\x^{cd}(y)\}\not\approx 0,\eqno(6.23)$$
%
%
%
we just have to solve constraint equations (6.16) and (6.20). As shown in
Chapter 4 of \cite{poona}, the most general solution to (6.16) is
$$\otw E^a{}_{IJ}=:2\otw E_{[I}^a n_{J]},\eqno(6.24)$$
for some unit timelike covariant normal $n_I$ ($n_I n^I=-1$) with $\otw
E_I^a$ invertible and $\otw E_I^a n^I=0$. By writing $\otw E^a{}_{IJ}$ in this
form, we see that the original 18 degrees of freedom per space point for $\otw
E^a{}_{IJ}$ has been reduced to 12. Note also that $\otw E_I^a \otw E^{bI}
=\otw{\otw q}{}^{ab}$, so $\otw E_I^a$ is in fact a (densitized) triad.

Given (6.24), the most convenient way to solve (6.20) is to gauge fix the
internal vector $n^I$. This will further reduce the number of degrees of
freedom of $\otw E^a {}_{IJ}$ to 9, since now only $\otw E_I^a$ will be
arbitrary. However, gauge fixing $n^I$ requires us to solve the boost part
of (6.19) relative to $n_I$ as well.\footnote{The {\it boost part} of any
anti-symmetric tensor $A_{IJ}$ relative to $n_I$ is defined to be $A_{IJ}n^J$.}
We can only keep that part of (6.19) which generates internal rotations
leaving $n^I$ invariant.

To solve these constraints, let us first define an internal connection 1-form
$K_{aI}{}^J$ via
$$\cD_a k_I=:D_a k_I + K_{aI}{}^J k_J,\eqno(6.25)$$
where $D_a$ is the unique, torsion-free generalized derivative operator on $\S$
compatible with the (densitized) triad $\otw E_I^a$ {\it and} the gauge fixed
internal vector $n^I$. Then constraint equation (6.20) and the boost part of
(6.19) become
$$\eqalignnotwo{&\x^{ab}=-4\e^{IJK}K_{cI}{}^L\otw E_L^{(a}\otw E_J^{b)}\otw
E_K^c\approx 0\quad{\rm and}&(6.26)\cr
&(\cD_a\otw E^a{}_{IJ})n^J=-K_{aM}{}^N\otw E_N^a q_I^M\approx 0,&(6.27)\cr}$$
where $q_J^I:=\d_J^I+n^I n_J$. By using the invertibility of the (densitized)
triad $\otw E_I^a$, one can then show (again, see Chapter 4 of \cite{poona})
that (6.26) and (6.27) imply that $K_a{}^{IJ}$ also be pure boost \wrt
$n_I$---i.e., that $K_a{}^{IJ}$ have the form
$$K_a{}^{IJ}=:2K_a^{[I}n^{J]},\eqno(6.28)$$
with $K_a^I n_I=0$. Since $D_a$ is determined completely by $\otw E_I^a$ and
$n_I$, the original 18 degrees of freedom for $A_a{}^{IJ}$ has also been
reduced to 9 degrees of freedom per space point. The information contained in
$A_a{}^{IJ}$ (which is independent of $\otw E_I^a$ and $n_I$) is completely
characterized by $K_a^I$. To emphasize the fact that $\otw E_I^a n^I=0$ and
$K_a^I n_I=0$, we will use a 3-dimensional abstract internal index $i$ and
write $\otw E_i^a$ and $K_a^i$ in what follows.

Thus, after eliminating the 2nd class constraints, the phase space
$(\G_P,\W_P)$ of the 3+1 \P theory is coordinatized by the pair $(\otw
E_i^a,K_a^i)$ and has the symplectic structure
$$\W_P=\int_\S\id K_a^i\ \iwedge\ \id\otw E_i^a.\eqno(6.29)$$
The Hamiltonian is given by
$$\eqalign{H_P(\otw E,K)=\int_\S {1\over 2}\utw N\Big(-q\cR-&2\otw E_{[i}^a
\otw E_{j]}^b K_a^i K_b^j\Big)-2N^a\otw E_i^b D_{[a}K_{b]}^i\cr
&+(\4 A\cdot t)^{ij}\otw E_{[i}^a K_{aj]},\cr}\eqno(6.30)$$
where $\cR$ denotes the scalar curvature of $D_a$. This is just a sum of the
1st class constraints functions associated with
$$\eqalignnotwo{&\otw{\otw C}(\otw E,K):=-q\cR-2\otw E_{[i}^a\otw E_{j]}^b
K_a^i K_b^j \approx 0,&(6.31)\cr
&\otw C_a(\otw E,K):=4\otw E_i^b D_{[a}K_{b]}^i\approx 0,\quad{\rm and}&(6.32)
\cr
&\otw G_{ij}(\otw E,K):=-\otw E_{[i}^a K_{aj]}\approx 0.&(6.33)\cr}$$
(The overall numerical factors have been chosen in order to facilitate the
comparison with the standard \EH theory.) Note that constraint equations
(6.31), (6.32), and (6.33) are the remaining constraints (6.17), (6.18), and
the rotation part of (6.19) relative to $n_I$ expressed in terms of the phase
space variables $(\otw E_i^a,K_a^i)$.\footnote{The {\it rotation part} of any
anti-symmetric tensor $A_{IJ}$ relative to $n_I$ is given by $q_I^M q_J^N\
A_{MN}$ where $q_J^I:=\d_J^I+n^I n_J$.}  We will call (6.31), (6.32), and
(6.33) the {\it scalar}, {\it vector}, and {\it Gauss constraints} for the 3+1
\P theory.

Note that as a consequence of eliminating the 2nd class constraints by solving
(6.16), (6.20), and the boost part of (6.19) relative to $n_I$, the constraint
equations (6.31)-(6.33) (and hence the evolution equations generated by the
Hamiltonian) are now {\it non-polynomial} in the canonically conjugate pair
$(\otw E_i^a,K_a^i)$. This is due to the dependence of $\cR$ on the inverse of
$\otw E_i^a$. In fact, since $\otw E_i^a$ must be invertible, we are forced to
take $\otw E_i^a$ as the configuration variable of the theory. We are led back
to a geometrodynamical description of 3+1 gravity. Thus, the Hamiltonian
formulation of the 3+1 \P theory has the same drawback as the Hamiltonian
formulation of standard \EH theory. As we shall see in the next subsection,
these theories are effectively the same.
\vskip .5cm

\noindent{\sl 6.3 Relationship to the \EH theory}
\medskip

In this subsection, we will not explicitly evaluate the \Pb algebra of the
constraint functions for the 3+1 \P theory. Rather, we will describe the
relationship between the constraint equations (6.31)-(6.33) and those of the
standard \EH theory. We shall see that if we solve the first class constraint
(6.33) by passing to a reduced phase space, we recover the Hamiltonian
formulation of the standard \EH theory in terms of the induced metric $q_{ab}$
and its canonically conjugate momentum $\otw p^{ab}$.

To do this, let us first define a tensor field $\utw M{}_{ab}$ (of density
weight -1 on $\S$) via
$$\utw M{}_{ab}:=K_a^i\utw E{}_{bi},\eqno(6.34)$$
where $\utw E{}_a^i$ is the inverse of the (densitized) triad $\otw E_i^a$.
Then in terms of $\utw M{}_{ab}$, one can show that constraint equation (6.33)
is equivalent to
$$\utw M{}_{[ab]}\approx 0.\eqno(6.35)$$
Thus, the constraint surface in $\G_P$ defined by (6.33) will be coordinatized
in part by the symmetric part of $\utw M{}_{ab}$---i.e., by $\utw
K{}_{ab}:=\utw M{}_{(ab)}$.

But we are not yet finished. Since (6.33) is a 1st class constraint, we
must also factor-out the constraint surface by the orbits of the Hamiltonian
vector field associated with the constraint function
$$G(\L):=\int_\S\otw G_{ij}(\otw E,K)\L^{ij} \ \ \left(=\int_\S -\otw E_i^a
K_{aj}\L ^{ij}\right).\eqno(6.36)$$
(Here $\L^{ij}=\L^{[ij]}$ denotes an arbitrary anti-symmetric test field on
$\S$.) Since it is fairly easy to show that $G(\L)$ generates (gauge) rotations
of the internal indices (i.e., $\otw E_i^a\mapsto\otw E_i^a+\e\L_i{}^j \otw
E_j^a+O(\e^2)$ and $K_a^i\mapsto K_a^i-\e\L_j{}^i K_a^j+O(\e^2)$), the factor
space will be coordinatized by $\utw K{}_{ab}$ and the gauge invariant
information  contained in $\otw E_i^a$. This is precisely $\otw{\otw q}{}^{ab}
=\otw E_i^a\otw E^{bi}$. Thus, the {\it reduced phase space} $(\hat\G_P,
\hat\W_P)$ is coordinatized by the pair $(\otw{\otw q}{}^{ab}, \utw
K{}_{ab})$ and has symplectic structure
$$\hat\W_P={1\over 2}\int_\S\id\utw K{}_{ab}\ \iwedge \ \id\otw{\otw q}{}^
{ab}.\eqno(6.37) $$

All we must do now is make contact with the usual canonical variables of the
standard \EH theory. To do this, let us work with the  undensitized fields
$q^{ab}$ and $K_{ab}$, and lower and raise their indices, respectively. Then in
terms of $\otw p^{ab}$ defined by
$$\otw p^{ab}:={\sqrt q}(K^{ab}-K q^{ab}),\eqno(6.38)$$
one can show that
$$\hat\W_P=-{1\over 2}\int_\S\id\otw p^{ab}\ \iwedge \ \id q_{ab}\eqno(6.39)$$
and
$$\eqalignnotwo{&\otw{\otw C}(q,\otw p)=-q\cR+(\otw p^{ab}\otw p{}_{ab}-{1\over
2}\otw p^2) \approx 0,&(6.40)\cr
&\otw C_a(q,\otw p)=-2 q_{ab} D_c\otw p^{bc}\approx 0.&(6.41)\cr}$$
Up to overall factors, these are just the symplectic structure $\W_{EH}$ and
scalar and vector constraint equations of the standard \EH theory described in
Section 2. The factor of $-1/2$ in the symplectic structure is due to the
combination of using an action which is $1/2$ the standard \EH action and using
a (densitized) triad $\otw E_i^a$ instead of a covariant metric $q_{ab}$ as
our basic dynamical variables. Thus, we see that the Hamiltonian formulation of
the 3+1 \P theory is nothing more than the familiar geometrodynamical
description of general relativity.
\vskip 1cm

\noindent{\bf 7. Self-dual theory}
\vskip .5cm

In this section, we will describe the \sd theory of 3+1 gravity. This theory is
similar in form to the 3+1 \P theory described in the previous section;
however, it uses a {\it \sd} connection 1-form as one of its basic variables.
We define the \sd action for complex 3+1 gravity and show that we still
recover the standard vacuum Einstein's equation even though we are using only
half of a Lorentz connection. We then perform a Legendre transform to put the
theory in Hamiltonian form. In terms of the resulting complex phase space
variables, all equations of the theory are {\it polynomial}. This
simplification gives the \sd theory a major advantage over the 3+1 \P theory.
As noted in the previous section, the Hamiltonian formulation of the 3+1 \P
theory reduces to that of the standard \EH theory with its troublesome
non-polynomial constraints.

As mentioned in footnote 3 in Section 1, to obtain the phase space variables
for the real theory, we must impose {\it reality conditions} to select a
real section of the original complex phase space.  At the end of subsection
7.2 we will describe these conditions and discuss how they are implemented. The
need to use reality conditions is a necessary consequence of using an action
principle to obtain the new variables for real 3+1 gravity. Although we mention
here that it is possible to stay within the confines of the real theory by
performing a canonical transformation on the standard phase space of real
general relativity, we will not follow that approach in this paper. (Interested
readers should see \cite{hf} for a detailed discussion of Ashtekar's original
approach.) Rather, we will start with an action for the complex theory and
obtain the new variables as outlined above. Henceforth, the co-tetrad $\4
e_a^I$ will be assumed to be complex unless explicitly stated otherwise.
\vskip .5cm

\noindent{\sl 7.1 \EL equations of motion}
\medskip

\def\T{T^{a\cdots b}{}_{c\cdots d}{}}

To write down the \sd action for  complex 3+1 gravity, all we have to do
is replace the (Lorentz) connection 1-form $\4 A_{aI}{}^J$ of the 3+1 \P theory
by a {\it \sd} connection 1-form $\+\4 A_{aI}{}^J$ and let the co-tetrad $\4
e_a^I$ become complex.  We define the {\it \sd action} to be
$$S_{SD}(\4 e,\+\4 A):={1\over 4}\int_M\otw\n^{abcd}\e_{IJKL}\ \4 e_a^I\ \4
e_b^J\ \+\4 F_{cd}{}^{KL},\eqno(7.1)$$
where $\+\4 F_{abI}{}^J=2\pd_{[a}\+\4 A_{b]I}{}^J + [\+\4 A_a,\+\4 A_b]_I{}^J$
is the internal curvature tensor of the \sd generalized derivative operator
$\+\4\cD_a$ defined by
$$\+\4\cD_a k_I:=\pd_a k_I + \+\4 A_{aI}{}^J k_J.\eqno(7.2)$$

Some remarks are in order:

\begin{enumerate}
\item We will always take our \st manifold $M$ to be a  real 4-dimensional
manifold. Complex tensors at a point $p\in M$ will take values in the
appropriate tensor product of the {\it complexified} tangent and cotangent
spaces to $M$ at $p$. The fixed internal space will also be complexified;
however, the internal Minkowski metric $\n_{IJ}$ will remain real. Since the
co-tetrad $\4 e_a^I$ are allowed to be complex, the \st metric $g_{ab}$
defined by $g_{ab}:=\4 e_a^I\ \4 e_b^J\n_{IJ}$ will also be complex.
\item Although we can no longer talk about the signature of a complex metric
$g_{ab}$, compatibility with a complex co-tetrad $\4 e_a^I$ still defines a
unique, torsion-free generalized derivative operator $\grad_a$. Thus, the
complex Einstein tensor $G_{ab}:= R_{ab}-{1\over 2}R g_{ab}$ is well-defined;
whence the complex \eom $G_{ab}=0$ makes sense. It is this equation that
defines for us the complex theory of 3+1 gravity.
\item When we say that the connection 1-form $\+\4 A_{aI}{}^J$ (or any other
generalized tensor field) is self-dual, we will always mean \wrt its {\it
internal} indices. Thus, the notion of self-duality makes sense only in 3+1
dimensions and applies only to generalized tensor fields with a pair of
skew-symmetric internal indices, $\T_{IJ}=\T_{[IJ]}$. The {\it dual} of
$\T_{IJ}$, denoted by $\*\T_{IJ}$, is defined to be
$$\*\T_{IJ}:={1\over 2}\e_{IJ}{}^{KL}\T_{KL},\eqno(7.3)$$
where the internal indices of $\e_{IJKL}$ are raised with the internal metric
$\n^{IJ}$. Since $\n_{IJ}$ has signature $(-+++)$, it follows that the
square of the duality operator is  minus the identity. Hence, our definition of
self-duality involves the complex number $i$. We define $\T_{IJ}$ to be {\it
\sd} if and only if\footnote{$\T_{IJ}$ is defined to be {\it \asd} if and only
if $\*\T_{IJ}=-i\T_{IJ}$. The choice of $+i$ for \sd and $-i$ for \asd is
purely convention. I have chosen our conventions to agree with those of
\cite{poona}.}
$$\*\T_{IJ}=i\T_{IJ}.\eqno(7.4)$$
Thus, \sd fields in Lorentzian 3+1 gravity are necessarily complex.
\item
Given any generalized tensor field $\T_{IJ}=\T_{[IJ]}$, we can always
decompose it as
$$\T_{IJ}=\+\T_{IJ} + \-\T_{IJ},\eqno(7.5)$$
where
$$\eqalignnotwo{&\+\T_{IJ}:={1\over 2}(\T_{IJ}-i\*\T_{IJ})\quad{\rm and}&(7.6a)
\cr
&\-\T_{IJ}:={1\over 2}(\T_{IJ}+i\*\T_{IJ}).&(7.6b)\cr}$$
Since $\*(\+\T_{IJ})=i\+\T_{IJ}$ and $\*(\-\T_{IJ})=-i\-\T_{IJ}$, it follows
that $\+\T_{IJ}$ and $\-\T_{IJ}$ are the {\it \sd} and {\it \asd parts} of
$\T_{IJ}$. Equations ($7.6a$) and ($7.6b$) define the self-duality and anti
self-duality operators ${}^+$ and ${}^-$.
\item
The generalized derivative operator $\+\4\cD_a$ is said to be \sd only in the
sense that it is defined in terms of a \sd connection 1-form $\+\4 A_{aI}{}^J$.
As in many of the previous theories, $\+\4\cD_a$ (as defined by (7.2)) knows
how to act only on internal indices. But as usual, we will often find it
convenient to consider a torsion-free extension of $\+\4\cD_a$ to spacetime
tensor fields. All calculations and all results will be independent of this
choice of extension. Note also that the internal curvature tensor of the
generalized derivative operator $\+\4\cD_a$ is given by
$$\+\4 F_{abI}{}^J=2\pd_{[a}\+\4 A_{b]I}{}^J+[\+\4 A_a,\+\4 A_b]_I{}^J. \eqno
(7.7)$$
Since one can show that the (internal) commutator of two \sd fields is also
self-dual, it follows that $\+\4 F_{abI}{}^J$ is \sd \wrt its internal indices.
\end{enumerate}

Given these general remarks, we are now ready to return to the \sd action (7.1)
and obtain the Euler-Lagrange equations of motion. Varying $S_{SD}(\4 e,\+\4
A)$ \wrt $\4 e_a^I$ gives
$$\otw\n^{abcd}\e_{IJKL}\ \4 e_b^J\ \+\4 F_{cd}{}^{KL}=0,\eqno(7.8)$$
while varying $S_{SD}(\4 e,\+\4 A)$ \wrt $\+\4 A_a{}^{IJ}$ gives
$$\+\4\cD_b\left((\4 e)\ \+(\4 e_I^{[a}\ \4 e_J^{b]})\right)=0.\eqno(7.9)$$
To obtain (7.9), we used the fact that $\otw\n^{abcd}\e_{IJKL}\ \4 e_c^K\
\4 e_d^L=4(\4 e)\ \4 e_I^{[a}\ \4 e_J^{b]}$ (where $(\4 e):=\sqrt{-g}$). Note
also that we are forced to take the \sd part of $\4 e_I^{[a}\ \4 e_J^{b]}$
since the variations $\d\+\4 A_a{}^{IJ}$ are required to be self-dual. This is
the distinguishing feature between the \sd and 3+1 \P equations of motion.
Finally note that although (7.9) requires a torsion-free extension of
$\+\4\cD_a$ to spacetime tensor fields, it is independent of this choice since
the left hand side is the divergence of a skew spacetime tensor density of
weight +1 on $M$.

We would now like to show that (7.9) implies that $\+\4\cD_a$ is the \sd part
of the unique, torsion-free generalized derivative operator $\grad_a$
compatible with $\4 e_a^I$. But since we are working with \sd fields, the
argument used for the 2+1  and 3+1 \P theories does not yet apply. We will have
to do some preliminary work before we can use those results.

If $\G_{aI}{}^J$ denotes the internal Christoffel symbol of $\grad_a$, we
define the {\it \sd part} $\+\grad_a$ of $\grad_a$ by
$$\+\4\grad_a k_I:=\pd_a k_I + \+\G_{aI}{}^Jk_J,\eqno(7.10)$$
where $\+\G_{aI}{}^J$ is the \sd part of $\G_{aI}{}^J$. The difference between
$\+\4\cD_a$ and $\+\grad_a$ is then characterized by a generalized tensor field
$\+\4 C_{aI}{}^J$ defined by
$$\+\4\cD_a k_I=:\+\grad_a k_I + \+\4 C_{aI}{}^J k_J.\eqno(7.11)$$
Note that $\+\4 C_{aI}{}^J$ is \sd as the notation suggests. In fact,
$\+\4 C_{aI}{}^J=\+\4 A_{aI}{}^J-\+\G_{aI}{}^J$. Now let us write (7.9) in
terms of $\+\grad_a$ and $\+\4 C_{aI}{}^J$. Using (7.11) to expand the left
hand side of (7.9), we get
$$\+\grad_b\left((\4 e)\ \+(\4 e_I^{[a}\ \4 e_J^{b]})\right)
+(\4 e)\left(\+\4 C_{bI}{}^K\ \+(\4 e_K^{[a}\ \4 e_J^{b]}) + \+\4 C_{bJ}{}^K
\ \+(\4 e_I^{[a}\ \4 e_K^{b]})\right)=0.\eqno(7.12)$$
Since $\+\grad_a k_I=\grad_a k_I - \-\G_{aI}{}^J k_J$ (where $\-\G_{aI}{}^J$ is
the \asd part of the internal Christoffel symbol $\G_{aI}{}^J$), and since the
last two terms of (7.12) can be written as the (internal) commutator of $\+\4
C_{aIJ}$ and $\+(\4 e_I^{[a}\ \4 e_J^{b]})$, we get
$$-\left[\-\G_b,\+(\4 e^{[a}\ \4 e^{b]})\right]_{IJ}+\left[\+\4 C_b,
\+(\4 e^{[a}\ \4 e^{b]})\right]_{IJ}=0.\eqno(7.13)$$
The first commutator vanishes since $\-\G_{bIJ}$ is \asd while $\+(\4 e_I
^{[a}\ \4 e_J^{b]})$ is self-dual; in the second commutator, $\+(\4
e_I^{[a} \ \4 e_J^{b]})$ can be replaced by $\4 e_I^{[a}\ \4 e_J^{b]}$. Thus,
(7.13) reduces to
$$\left[\+\4 C_b,(\4 e^{[a}\ \4 e^{b]})\right]_{IJ}=0.\eqno(7.14)$$
This is exactly the form of the equation found in the 2+1 \P theory with
$\+\4 C_{bI}{}^J$ replacing $\3 C_{bI}{}^J$. (See equation (3.20).) We can now
follow the argument given there to conclude that $\+\4 C_{aI}{}^J=0$. Thus,
$\+\4 A_{aI}{}^J=\+\G_{aI}{}^J$ as desired.

By substituting the solution $\+\4 A_{aI}{}^J=\+\G_{aI}{}^J$ into the
remaining equation of motion (7.8), we get
$$\otw\n^{abcd}\e_{IJKL}\ \4 e_b^J\ \+ R_{cd}{}^{KL}=0\eqno(7.15)$$
where $\+ R_{cd}{}^{KL}$ is the \sd part of the internal curvature tensor
$R_{cd}{}^{KL}$ of $\grad_a$.  Then by using the definition of $\+ R_{cd}
{}^{KL}$, we see that equation (7.15) becomes
$$\eqalign{0&={1\over 2}\otw\n^{abcd}\e_{IJKL}\ \4 e_b^J (R_{cd}{}^{KL}-{i
\over 2} \e^{KL}{}_{MN}R_{cd}{}^{MN})\cr
&={1\over 2}\otw\n^{abcd}\e_{IJKL}\ \4 e_b^J R_{cd}{}^{KL},\cr}\eqno(7.16)$$
where the second term on the first line vanishes by the Bianchi identity
$R_{[abc]d}=0$. When (7.16) is contracted with $\4 e^{eI}$, we get $G^{ae}=0$.
Thus, the \sd action (7.1) reproduces the vacuum Einstein's equation for
complex 3+1 gravity.

Since this is an important---yet somewhat suprising---result, it is perhaps
worthwhile to repeat the above argument from a slightly different perspective.
First note that the \sd action (7.1) and the 3+1 \P action (6.7) differ by a
term involving the dual of the curvature tensor $\4 F_{cd}{}^{KL}$. This extra
term in the \sd action is not a total divergence and thus gives rise to an
additional equation of motion that is not present in the 3+1 \P theory.  This
equation of motion also involves the dual of the curvature tensor. (Compare
equations (7.8) and (6.8).) However, as we showed above, if we solve (7.9) for
$\+\4 A_a{}^{IJ}$ and substitute the solution $\+\4 A_a{}^{IJ}= \+\G_a{}^{IJ}$
back into (7.8), the additional equation of motion is automatically satisfied
as a consequence of the Bianchi identity $R_{[abc]d}=0$.  Hence, there are no
``spurious" equations of motion. Moreover, since the \sd and 3+1 \P actions
differ by a term that is not a total divergence, the canonically conjugate
variables for the two theories will disagree. As we shall see in the following
section, it is this difference that will allow us to construct a Hamiltonian
formulation of 3+1 gravity with a connection 1-form as the basic configuration
variable.

Finally, we conclude this subsection by showing the relationship between the
\sd and standard \EH actions. To do this, note that since the equation of
motion (7.9) for $\+\4 A_{aI}{}^J$ could be solved uniquely for $\+\4
A_{aI}{}^J$ in terms of the remaining basic variables $\4 e_a^I$, we can
pull-back the \sd action $S_{SD}(\4 e,\+\4 A)$ to the solution space $\+\4
A_{aI}{}^J =\+\G_{aI}{}^J$ and obtain a new action $\und S_{SD}(\4 e)$. Doing
this, we find
$$\eqalign{\und S_{SD}(\4 e)&={1\over 4}\int_M\otw\n^{abcd}\e_{IJKL}\ \4 e_a^I\
\4 e_b^J\ \+ R_{cd}{}^{KL}\cr
&={1\over 8}\int_M\otw\n^{abcd}\e_{IJKL}\ \4 e_a^I\ \4 e_b^J R_{cd}{}^{KL},\cr}
\eqno(7.17)$$
where we expanded $\+ R_{cd}{}^{KL}$ and used the Bianchi identity
$R_{[abc]d}=0$ to get the last line of (7.17). Thus, $\und S_{SD}(\4 e)$ is
just $1/2$ times the standard \EH action $S_{EH}(\4 e)$ viewed as a functional
of a complex co-tetrad $\4 e_a^I$. (See equation (6.5).) In fact, $\und
S_{SD}(\4 e)=\und S_P(\4 e)$, where $\und S_P(\4 e)$ is the pull-back of the
3+1 \P action.  It was precisely to obtain this last equality that we defined
the \sd action (7.1) with an overall factor of $1/4$ rather than $1/8$.
\vskip .5cm

\noindent{\sl 7.2 Legendre transform}
\medskip

To put the \sd theory for complex 3+1 gravity in Hamiltonian form, we will
basically proceed as we did in Section 6 for the 3+1 \P theory. However, since
the \st metric $g_{ab}:=\4 e_a^I\ \4 e_b^J\n_{IJ}$ is now complex, we can only
assume that $M$ is topologically $\S\times R$ for some submanifold $\S$ and
assume that there exists a real function $t$ whose $t=\const$ surfaces foliate
$M$. (We cannot assume that $\S$ is spacelike, since the signature of a complex
metric is not defined.) We can still introduce a real flow vector field $t^a$
(satisfying $t^a(dt)_a=1$) and a unit covariant normal $n_a$ to the $t=\const$
surfaces satisfying $n_a n^a=-1$. (We are free to choose $-1$ for the
normalization of $n_a$ since $n_a$ is allowed to be complex.) \
$n^a:=g^{ab}n_b$ is the vector field associated to $n^a$, and is related to
$t^a$ by a complex lapse $N$ and complex shift $N^a$ via $t^a=Nn^a+N^a$, with
$N^an_a=0$. Finally, the induced metric $q_{ab}$ on $\S$ is given by
$q_{ab}=g_{ab}+{n_a n_b}$.

Following the same steps that we used in the previous section for the 3+1 \P
theory, we find that (modulo a surface integral) the Lagrangian $L_{SD}$ of
the \sd theory is given by
$$\eqalign{L_{SD}=\int_\S -\utw N\Tr&(\+\otw E^a\ \+\otw E^b\ \+ F_{ab})+N^a
\Tr(\+\otw E^b\ \+ F_{ab})\cr
&+\ (\+\otw E^a{}_{IJ})\cL_{\vec t} \ \+ A_a{}^{IJ} +
(\+\cD_a\ \+\otw E^a{}_{IJ})(\+\4 A\cdot t)^{IJ}.\cr}\eqno(7.18)$$
Here $\+\otw E^a{}_{IJ}$ denotes the \sd part of $\otw E^a{}_{IJ}:={1\over 2}
\e_{IJKL}\otw\n^{abc}\ \4 e_b ^K\ \4 e_c^L$.  Note that the transition from
the 3+1 \P Lagrangian to the \sd Lagrangian can be made by simply replacing
all of the real fields by their \sd parts.  The configuration variables of the
theory are $(\+\4 A\cdot t)^{IJ}$, $\utw N$, $N^a$, $\+ A_a{}^{IJ}$, and
$\+\otw E^a{}_{IJ}$.

Now recall that for the real 3+1 \P theory, the configuration variable $\otw
E^a {}_{IJ}$ was not free to take on arbitrary values. From its definition in
terms of the co-tetrad $\4 e_a^I$, we saw that
$$\otw{\otw\phi}{}^{ab}:=\e^{IJKL}\otw E^a{}_{IJ}\otw E^b{}_{KL}=0\quad{\rm
and}\quad\Tr(\otw E^a\otw E^b)>0.\eqno(7.19)$$
The second condition followed from the fact that $\Tr(\otw E^a\otw E^b)=2
\otw{\otw q}{}^{ab}\ (=2 q q^{ab})$, where $q^{ab}$ was the inverse of the
induced positive-definite metric $q_{ab}$ on $\S$. Taking the primary
constraint (7.19) together with the 3+1 \P Lagrangian as the starting point for
the Legendre transform, we found that the standard Dirac constraint analysis
gave rise to additional constraints---one of which was 2nd class \wrt
$\otw{\otw\phi}{}^{ab}=0$. By solving this 2nd class pair, the remaining (1st
class) constraints became non-polynomial and we were forced back to the usual
geometrodynamical description of real 3+1 gravity.\footnote{It is fairly easy
to see that all of the above statements---except for the non-holonomic
constraint which would now say that $\Tr(\otw E^a\otw E^b)$ be
non-degenerate---apply to the complex 3+1 \P theory as well. \
$\otw{\otw\phi}{}^{ab}=0$ is a primary constraint on the complex configuration
variable $\otw E^a{}_{IJ}$, and it must be included when performing the
Legendre transform. The standard Dirac constraint analysis leads to a pair of
2nd class constraints which, when solved, gives back the usual
geometrodynamical description of complex 3+1 gravity.}

Similarly, we must check to see if there are any primary constraints on the
configuration variables of the \sd theory. It turns out that although
$\otw{\otw\phi}{}^{ab}:=\e^{IJKL}\otw E^a{}_{IJ}\otw E^b{}_{KL}=0$ still
follows from the definition of $\otw E^a{}_{IJ}$ in terms of the complex
co-tetrad $\4 e_a^I$, it does not imply a constraint on the {\it \sd} field
$\+\otw E^a{}_{IJ}$.  Equation (7.19) may be viewed, instead, as a constraint
on the \asd field $\-\otw E^a{}_{IJ}$. (Recall that for complex
$\otw E^a{}_{IJ}$, \  $\-\otw E^a{}_{IJ}$ is not necessarily
the complex conjugate of $\+\otw E^a{}_{IJ}$). Thus, $\+\otw E^a {}_{IJ}$ is
free to take on arbitrary values, and the Legendre transform for the complex
\sd theory is actually fairly simple. By following the standard Dirac
constraint analysis, we find that $\+\otw E^a{}_{IJ}$ is the momentum
canonically conjugate to $\+ A_a{}^{IJ}$, while $(\+\4 A\cdot t)^{IJ}$, $\utw
N$, and $N^a$ play the role of Lagrange multipliers.  The complex phase space
$({}^C\G_{SD},{}^C\W_{SD})$ is coordinatized by the pair of complex fields $(\+
A_a{}^{IJ},\ \+\otw E^a{}_{IJ})$ and has the natural complex symplectic
structure\footnote{Note that in terms of the \Pb $\{\ ,\ \}$ defined by
${}^C\W_{SD}$, we have $\{\+ A_a{}^{IJ}(x),\+\otw E^b{}_{KL}(y)\}={1\over
2}\d_a^b\d(x,y) (\d_K^{[I}\d_L^{J]}-{i\over 2}\e_{KL}{}^{MN}\d_M^I\d_N^J)$. The
``extra" term on the \RHS is needed to make the \Pb \sd in the $IJ$ and $KL$
pairs of indices.}
$${}^C\W_{SD}=\int_\S\id\+\otw E^a{}_{IJ}\ \iwedge\ \id\+ A_a{}^{IJ}.
\eqno(7.20)$$
The Hamiltonian is given by
$$\eqalign{H_{SD}(\+ A,\+\otw E)=\int_\S \utw N\Tr(\+\otw E^a\ \+\otw E^b\
&\+ F_{ab})-N^a \Tr(\+\otw E^b\ \+ F_{ab})\cr
&-(\+\cD_a\ \+\otw E^a{}_{IJ})(\+\4 A\cdot t)^{IJ}.\cr}\eqno(7.21)$$
As we shall see in the next subsection, this is just a sum of 1st class
constraint functions associated with
$$\eqalignnotwo{&\Tr(\+\otw E^a\ \+\otw E^b\ \+ F_{ab})\approx 0,&(7.22)\cr
&\Tr(\+\otw E^b\ \+F_{ab})\approx 0,\quad{\rm and}&(7.23)\cr
&\+\cD_a\ \+\otw E^a{}_{IJ}\approx 0.&(7.24)\cr}$$
Note that all the constraints (and hence the evolution equations) are {\it
polynomial} in the canonically conjugate pair $(\+ A_a{}^{IJ},\+\otw
E^a{}_{IJ})$.  This is a simplification that we found in the 2+1 \P theory, but
lost in the 3+1 \P theory when we solved the 2nd class constraints.  In fact,
since the constraint equations never involve the inverse of $\+\otw
E^a{}_{IJ}$, the above Hamiltonian formulation is well-defined even if $\+\otw
E^a{}_{IJ}$ is non-invertible. Thus, we have a slight extension of complex
general relativity. The \sd theory makes sense even when the induced metric
$\otw{\otw q}{}^{ab}=\Tr(\+\otw E^a\ \+\otw E^b)$ becomes degenerate.

In order to make contact with the notation used in the literature (see, e.g.,
\cite{poona}), let us use the fact that the covariant normal $n_a$ to $\S$
defines a unit internal vector $n_I$ via $n_I:=n_a\ \4 e_I^a$. One can then
show that
$$\+ b^K{}_{IJ}:=-{1\over 2}\e^K{}_{IJ}+iq_{[I}^K n_{J]}\eqno(7.25)$$
is an isomorphism from the \sd sub-Lie algebra of the complexified Lie algebra
of $SO(3,1)$ to the complexified tangent space of $\S$. (Here
$\e_{JKL}:=n^I\e_{IJKL}$, $q_I^K:=\d_I^K+n_I n^K$, and $n_I n^I=-1$.) It
satisfies
$$\eqalignnotwo{&\*(\+ b^K{}_{IJ}):={1\over 2}(\+ b^K{}_{IJ}-{i\over 2}\e_{IJ}
{}^{MN}\ \+ b^K{}_{MN})=i\ \+ b^K{}_{IJ},&(7.26)\cr
&[\+ b^I,\+ b^J]_{MN}=\e^{IJ}{}_K\ \+ b^K{}_{MN},\quad{\rm and}&(7.27)\cr
&\Tr(\+ b^I\ \+ b^J):=-\ \+ b^I{}_{MN}\ \+ b^{JMN}=-q^{IJ}.&(7.28)\cr}$$
The inverse of $\+ b^K{}_{IJ}$ will be denoted by $\+ b_K{}^{IJ}$, and is
obtained by simply raising and lowering the indices of $\+ b^K{}_{IJ}$ with
the internal metric $\n_{IJ}$. Since $n_K\+ b^K{}_{IJ}=0$, we will use a
3-dimensional abstract internal index $i$ and write $\+ b^i{}_{IJ}$ and
$\+ b_i{}^{IJ}$ in what follows. From property (7.27), it follows that
$\+ b^i{}_{IJ}$ can actually be thought of as an isomorphism from the \sd
sub-Lie algebra of the complexified Lie algebra of $SO(3,1)$ to the
complexified Lie algebra of $SO(3)$.

Given this isomorphism, we can now define a $C\cL_{SO(3)}$-valued connection
1-form $A_a^i$ and a $C\cL_{SO(3)}^*$-valued vector density $\otw E_i^a$ via
$$\+ A_a{}^{IJ}=:A_a^i\ \+ b_i{}^{IJ}\quad{\rm and}\quad\ \+\otw E^a{}_{IJ}
=:-i\otw E_i^a\ \+ b^i{}_{IJ}.\eqno(7.29)$$
A straightforward calculation then shows that\footnote{To obtain equation
(7.30), I assume that the fiducial derivative operator $\pd_a$ has been
extended to act on $C\cL_{SO(3)}$-indices in such a way that $\pd_a\+
b_i{}^{IJ}=0$.}
$$\eqalignnotwo{&\+ F_{ab}{}^{IJ}=(2\pd_{[a}A_{b]}^i+\e^i{}_{jk}A_a^j A_b^k)
\ \+b_i{}^{IJ}=: F_{ab}^i\ \+ b_i{}^{IJ}\quad{\rm and}&(7.30)\cr
&\Tr(\+\otw E^a\ \+\otw E^b)=\otw E_i^a\otw E^{bi}=\otw{\otw q}{}^{ab}.&(7.31)
\cr}$$
Thus, $\otw E_i^a$ is a  complex (densitized) triad and $F_{ab}^i$ is the Lie
algebra-valued curvature tensor of the generalized derivative operator $\cD_a$
defined by $\cD_a v^i:=\pd_a v^i+\e^i{} _{jk}A_a^j v^k$.

In terms of $A_a^i$ and $\otw E_i^a$, the complex symplectic structure
${}^C\W_{SD}$ becomes
$${}^C\W_{SD}=-i\int_\S \id\otw E_i^a\ \iwedge \ \id A_a^i,\eqno(7.32)$$
so $-i\otw E_i^a$ is the momentum canonically conjugate to $A_a^i$. The
Hamiltonian (7.21) can be written as
$$H_{SD}(A,\otw E)=\int_\S {1\over 2}\utw N\e^{ijk}\otw E_i^a\otw E_j^b F_{abk}
-i N^a\otw E_i^b F_{ab}^i+i(\cD_a\otw E_i^a)(\4 A\cdot t)^i,\eqno(7.33)$$
while the constraint equations (7.22)-(7.24) can be written as
$$\eqalignnotwo{&\e^{ijk}\otw E_i^a\otw E_j^b F_{abk}\approx 0,&(7.34)\cr
&\otw E_i^b F_{ab}^i\approx 0,\quad{\rm and}&(7.35)\cr
&\cD_a\otw E_i^a\approx 0.&(7.36)\cr}$$
We will take the constraint equations in this form when we analyze the \Pb
algebra of the corresponding constraint functions in the following section.

So far, all of the discussion in this section has dealt with complex 3+1
gravity. In order to recover the real theory, we must now impose {\it reality
conditions} on the complex phase space variables $(A_a^i,\otw E_i^a)$ to select
a real section of $({}^C\G_{SD},{}^C\W_{SD})$. To do this, recall that in terms
of the standard geometrodynamical variables $(q_{ab},\otw p^{ab})$, one
recovers real \gr from the complex theory by requiring that $q_{ab}$ and $\otw
p^{ab}$ both be real. Since equation (7.31) tells us that $\otw E_i^a\otw
E^{bi}=\otw{\otw q}{}^{ab}\ (=q q^{ab})$, the condition that $q_{ab}$ be real
can be conveniently expressed in terms of $\otw E_i^a$ as
$$\otw E_i^a\otw E^{bi}\quad{\rm be \ real.}\eqno(7.37)$$
Since we will want to ensure that this reality condition be preserved under the
dynamical evolution generated by the Hamiltonian, we must also demand that
$$(\otw E_i^a\otw E^{bi})^\bullet\quad{\rm be \ real.}\eqno(7.38)$$
Since in a 4-dimensional solution of the field equations $\otw p^{ab}$ is
effectively the time derivative of $q_{ab}$, requirement (7.38) is equivalent
to the condition that $\otw p^{ab}$ be real. In addition, since the Hamiltonian
of the theory is just a sum of the constraints (7.34)-(7.36) (all of which are
polynomial in the canonically conjugate variables), the reality conditions
(7.37) and (7.38) are also polynomial in $(A_a^i,\otw E_i^a)$.

Finally, to conclude this subsection, I should point out that the \sd action
(7.1) viewed as a functional of a \sd connection 1-form $\+\4 A_{aI}{}^J$ and
a {\it real} co-tetrad $\4 e_a^I$ does not yield the new variables for
real 3+1 gravity when one performs a 3+1 decomposition. The definition of
the configuration variable $\+\otw E^a{}_{IJ}$ in terms of a real co-tetrad
$\4 e_a^I$  gives rise to a primary constraint. Although the non-holonomic
constraint can be expressed in terms of $\+\otw E^a{}_{IJ}$ as
$$\Tr(\+\otw E^a\ \+\otw E^b)>0,\eqno(7.39)$$
the holonomic constraint $\otw{\otw\phi}{}^{ab}=0$ cannot be expressed solely
in terms of $\+\otw E^a{}_{IJ}$. For real $\otw E^a{}_{IJ}$ we have that
$\-\otw E^a{}_{IJ}$ equals the complex conjugate of $\+\otw E^a{}_{IJ}$, so
$$\otw{\otw\phi}{}^{ab}=\e^{IJKL}(\+\otw E^a{}_{IJ} + \ovr{\+\otw E^a{}_{IJ}})
(\+\otw E^b{}_{KL} + \ovr{\+\otw E^b{}_{KL}})=0.\eqno(7.40)$$
But by writing $\otw{\otw\phi}{}^{ab}=0$ in this way, we have destroyed the
possibility of completing the standard Dirac constraint analysis. For nowhere
in the analysis have we been told how to take Poisson brackets of the complex
conjugate fields. The Legendre transform of the \sd Lagrangian for real 3+1
gravity breaks down when we try to incorporate the primary constraints into the
analysis.
\vskip .5cm

\noindent{\sl 7.3 Constraint algebra}
\medskip

Given constraint equations (7.34)-(7.36) for the complex \sd theory, we would
now like to verify the claim that their associated constraint functions form a
1st class set. To do this, let $v^i$ (which takes values in
$C\cL_{SO(3)}$), $\utw N$, and $N^a$ be arbitrary complex-valued test
fields on $\S$ and define
$$\eqalignnotwo{&C(\utw N):={1\over 2}\int_\S\utw N\e^{ijk}\otw E_i^a\otw E_j^b
F_{abk},&(7.41)\cr
&C'(\vec N):=-i\int_\S N^a\otw E_i^b F_{ab}^i,\quad{\rm and}&(7.42)\cr
&G(v):=-i\int_\S v^i(\cD_a\otw E_i^a).&(7.43)\cr}$$
These will be called the {\it scalar}, {\it vector}, and {\it Gauss constraint
functions}.  As the names and notation suggest, these constraint functions
will play a similar role to the constraint functions defined in subsection 5.3.
Many of the calculations and results found there will apply here as well.

As usual, it is fairly easy to show that the Gauss constraint functions
generate the standard gauge transformations of the connection 1-form and
rotation of internal indices. Since
$${\d G(v)\over\d\otw E_i^a}=i\cD_a v^i\quad{\rm and}\quad{\d G(v)\over
\d A_a^i}=-i\{v,\otw E^a\}_i\ \left(:=-i\e^k{}_{ji}v^j\otw E_k^a\right),
\eqno(7.44)$$
it follows that $A_a^i\mapsto A_a^i-\e\cD_a v^i+O(\e^2)$ and $\otw E_i^a
\mapsto \otw E_i^a-\e\{v,\otw E^a\}_i+O(\e^2)$. It also follows that
$$\{G(v),G(w)\}=G([v,w]),\eqno(7.45)$$
where $[v,w]^i:=\e^i{}_{jk}v^j w^k$ is the Lie bracket of $v^i$ and $w^i$.
Thus, the mapping $v\mapsto G(v)$ is a representation of the Lie algebra
$C\cL_{SO(3)}$. Furthermore, given its geometrical interpretation as the
generator of internal rotations, we have
$$\eqalignnotwo{&\{G(v),C(\utw N)\}=0\quad{\rm and}&(7.46)\cr
&\{G(v),C'(\vec N)\}=0,&(7.47)\cr}$$
as well.

Since it is possible to show that the vector constraint function does not by
itself have any direct geometrical interpretation (see, e.g., \cite{AMT})
we will define a new constraint function, $C(\vec N)$, by taking a linear
combination of the vector and Gauss constraints. We define
$$C(\vec N):=C'(\vec N)-G(N),\eqno(7.48)$$
where $N^i:=N^a A_a^i$. We will call $C(\vec N)$ the {\it diffeomorphism
constraint function} since the motion it generates on phase space corresponds
to the 1-parameter family of diffeomorphisms on $\S$ associated with the
vector field $N^a$. To see this, we can write
$$\eqalign{C(\vec N):&=C'(\vec N)-G(N)\cr
&=-i\int_\S N^a\otw E_i^b F_{ab}^i+i\int_\S N^i(\cD_a\otw E_i^a)\cr
&=-i\int_\S N^a(\otw E_i^b F_{ab}^i-A_a^i\cD_b\otw E_i^b)\cr
&=-i\int_\S \otw E_i^a\cL_{\vec N} A_a^i,\cr}
\eqno(7.49)$$
where the Lie derivative \wrt $N^a$ treats fields having only internal indices
as scalars. To obtain the last line of (7.49), we ignored a surface integral
(which would vanish anyways for $N^a$ satisfying the appropriate boundary
conditions). By inspection, it follows that $A_a^i\mapsto A_a^i+\e\cL_{\vec N}
A_a^i+O(\e^2)$, etc. Using this geometric interpretation of $C(\vec N)$, it
follows that
$$\eqalignnotwo{&\{C(\vec N),G(v)\}=G(\cL_{\vec N}v),&(7.50)\cr
&\{C(\vec N),C(\utw M)\}=C(\cL_{\vec N}\utw M),\quad{\rm and}&(7.51)\cr
&\{C(\vec N),C(\vec M)\}=C([\vec N,\vec M]).&(7.52)\cr}$$

We are left to evaluate the \Pb $\{C(\utw N),C(\utw M)\}$ of two scalar
constraints. Using
$${\d C(\utw N)\over\d\otw E_i^a}=\utw N\e^{ijk}\otw E_j^b F_{abk}\quad
{\rm and}\quad{\d C(\utw N)\over\d A_a^i}=\e_i{}^{jk}\cD_b(\utw N\otw
E_j^a\otw E_k^b),\eqno(7.53)$$
it follows that
$$\eqalign{\{C(\utw N),C(\utw M)\}&=\int_\S{\d C(\utw N)\over\d A_a^i}
{\d C(\utw M)\over\d(-i\otw E_i^a)}-(\utw N\leftrightarrow\utw M)\cr
&=\int_\S i\e_i{}^{mn}\cD_c(\utw N\otw E_m^a\otw E_n^c)\utw M\e^{ijk}\otw
E_j^b F_{abk}-(\utw N\leftrightarrow\utw M)\cr
&=\int_\S i\e^{ijk}\e_i{}^{mn}\otw E_m^a\otw E_n^c\otw E_j^b(\utw M\pd_c
\utw N-\utw N\pd_c\utw M) F_{abk}.}\eqno(7.54)$$
If we now use the fact that
$$\e^{ijk}\e_i{}^{mn}=(\d^{jm}\d^{kn}-\d^{jn}\d^{km})\eqno(7.55)$$
(which is a property of the structure constants of $SO(3)$), we get
$$\{C(\utw N),C(\utw M)\}=C'(\vec K) \ \ \left(=C(\vec K)+G(K)\right),
\eqno(7.56)$$
where $K^a:=\otw{\otw q}{}^{ab}(\utw N\pd_b\utw M-\utw M\pd_b\utw N)$ and
$\otw{\otw q}{}^{ab}=\otw E_i^a\otw E^{bi}$. Thus, the constraint functions are
closed under Poisson bracket---i.e., they form a 1st class set. Note, however,
that since the vector field $K^a$ depends on the phase space variable $\otw
E_i^a$, the \Pb (7.56) involves {\it structure functions}. The constraint
functions do not form a Lie algebra.
\vskip 1cm

\noindent{\bf 8. 3+1 matter couplings}
\vskip .5cm

In this section, we will couple various matter fields to 3+1 gravity. We will
repeat much of what we did in Section 5, but this time in the context of the
3+1 theory, and for a \YM field instead of a massless scalar field. In
subsections 8.1 and 8.2, we couple a \cc $\L$ and a \YM field to complex
3+1 gravity using an action principle and the \sd action as our starting point.
We shall show that the inclusion of these matter fields does not destroy the
polynomial nature of the constraint equations. This is the main result. (As
usual, reality conditions should be included to recover the real theory.) As I
mentioned for the 2+1 theory, it is possible to couple other fundamental matter
fields (e.g., scalar and Dirac fields) to 3+1 gravity in a similar fashion and
obtain the same basic results. For a more detailed discussion of this and
related issues, interested readers should see, e.g., \cite{matter}.
\vskip .5cm

\noindent{\sl 8.1 \Sd theory coupled to a cosmological constant}
\medskip

To couple a \cc $\L$ to complex 3+1 gravity via the \sd action, we will
start with the action
$$S_\L(\4 e,\+\4 A):={1\over 4}\int_M\otw\n^{abcd}\e_{IJKL}\ \4 e_a^I\ \4
e_b^J(\+\4 F_{cd}{}^{KL}-{\L\over 3!}\ \4 e_c^K\4 e_d^L).\eqno(8.1)$$
Here $\+\4 F_{abI}{}^J=2\pd_{[a}\+\4 A_{b]I}{}^J + [\+\4 A_a,\+\4 A_b]_I{}^J$
is the internal curvature tensor of the \sd generalized derivative operator
$\+\4\cD_a$ defined by the \sd connection 1-form $\+\4 A_{aI}{}^J$, and $\4
e_a^I$ is a  complex co-tetrad which defines a \st metric $g_{ab}$ via
$g_{ab}:=\4 e_a^I\ \4 e_b^J\n_{IJ}$.  Note that $S_\L(\4 e,\+\4 A)$ is just a
sum of the \sd action
$$S_{SD}(\4 e,\+\4 A):={1\over 4}\int_M\otw\n^{abcd}\e_{IJKL}\ \4 e_a^I\ \4
e_b^J\+\4 F_{cd}{}^{KL}\eqno(8.2)$$
and a term proportional to the volume of the spacetime. In fact,
$${\L\over 4!}\int_M\otw\n^{abcd}\e_{IJKL}\ \4 e_a^I\ \4 e_b^J\ \4 e_c^K\ \4
e_d^L=\L\int_M\sqrt{-g},\eqno(8.3)$$
where $g$ is the determinant of the covariant metric $g_{ab}$.

To show that (8.1) reproduces the standard equation of motion,
$$G_{ab}+\L g_{ab}=0,\eqno(8.4)$$
for gravity coupled to the cosmological constant $\L$, we will first vary (8.1)
\wrt the \sd connection 1-form $\+\4 A_a{}^{IJ}$. Since the second term (8.3)
is independent of $\+\4 A_a{}^{IJ}$, we get
$$\+\4\cD_b\left((\4 e)\ \+(\4 e_I^{[a}\ \4 e_J^{b]})\right)=0,\eqno(8.5)$$
which is exactly the \eom we obtained in Section 7 for the vacuum case. Thus,
just as we saw in subsection 7.1, \ $\+\4 A_{aI}{}^J=\+\G_{aI}{}^J$ where
$\+\G_{aI}{}^J$ is the \sd part of the internal Christoffel symbol of $\grad_a$
(the unique, torsion-free generalized derivative operator compatible with the
co-tetrad.) Since (8.5) can be solved uniquely for $\+\4 A_{aI}{}^J$ in terms
of the remaining basic variables $\4 e_a^I$, we can pull-back $S_\L(\4 e,\+\4
A)$ to the solution space $\+\4 A_{aI}{}^J=\+\G_{aI}{}^J$. We obtain a new
action
$$\und S{}_\L(\4 e)={1\over 4}\int_M\otw\n^{abcd}\e_{IJKL}\ \4 e_a^I\ \4
e_b^J(\+ R_{cd}{}^{KL}-{\L\over 3!}\ \4 e_c^K\4 e_d^L),\eqno(8.6)$$
where $\+ R_{cd}{}^{KL}$ is the \sd part of the internal curvature tensor
defined by $\grad_a$. Then by using the Bianchi identity $R_{[abc]d}=0$ for the
first term and (8.3) for the second, we get
$$\und S{}_\L(\4 e)={1\over 2}\int_M\sqrt{-g}(R-2\L).\eqno(8.7)$$
As mentioned in subsection 5.1, this is (up to an overall factor of $1/2$) the
action that one uses to obtain (8.4) starting from an action principle. This is
the desired result.

To put this theory in Hamiltonian form, we proceed as in subsection 7.2.
Recall that (modulo a surface integral) the Lagrangian $L_{SD}$ of the \sd
theory is given by
$$\eqalign{L_{SD}=\int_\S -\utw N\Tr&(\+\otw E^a\ \+\otw E^b\ \+ F_{ab})+N^a
\Tr(\+\otw E^b\ \+ F_{ab})\cr
&+\ (\+\otw E^a{}_{IJ})\cL_{\vec t} \ \+ A_a{}^{IJ} + (\+\cD_a\ \+
\otw E^a{}_{IJ})(\+\4 A\cdot t)^{IJ},\cr}\eqno(8.8)$$
where $\+\otw E^a{}_{IJ}$ denotes the \sd part of $\otw E^a{}_{IJ}:={1\over
2}\e_{IJKL}\otw\n^{abc}\ \4 e_b ^K\ \4 e_c^L$.  By using the isomorphism
between the \sd sub-Lie algebra of the complexified Lie algebra of $SO(3,1)$
and the complexified Lie algebra of $SO(3)$, we can rewrite (8.8) as
$$L_{SD}=\int_\S -{1\over 2}\utw N\e^{ijk}\otw E_i^a\otw E_j^b F_{abk}+iN^a
\otw E_i^b F_{ab}^i -i\otw E_i^a\cL_{\vec t} \ A_a^i -i(\cD_a\otw E_i^a)
(\4 A\cdot t)^i,\eqno(8.9)$$
where $\otw E_i^a$ is a complex (densitized) triad (i.e., $\otw E_i^a\otw
E^{bi}= \otw{\otw q}{}^{ab}\ (=q q^{ab})$) and $A_a^i$ is a connection 1-form
on $\S$ that takes values in the complexified Lie algebra of $SO(3)$.

By using the decomposition $\sqrt{-g}=N\sqrt{q}\ dt$ together with the fact
that
$${1\over 3!}\utw\n{}_{abc}\e^{ijk}\otw E_i^a\otw E_j^b\otw E_k^c=q,\eqno(8.10)
$$
one can similarly show that
$${\L\over 4!}\int_M\otw\n^{abcd}\e_{IJKL}\ \4 e_a^I\ \4 e_b^J\ \4 e_c^K\ \4
e_d^L={\L\over 3!}\int dt\int_\S \utw N\utw\n{}_{abc}\e^{ijk}\otw E_i^a\otw
E_j^b\otw E_k^c.\eqno(8.11)$$
Thus, the Lagrangian $L_\L$ for 3+1 gravity coupled to the \cc $\L$ via the
\sd action is given by
$$\eqalign{L_\L=\int_\S -\utw N&({1\over 2}\e^{ijk}\otw E_i^a\otw E_j^b F_{abk}
+{\L\over 3!}\utw\n{}_{abc}\e^{ijk}\otw E_i^a\otw E_j^b\otw E_k^c)\cr
&+iN^a\otw E_i^b F_{ab}^i -i\otw E_i^a\cL_{\vec t} \ A_a^i-i(\cD_a\otw E_i^a)
(\4 A\cdot t)^i.\cr}\eqno(8.12)$$
The configuration variables of the theory are $(\4 A\cdot t)^i$, $\utw N$,
$N^a$, $A_a^i$, and $\otw E_i^a$.

By following the standard Dirac constraint analysis, we find (as in the vacuum
case) that $-i\otw E_i^a$ is the momentum canonically conjugate to $ A_a^i$
while $(\4 A\cdot t)^i$, $\utw N$, and $N^a$ play the role of Lagrange
multipliers. The complex phase space and complex symplectic structure are the
same as those found for the \sd theory with $\L=0$, while the Hamiltonian is
given by
$$\eqalign{H_\L(A,\otw E)=\int_\S \utw N\Big({1\over 2}\e^{ijk}\otw E_i^a\otw
E_j^b &F_{abk}+{\L\over 3!}\utw\n{}_{abc}\e^{ijk}\otw E_i^a\otw E_j^b\otw
E_k^c\Big)\cr
&-iN^a\otw E_i^b F_{ab}^i+i(\cD_a\otw E_i^a)(\4 A \cdot t)^i.\cr}\eqno(8.13)$$
We shall see that this is just a sum of 1st class constraint functions
associated with
$$\eqalignnotwo{&{1\over 2}\e^{ijk}\otw E_i^a\otw E_j^b F_{abk}+{\L\over
3!}\utw\n{}_{abc} \e^{ijk}\otw E_i^a\otw E_j^b\otw E_k^c\approx 0,&(8.14)\cr
&\otw E_i^b F_{ab}^i\approx 0,\quad{\rm and}&(8.15)\cr
&\cD_a\otw E_i^a\approx 0.&(8.16)\cr}$$
These are the constraint equations associated with the Lagrange multipliers
$\utw N$, $N^a$, and $(\4 A\cdot t)^i$, respectively. Note that they are
polynomial in the canonically conjugate variables $(A_a^i,\otw E_i^a)$ even
when $\L\not =0$. In fact, only constraint equation (8.14) differs from its
$\L=0$ counterpart.

To conclude this subsection, we will verify the claim that the constraint
functions associated with (8.14)-(8.16) form a 1st class set. Since the Gauss
and diffeomorphism constraint functions associated with (8.16) and (8.15) will
be the same as in subsection 7.3, we need only concentrate on the {\it scalar
constraint function}
$$C(\utw N):=\int_\S\utw N\Big({1\over 2}\e^{ijk}\otw E_i^a\otw E_j^b F_{abk}+
{\L\over 3!}\utw\n{} _{abc}\e^{ijk}\otw E_i^a\otw E_j^b\otw E_k^c\Big).\eqno
(8.17)$$
Since
$$\eqalignnotwo{&{\d C(\utw N)\over\d\otw E_i^a}=\utw N(\e^{ijk}\otw E_j^b
F_{abk}+{\L\over 2}\utw\n{}_{abc}\e^{ijk}\otw E_j^b\otw E_k^c)\quad{\rm and}
&(8.18a)\cr
&{\d C(\utw N)\over\d A_a^i}=\e_i{}^{jk}\cD_b(\utw N\otw E_j^a\otw E_k^b),
&(8.18b)\cr}$$
it follows that
$$\eqalign{\{C(\utw N),C(\utw M)\}&=\int_\S{\d C(\utw N)\over\d A_a^i}
{\d C(\utw M)\over\d(-i\otw E_i^a)}-(\utw N\leftrightarrow\utw M)\cr
&=\int_\S i\e_i{}^{mn}\cD_c(\utw N\otw E_m^a\otw E_n^c) \utw M(\e^{ijk}\otw
E_j^b F_{abk}+{\L\over 2}\utw\n{}_{abd}\e^{ijk}\otw E_j^b\otw E_k^d)-(\utw N
\leftrightarrow\utw M)\cr
&=\int_\S i\e^{ijk}\e_i{}^{mn}(\utw M\pd_c\utw N-\utw N\pd_c\utw M)
(\otw E_m^a\otw E_n^c\otw E_j^b F_{abk}+{\L\over 2}q\e_{mjk}\otw E_n^c).\cr}
\eqno(8.19)$$
If we again use the fact that the structure constants of $SO(3)$ satisfy
$$\e^{ijk}\e_i{}^{mn}=(\d^{jm}\d^{kn}-\d^{jn}\d^{km}),\eqno(8.20)$$
we get
$$\{C(\utw N),C(\utw M)\}=C'(\vec K) \ \ \left(=C(\vec K)+G(K)\right),
\eqno(8.21)$$
where $K^a:=\otw{\otw q}{}^{ab}(\utw N\pd_b\utw M-\utw M\pd_b\utw N)$ and
$\otw{\otw q}{}^{ab}=\otw E_i^a\otw E^{bi}$ as before. Thus, the constraint
functions are closed under Poisson bracket---i.e., they form a 1st class set.
The \Pb algebra of the constraint functions is exactly the same as it was for
the $\L=0$ case.  In particular, since the vector field $K^a$ depends on the
phase space variable $\otw E_i^a$, the constraint functions again do not form a
Lie algebra.
\vskip .5cm

\noindent{\sl 8.2 \Sd theory coupled to a \YM field}
\medskip

To couple a \YM field (with gauge group $\bG$) to complex 3+1 gravity via
the \sd action, we will start with the {\it total action}
$$S_T(\4 e,\4 A,\4\bA):=S_{SD}(\4 e,\+\4 A)+{1\over 2}S_{YM}(\4 e,\4\bA),
\eqno(8.22)$$
where $S_{SD}(\4 e,\+\4 A)$ is the \sd action (8.2) and $S_{YM}(\4 e,\4 \bA)$
is the usual \YM action
$$S_{YM}(\4 e,\4\bA):=-\int_M\Tr(\sqrt{-g}\ g^{ac} g^{bd}\ \4\bF_{ab}
\ \4\bF_{cd}).\eqno(8.23)$$
Here $S_{YM}(\4 e,\4\bA)$ is to be viewed as a functional of a co-tetrad $\4
e_a^I$ and a connection 1-form $\4\bA_a$ which takes values in the Lie algebra
of the gauge group $\bG$.\footnote{\YM fields will be denoted by bold face stem
letters and their (internal) Lie algebra indices will be suppressed.
Throughout, we will assume that we have a representation of the \YM Lie algebra
$\cL_{\bG}$ by linear operators (on some vector space $V$) with the trace
operation $\Tr$ playing the role of an invariant, non-degenerate bilinear form
$\bk$.} \ $\Tr$ denotes the trace operation in some representation of the \YM
Lie algebra, and $\4\bF_{ab}=2\pd_{[a}\4\bA_{b]}+ [\4\bA_a,\4\bA_b]$ is the
(internal) curvature tensor of the generalized derivative operator $\4\bD_a$
defined by $\4\bA_a$. The additional factor of $1/2$ is needed in front of
$S_{YM}(\4 e,\4\bA)$ so that the above definition of the total action will be
consistent with the definition of $S_{SD}(\4 e,\4 A)$. The \YM action depends
on the co-tetrad $\4 e_a^I$ through its dependence on $\sqrt{-g}$ and $g^{ab}$,
but is independent of the \sd connection 1-form $\+\4 A_{aI}{}^J$.  As
mentioned in Section 5, out of all the fundamental matter couplings, only the
action for the Dirac field would depend on $\+\4 A_{aI} {}^J$.

To show that (8.22) reproduces the standard \YM coupled to gravity \eoms
$$\4\bD_b(\sqrt{-g}\ \4\bF^{ab})=0\quad{\rm and}\quad G^{ad}=8\pi T^{ad}(YM),
\eqno(8.24)$$
where
$$T_{ab}(YM):={1\over 4\pi}\Tr(\4\bF_a{}^c\ \4\bF_{bc}-{1\over 4}g_{ab}\4
\bF_{cd}\ \4\bF^{cd})\eqno(8.25)$$
is the stress-energy tensor of the \YM field, we proceed as we did in the
previous subsection. Since $S_{YM}(\4 e,\4\bA)$ is independent of $\+\4
A_a{}^{IJ}$, the variation of (8.22) \wrt $\+\4 A_a{}^{IJ}$ implies
$$\+\4\cD_b\left((\4 e)\ \+(\4 e_I^{[a}\ \4 e_J^{b]})\right)=0.\eqno(8.26)$$
As before, this tells us that $\+\4 A_{aI}{}^J=\+\G_{aI}{}^J$. Recalling that
the Bianchi identity $R_{[abc]d}=0$ implies that the pull-back of $S_{SD}(\4
e,\+\4 A)$ to the solution space $\+\4 A_{aI}{}^J=\+\G_{aI}{}^J$ is just $1/2$
times the standard \EH action $S_{EH}(\4 e)$ for complex 3+1 gravity, we
obtain
$$\und S{}_T(\4 e,\4\bA)={1\over 2}\Big(S_{EH}(\4 e)+ S_{YM}(\4 e,\4\bA)
\Big).\eqno(8.27)$$
This is (up to an overall factor of $1/2$) the usual total action that one uses
to couple a \YM field to gravity. If we now vary $\und S{}_T(\4 e, \4\bA)$ \wrt
$\4\bA_a$ and $\4 e_a^I$, and contract the second equation with $\4 e_I^d$, we
recover (8.24). Note that to write the first equation in (8.24), we had to
consider a torsion-free extension of $\4\bD_a$ to \st tensor fields. But since
the left hand side is the divergence of  a skew \st tensor density of weight +1
on $M$, it is independent of this choice.

To put this theory in Hamiltonian form, we need only decompose the \YM action
$S_{YM}(\4 e,\4\bA)$ since the \sd Lagrangian $L_{SD}$ is given by (8.9). Using
$g^{ab}=q^{ab}-n^a n^b$ and $\sqrt{-g}=N\sqrt{q}\ dt$ it follows that
$$\eqalign{S_{YM}(\4 e,&\4\bA)=\int dt\int_\S\Tr\Big\{-\utw N
q^{-1}\otw{\otw q} {}^{ac}\otw{\otw q}{}^{bd}\bF_{ab}\bF_{cd}+2\otw{\otw
q}{}^{ab}\utw N{}^{-1} q^{-1}\X\cr
&\X(\cL_{\vec t} \ \bA_a-\bD_a(\4\bA\cdot t)+N^c \bF_{ac})(\cL_{\vec t} \
\bA_b-\bD_b(\4\bA\cdot t)+N^d\bF_{bd})\Big\},\cr}\eqno(8.28)$$
where $\otw{\otw q}{}^{ab}:=q q^{ab}\ (=\otw E_i^a\otw E^{bi})$, $(\4\bA\cdot
t):=t^a\ \4\bA_a$, and $\bA_a:=q_a^b\ \4\bA_b$. Here $\bF_{ab}:=q_a^c\ q_b^d
\4\bF_{cd}$ is the curvature tensor of the generalized derivative operator
$\bD_a \ (:=q_a^b \ \4\bD_b)$ on $\S$ associated with $\bA_a$. If we now define
the ``magnetic field'' of $\bA_a$ to be $\bB_{ab}:=2\bF_{ab}\ (=2 q_a^c\ q_b^d\
\4\bF_{cd})$, we see that the \YM Lagrangian $L_{YM}$ is given by
$$\eqalign{L_{YM}=&\int_\S\Tr\Big\{-{1\over 4}\utw N q^{-1}\otw{\otw
q}{}^{ac}\otw {\otw q}{}^{bd}\bB_{ab}\bB_{cd}+2\otw{\otw q}{}^{ab} \utw
N{}^{-1} q^{-1}\X\cr
&\X(\cL_{\vec t} \ \bA_a-\bD_a(\4\bA\cdot t)+{1\over 2}N^c\bB_{ac})
(\cL_{\vec t} \ \bA_b-\bD_b(\4\bA\cdot t)+{1\over 2}N^d\bB_{bd})\Big\}.
\cr}\eqno(8.29)$$
The total Lagrangian $L_T$ is the sum  $L_T=L_{SD}+{1\over 2}L_{YM}$ and is to
be viewed as a functional of the configuration variables $(\4\bA\cdot t)$, $(\4
A\cdot t)^i$, $\utw N$, $N^a$, $A_a^i$, $\otw E_i^a$, $\bA_a$ and their first
time derivatives.

Following the standard Dirac constraint analysis, we find that
$$\otw\bE^a:={\d L_T\over\d(\cL_{\vec t} \ \bA_a)}=2\otw{\otw q}{}^{ab}
\utw N{}^{-1}
q^{-1}(\cL_{\vec t} \ \bA_b-\bD_b(\4\bA\cdot t)+{1\over 2}N^d\bB_{bd})
\eqno(8.30)$$
is the momentum (or ``electric field'') canonically conjugate to $\bA_a$.
Since this equation can be inverted to give
$$\cL_{\vec t} \ \bA_a={1\over 2}q_{ab}\utw N\otw\bE^b+\bD_a(\4\bA\cdot t)-
{1\over 2}N^c\bB_{ac},\eqno(8.31)$$
it does not define a constraint. On the other hand, $-i\otw E_i^a$ is
constrained to be the momentum canonically conjugate to $A_a^i$, while
$(\4\bA\cdot t)$, $(\4 A\cdot t)^i$, $\utw N$, and $N^a$ play the role of
Lagrange multipliers. The resulting complex total phase space
$({}^C\G_T,{}^C\W_T)$ is coordinatized by the pairs of fields $(A_a^i,\otw
E^a_i)$ and $(\bA_a, \otw\bE^a)$ with symplectic structure
$${}^C\W_T=\int_\S-i\id\otw E_i^a\ \iwedge\ \id A_a^i+\Tr(\id\otw\bE^a\
\iwedge\ \id\bA_a).\eqno(8.32)$$
The Hamiltonian is given by
$$\eqalign{H_T&(A,\otw E,\bA,\otw\bE)=\int_\S\utw N\Big({1\over
2}\e^{ijk}\otw E_i^a\otw E_j^b F_{abk}+{1\over 8}q^{-1}\otw{\otw
q}{}^{ac}\otw{\otw q}{}^ {bd}\Tr(\bB_{ab}\bB_{cd}+\bE_{ab}\bE_{cd})\Big)\cr
&+N^a\Big(-i\otw E_i^b F_{ab}^i+\Tr(\otw\bE^b\bF_{ab})\Big)+i(\cD_a\otw
E_i^a)(\
4 A
\cdot t)^i-\Tr((\4\bA\cdot t)\bD_a\otw\bE^a),\cr}\eqno(8.33)$$
where $\bE_{ab}:=\utw\n{}_{abc}\otw\bE^c$ is the dual to the \YM ``electric
field'' $\otw\bE^a$. We shall see that this is just a sum of 1st class
constraint functions associated with
$$\eqalignnotwo{&{1\over 2}\e^{ijk}\otw E_i^a\otw E_j^b F_{abk}+{1\over
8}q^{-1}
\otw{\otw q}{}^{ac}\otw{\otw q}{}^{bd}\Tr(\bB_{ab}\bB_{cd}+\bE_{ab}\bE_{cd})
\approx 0,&(8.34)\cr
&-i\otw E_i^b F_{ab}^i+\Tr(\otw\bE^b\bF_{ab})\approx 0,&(8.35)\cr
&\cD_a\otw E_i^a\approx 0,\quad{\rm and}\quad \bD_a\otw\bE^a\approx 0.&(8.36)
\cr}$$
These are the constraint equations associated with the Lagrange multipliers
$\utw N$, $N^a$, $(\4 A\cdot t)^I$, and $(\4\bA\cdot t)$, respectively.

Note that by inspection (8.35) and (8.36) are polynomial in the canonically
conjugate variables. However, constraint equation (8.34)  fails to be
polynomial due to the presence of the non-polynomial multiplicative factor
$q^{-1}$. But since $q={1\over 3!}\utw\n{}_{abc}\e^{ijk}\otw E_i^a\otw
E_j^b\otw E_k^c$ is polynomial in $\otw E_i^a$, we can multiply (8.34) by $q$
and restore the polynomial nature of all the constraints. Thus, to couple
a \YM field to 3+1 gravity via the \sd action, we are led to a scalar
constraint with density weight +4. This implies that the associated constraint
function will be labeled by a test field (i.e., lapse function) having density
weight $-3$.

To verify the claim that the constraint functions associated with (8.34)-(8.36)
form a 1st class set, let $v^i$ and $\bv$ (which take values in complexified
Lie algebra of $SO(3)$ and the representation of the Lie algebra of the \YM
gauge group $\bG$), \ $\utw N$, and $N^a$ be arbitrary complex-valued test
fields on $\S$.  Then define
$$\eqalignnotwo{&C(\utw N):=\int_\S\utw N\Big({1\over 2}\e^{ijk}\otw E_i^a\otw
E_j^b F_{abk}+ {1\over 8}q^{-1}\otw{\otw q}{}^{ac}\otw{\otw
q}{}^{bd}\Tr(\bB_{ab} \bB_{cd}+ \bE_{ab}\bE_{cd})\Big), &(8.37)\cr
&C'(\vec N):=\int_\S N^a(-i\otw E_i^b F_{ab}^i+\Tr(\otw\bE^b \bF_{ab})),\quad
{\rm and}&(8.38)\cr
&G(\bv,v):=\int_\S\Tr(\bv\bD_a\otw\bE^a)-i v^i(\cD_a\otw E_i^a),
&(8.39)\cr}$$
to be the {\it scalar}, {\it vector}, and {\it Gauss constraint functions}.

As usual, it is fairly easy to show that the Gauss constraint functions
generate the standard gauge transformations of the connection 1-forms and
rotations of internal indices. Using this information, we find that
$$\eqalignnotwo{&\{G(\bv,v),G(\bw,w)\}=G([\bv,\bw],[v,w]),&(8.40)\cr
&\{G(\bv,v),C(\utw N)\}=0,\quad{\rm and}&(8.41)\cr
&\{G(\bv,v),C'(\vec N)\}=0,&(8.42)\cr}$$
where $[\bv,\bw]$ and $[v,w]^i$ are the Lie brackets in $\cL_{\bG}$ and
$C\cL_{SO(3)}$.  Thus, the subset of Gauss constraint functions form a Lie
algebra \wrt Poisson bracket. In fact, the mapping $(\bv,v)\mapsto G(\bv,v)$ is
a representation of the direct sum Lie algebra $\cL_{\bG}\dsum C\cL_
{SO(3)}$.

Again, the the vector constraint function will not have any direct geometrical
interpretation, so we define the {\it diffeomorphism constraint function}
$C(\vec N)$ by taking a linear combination of the vector and Gauss law
constraints. Setting
$$C(\vec N):=C'(\vec N)-G(\bN,N),\eqno(8.43)$$
where $\bN:=N^a\bA_a$ and $N^i:=N^a A_a^i$, we can show that
$$C(\vec N)=\int_\S -i\otw E_i^a\cL_{\vec N} A_a^i+\Tr(\otw\bE^a\cL_{\vec N}
\bA_a),\eqno(8.44)$$
where the Lie derivative \wrt $N^a$ treats fields having only internal indices
as scalars. By inspection, $A_a^i\mapsto A_a^i+\e\cL_{\vec N} A_a^i+O(\e^2)$, \
etc., so the motion on phase space generated by $C(\vec N)$ corresponds to the
1-parameter family of diffeomorphisms on $\S$ associated with $N^a$.  From
this geometric interpretation of $C(\vec N)$, it follows that
$$\eqalignnotwo{&\{C(\vec N),G(\bv,v)\}=G(\cL_{\vec N}\bv,\cL_{\vec
N}v),&(8.45)
\cr
&\{C(\vec N),C(\utw M)\}=C(\cL_{\vec N}\utw M),\quad{\rm and}&(8.46)\cr
&\{C(\vec N),C(\vec M)\}=C([\vec N,\vec M]).&(8.47)\cr}$$

Finally, we are left to evaluate the \Pb $\{C(\utw N),C(\utw M)\}$ of two
scalar
constraint functions. After a fairly long but straightforward calculation that
uses the fact that the structure constants of $SO(3)$ satisfy
$$\e^{ijk}\e_i{}^{mn}=(\d^{jm}\d^{kn}-\d^{jn}\d^{km}),\eqno(8.48)$$
one can show that
$$\{C(\utw N),C(\utw M)\}=C'(\vec K) \ \ \left(=C(\vec K)+G(\bK,K)\right),
\eqno(8.49)$$
where $K^a:=\otw{\otw q}{}^{ab}(\utw N\pd_b\utw M-\utw M\pd_b\utw N)$ and
$\otw{\otw q}{}^{ab}=\otw E_i^a\otw E^{bi}$. Thus, the constraint functions are
again closed under Poisson bracket---i.e., they form a 1st class set. Just as
we saw in subsection 8.1 for the \cc $\L$, the \Pb algebra of the constraint
functions for complex 3+1 gravity coupled to a \YM field via the \sd action
is exactly the same as it was for the vacuum case.
\vskip 1cm

\noindent{\bf 9. General relativity without-the-metric}
\vskip .5cm

To conclude this review, we will describe a theory of 3+1 gravity {\it without}
a metric. This will complete the transition from geometrodynamics to connection
dynamics in 3+1 dimensions. Although we saw in Section 7 that the Hamiltonian
formulation of the \sd theory for complex 3+1 gravity could be described in
terms of a connection 1-form $A_a^i$ and its canonically conjugate momentum (or
``electric field'') $\otw E_i^a$, the action for the \sd theory depended on
both a \sd connection 1-form $\+\4 A_a{}^{IJ}$ and a complex co-tetrad $\4
e_a^I$. Since the co-tetrad defines a \st metric $g_{ab}$ via $g_{ab}:=\4
e_a^I\ \4 e_b^J\n_{IJ}$, the \sd action had an implicit dependence on $g_{ab}$.
The purpose of this section is to show that (modulo an important degeneracy)
complex 3+1 gravity can be described by an action which does not depend on a
\st metric in any way whatsoever. We shall see in subsection 9.1 that this
action depends only on a connection 1-form $\4 A_a^i$ (which takes values in
the complexified Lie algebra of $SO(3)$) and a scalar density $\utw\Phi$ of
weight -1 on $M$. Hence we obtain a {\it pure spin-connection formulation} of
gravity. We shall also see how this pure spin-connection action is related to
the \sd action in the non-degenerate case.

In subsection 9.2, we will analyze the constraint equations for this theory.
Since we will have shown in subsection 9.1 that the \sd action and the pure
spin-connection action are equivalent when the \sd part of the Weyl tensor
is non-degenerate, the constraint equations of this theory are the same as the
the constraint equations for the \sd theory found in subsection 7.2. However,
we will now be able to write down the most general solution to the four
diffeomorphism constraint equations (the scalar and vector constraints of the
\sd theory) when the ``magnetic field'' $\otw B_i^a$ associated with the
connection 1-form $A_a^i$ is non-degenerate. This is a new result for the
Hamiltonian formulation of the 3+1 theory that was made manifest by working
in the pure spin-connection formalism.

I should emphasize here that all of the results in this section are taken
from previous work of Capovilla, Dell, Jacobson, Mason, and Plebanski. I am
not adding anything new in this section; rather, I am reporting their results
to bring the discussion of geometrodynamics versus connection dynamics for 3+1
gravity to it logical conclusion. Readers interested in a more detailed
discussion of the general relativity without-the-metric theory (including
matter couplings and an extension of this theory to a class of generally
covariant gauge theories) should see \cite{Plebanski,grwm,sdtf,scfg,gcgt}
and references mentioned therein. In addition, Peld\'an has recently
provided a similar pure spin-connection formulation of 2+1 gravity. Interested
readers should see \cite{Peldan}.
\vskip .5cm

\noindent{\sl 9.1 A pure spin-connection formulation of 3+1 gravity}
\medskip

The {\it pure spin-connection action} for complex 3+1 gravity is defined
to be
$$S(\utw\Phi,\4 A):={1\over 8}\int_M\utw\Phi(\otw\n\cdot\ \4 F^i\wedge\
\4 F^j)(\otw\n\cdot\ \4 F^k\wedge\ \4 F^l)h_{ijkl},\eqno(9.1)$$
where $\4 A_a^i$ is a connection 1-form which takes values in the complexified
Lie algebra of $SO(3)$, $\utw\Phi$ is a scalar density of weight -1 on $M$,
and $h_{ijkl}$ and $(\otw\n\cdot\ \4 F^i\wedge\ \4 F^j)$ are shorthand
notations for
$$\eqalignnotwo{&h_{ijkl}:=(\d_{ik}\d_{jl}+\d_{il}\d_{jk}-\d_{ij}\d_{kl})\quad
{\rm and}&(9.2)\cr
&(\otw\n\cdot\ \4 F^i\wedge\ \4 F^j):=\otw\n^{abcd}\ \4 F_{ab}^i\ \4 F_{cd}^j.
&(9.3)\cr}$$
As usual, $\4 F_{ab}^i=2\pd_{[a}\4 A_{b]}^i + [\4 A_a,\4 A_b]^i$ is the Lie
algebra-valued curvature tensor of the generalized derivative operator
$\4\cD_a$ defined by
$$\4\cD_a v^i:=\pd_a v^i + [\4 A_a,v]^i,\eqno(9.4)$$
where $[\4 A_a,v]^i:=\e^i{}_{jk} \ \4 A_a^j v^k$ denotes the Lie bracket of $\4
A_a^i$ and $v^i$ in $C\cL_{SO(3)}$. Although $\4\cD_a$ defined by (9.4)
knows how to act only on internal indices, we will often find it convenient to
consider a torsion-free extension of $\4\cD_a$ to \st tensor fields. All
results and all calculations will be independent of this choice.

To show that the pure spin-connection action reproduces the standard results
of complex 3+1 gravity, one could vary (9.1) \wrt $\utw\Phi$ and $\4 A_a^i$
and analyze the resulting \EL equations of motion. Instead, we will start with
the \sd action
$$S_{SD}(\4 e,\+\4 A):={1\over 4}\int_M\otw\n^{abcd}\e_{IJKL}\ \4 e_a^I\ \4
e_b^J\ \+\4 F_{cd}{}^{KL}\eqno(9.5)$$
for complex 3+1 gravity and show that (modulo an important degeneracy) the \sd
action (9.5) is actually equivalent to (9.1). Basically, we will eliminate from
(9.5) the field variables which pertain to the \st metric by solving their
associated \EL equations of motion. This will require that a certain symmetric
trace-free tensor $\psi_{ij}$ be {\it invertible} as a $3\X 3$ matrix. By
substituting these solutions back into the original action (9.5), we will
eventually obtain (9.1). We should point out that, in a solution, $\psi_{ij}$
corresponds to the \sd part of the Weyl tensor associated with the connection
1-form $\4 A_a^i$. Thus, the equivalence between the two actions breaks down
whenever the \sd part of the Weyl tensor is degenerate. Note also that the pure
spin-connection action describes complex 3+1 gravity. To recover the real
theory, one would have to impose reality conditions similar to those used in
Section 7 for the \sd theory. For a detailed discussion of $\psi_{ij}$ and the
reality conditions see, e.g., \cite{grwm}.

Since the \sd action (9.5) depends on both a \sd connection 1-form $\+\4
A_a{}^{IJ}$ and a complex co-tetrad $\4 e_a^I$, it has an implicit dependence
on the \st metric $g_{ab}:=\4 e_a^I\ \4 e_b^J\n_{IJ}$. Thus, it should come as
no surprise that the first step in obtaining a  metric-independent action
for 3+1 gravity involves the elimination of $\4 e_a^I$ from (9.5). To do this,
let us define
$$\S_{abIJ}:=\e_{IJKL}\ \4 e_a^K\ \4 e_b^L\eqno(9.6)$$
and $\+\S_{abIJ}$ to be its \sd part.\footnote{Recall that the \sd part of
$\S_{abIJ}$ is defined by $\+\S_{abIJ}:= {1\over 2}(\S_{abIJ}-{i\over
2}\e_{IJ}{}^{KL} \S_{abKL})$.} Then we can write the \sd action as
$$S_{SD}(\4 e,\+\4 A)={1\over 4}\int_M\otw\n^{abcd}\ \+\S_{abIJ}\ \+\4
F_{cd}{}^{IJ},\eqno(9.7)$$
where we have used the fact that $\S_{abIJ}\ \+\4F_{cd}{}^{IJ}=\+\S_{abIJ}\
\+\4F_{cd}{}^{IJ}$. To simplify the notation somewhat, let us recall that the
\sd sub-Lie algebra of the complexified Lie algebra of $SO(3,1)$ is isomorphic
to the complexified Lie algebra of $SO(3)$. Using the isomorphism described in
Section 7, we can define a $C\cL_{SO (3)}$-valued connection 1-form $\4 A_a^i$
and a $C\cL_{SO(3)}^*$-valued 2-form $\S_{abi}$ via
$$\+\4 A_a{}^{IJ}=:\4 A_a^i\ \+ b_i{}^{IJ}\quad{\rm and}\quad\+\S_{abIJ}
=:\S_{abi}\ \+ b^i{}_{IJ}.\eqno(9.8)$$
Then
$$S_{SD}(\4 e,\4 A)={1\over 4}\int_M\otw\n^{abcd}\S_{abi}\ \4 F_{cd}^i,\eqno
(9.9)$$
where $\4 F_{ab}^i=2\pd_{[a}\ \4 A_{b]}^i+\e^i{}_{jk}\ \4 A_a^j\ \4 A_b^k$ is
the Lie algebra-valued curvature tensor of the generalized derivative operator
$\4\cD_a$ defined by $\4 A_a^i$. It is related to $\+\4 F_{ab}{}^{IJ}$ via
$\+\4 F_{ab}{}^{IJ}=\4 F_{ab}^i\ \+ b_i{}^{IJ}$.

Although the right hand side of (9.9) involves just $\S_{abi}$ and $\4 A_a^i$,
the action is still a functional of $\4 A_a^i$ and $\4 e_a^I$ since $\S_{abi}$
depends on $\4 e_a^I$ through equation (9.6). To eliminate $\4 e_a^I$ from the
action, we must use the result (see, e.g., \cite{sdtf}) that (9.6) holds
for some $\4 e_a^I$ \iff the trace-free part of $\S^i\wedge \S^j$ equals
zero---i.e., $\S_{abIJ}= \e_{IJKL}\ \4 e_a^K\ \4 e_b^L$ for some $\4 e_a^I$
\iff
$$\otw\n^{abcd}(\S_{ab}^i\S_{cd}^j-{1\over 3}\d^{ij}\S_{ab}^k\S_{cdk})=0.
\eqno(9.10)$$
Thus, the \sd action can be viewed as a functional of $\S_{abi}$ instead of $\4
e_a^I$ provided we include in the action a term which gives back (9.10) as one
of its \EL equations of motion. More precisely, let us define
$$S(\psi,\S,\4 A):={1\over 4}\int_M\otw\n^{abcd}(\S_{abi}\ \4 F_{cd}^i -
{1\over
2}\psi_{ij}\S_{ab}^i\S_{cd}^j),\eqno(9.11)$$
where $\psi_{ij}$ is a symmetric {\it trace-free} tensor which will play the
role of a Lagrange multiplier of the theory. Then the variation of
$S(\psi,\S,\4 A)$ \wrt $\psi_{ij}$ will yield (9.10). Solving this equation and
pulling-back the action (9.11) to this solution space gives back (9.9).

But instead of varying $S(\psi,\S,\4 A)$ \wrt $\psi_{ij}$, let us vary this
action \wrt $\S_{abi}$ and solve the resulting \EL \eom for $\S_{abi}$ in terms
of $\psi_{ij}$ and $\4 A_a^i$. Varying (9.11) \wrt $\S_{abi}$, we find
$$\4 F_{ab}^i-\psi^{ij}\S_{abj}=0,\eqno(9.12)$$
where $\psi^{ij}:=\d^{ik}\d^{jl}\psi_{kl}$. This equation can be solved for
$\S_{abi}$ in terms of the remaining field variables provided the inverse
$(\psi^{-1})_{ij}$ of $\psi_{ij}$ exists. Assuming that it does, we get
$$\S_{abi}=(\psi^{-1})_{ij}\ \4 F_{ab}^j.\eqno(9.13)$$
If we now pull-back (9.11) to the solution space defined by (9.13), the
resulting action becomes
$$S(\psi,\4 A)={1\over 8}\int_M\otw\n^{abcd}(\psi^{-1})_{ij}\ \4 F_{ab}^i\ \4
F_{cd}^j.\eqno(9.14)$$
This is to be viewed as a functional of only the symmetric trace-free tensor
$\psi_{ij}$ and the connection 1-form $\4 A_a^i$.

We are almost finished. What remains to be shown is that $\psi_{ij}$ can be
eliminated from the action (9.14) in lieu of a scalar density $\utw\Phi$ of
weight -1 on $M$. To do this, let us write the action in matrix notation and
introduce another Lagrange multiplier $\otw\u$ to guarantee that $\psi_{ij}$ is
trace-free.\footnote{By introducing $\otw\u$, we can consider arbitrary
symmetric variations of $\psi_{ij}$ rather than  symmetric and trace-free
variations.} Then (9.14) can be written as
$$S(\otw\u,\psi,\4 A)={1\over 8}\int_M \Tr(\psi^{-1}\otw M) + \otw\u\Tr\psi,
\eqno(9.15)$$
where $\psi_{ij}$ is now assumed to be only symmetric (and invertible) and
$\otw M^{ij}$ is defined by
$$\otw M^{ij}:=\otw\n^{abcd}\ \4 F_{ab}^i\ \4 F_{cd}^j.\eqno(9.16)$$
Varying $S(\otw\u,\psi,\4 A)$ \wrt $\psi_{ij}$, we find
$$-\psi^{-1}\otw M\psi^{-1}+\otw\u I=0.\eqno(9.17)$$
By multiplying on the left and right by $\psi$, we see that (9.17) is
equivalent to
$$\otw M=\otw\u\psi^2.\eqno(9.18)$$
This equation can be solved for $\psi_{ij}$ in terms of $\otw M_{ij}$ and
$\otw\u$ provided the square-root of $\otw M_{ij}$ exists. Then
$$\psi=\otw\u^{-1/2}\otw M^{1/2},\eqno(9.19)$$
so the action (9.15) pulled-back to this solution space equals
$$S(\otw\u,\4 A)={1\over 4}\int_M \otw\u^{1/2}\Tr\otw M^{1/2}.\eqno(9.20)$$
The variation of $S(\otw\u,\4 A)$ \wrt $\otw\u$ now implies that $\Tr\otw M
^{1/2}=0$. From (9.19) we see that this is nothing more than the tracelessness
of $\psi_{ij}$.

In order to write the action in its final form (9.1), recall that the
characteristic equation obeyed by any $3\X 3$ matrix is
$$B^3-(\Tr B)B^2+{1\over 2}\left((\Tr B)^2-\Tr B^2\right)B-(\det B)I=0.
\eqno(9.21)$$
Multiplying by $B$ and setting $B^2=\otw M$ (i.e., $B=\otw M^{1/2}$), we get
$$(\det B)B=\otw M^2-{1\over 2}(\Tr\otw M)\otw M,\eqno(9.22)$$
(Here we have used the fact that $\Tr B\ (=\Tr\otw M^{1/2})=0$.) Using $(\det
B)^2 =\det\otw M$ and assuming invertibility of $B$ (so that $\det B\not= 0$),
we can write this last equation as
$$B=(\det\otw M)^{-1/2}(\otw M^2-{1\over 2}(\Tr\otw M)\otw M).\eqno(9.23)$$
By substituting this expression for $B\ (=\otw M^{1/2})$ back into (9.20), we
find
$$S(\otw\u,\4 A)={1\over 4}\int_M \otw\u^{1/2}(\det\otw M)^{-1/2}\Tr(\otw M^2
-{1\over 2}(\Tr\otw M)\otw M).\eqno(9.24)$$
Finally, if we define
$$\utw\Phi=\otw\u^{1/2}(\det \otw M)^{-1/2}\eqno(9.25)$$
(which is a scalar density of weight -1 on $M$) and use the definition (9.16)
of $\otw M^{ij}$ in terms of $\4 F_{ab}^i$, we see that
$$S(\utw\Phi,\4 A)={1\over 8}\int_M\utw\Phi(\otw\n\cdot\ \4 F^i\wedge\
\4 F^j)(\otw\n\cdot\ \4 F^k\wedge\ \4 F^l)h_{ijkl}\eqno(9.26)$$
when viewed as a functional of $\utw\Phi$ and $\4 A_a^i$. Note that
$h_{ijkl}$ and $(\otw\n\cdot\ \4 F^i\wedge\ \4 F^j)$ are given as before
by equations (9.2) and (9.3). This is the desired result.
\vskip .5cm

\noindent{\sl 9.2 Solution of the diffeomorphism constraints}
\medskip

Given that the \sd and pure spin-connection actions are equivalent when the
\sd part of the Weyl tensor is non-degenerate, it follows that the constraint
equations of the theory can be written as
$$\eqalignnotwo{&\e^{ijk}\otw E_i^a\otw E_j^b F_{abk}\approx 0,&(9.27)\cr
&\otw E_i^b F_{ab}^i\approx 0,\quad{\rm and}&(9.28)\cr
&\cD_a\otw E_i^a\approx 0.&(9.29)\cr}$$
These are just the constraint equations that we found in subsection 7.2 when
we put the \sd theory in Hamiltonian form. As before, the canonically conjugate
variables consist of a pair of complex fields $(A_a^i,\otw E_i^a)$, where
$A_a^i$ is the pull-back of the connection 1-form $\4 A_a^i$ to the submanifold
$\S$ and $\otw E_i^a$ is a complex (densitized) triad which may or may not
define an invertible induced metric $\otw{\otw q}{}^{ab}:=\otw E_i^a\otw
E^{bi}$. However, by working in the pure spin-connection formalism, we will
obtain a new result. We will be able to write down the most general solution to
the four diffeomorphism constraints (9.27)-(9.28) when the ``magnetic field''
$\otw B^a_i$ associated with $A_a^i$ is non-degenerate.

To see this, recall that in the \sd theory
$$\+\otw E^a{}_{IJ}=:-i\otw E_i^a\ \+ b^i{}_{IJ},\eqno(9.30)$$
where $\+\otw E^a{}_{IJ}$ was the \sd part of
$$\otw E^a{}_{IJ}:={1\over 2}\e_{IJKL}\otw\n^{abc}\ \4 e_b^K\ \4 e_c^L.
\eqno(9.31)$$
Note that in terms of $\S_{abIJ}$ defined by (9.6), we have $\otw E^a{}_{IJ}
={1\over 2}\otw\n^{abc}\S_{bcIJ}$, so that
$$-i\otw E_i^a={1\over 2}\otw\n^{abc}\S_{bci}.\eqno(9.32)$$
If we now use the result that an invertible symmetric trace-free tensor
$\psi_{ij}$ implies
$$\S_{abi}=(\psi^{-1})_{ij}\ \4 F_{ab}^j,\eqno(9.33)$$
it follows that
$$\otw E_i^a=i(\psi^{-1})_{ij}\otw B^{aj},\eqno(9.34)$$
where $\otw B^{ai}:={1\over 2}\otw\n^{abc}\ \4 F_{bc}^i\ (={1\over 2}\otw
\n^{abc}F_{bc}^i)$ is the ``magnetic field'' of $A_a^i$. We will now show that
by taking $\otw E_i^a$ of this form, the four diffeomorphism constraints
(9.27)-(9.28) are automatically satisfied.

Substituting (9.34) into the vector constraint (9.28), we get
$$\otw E_i^b F_{ab}^i=i(\psi^{-1})_{ij}\otw B^{bj}\utw\n{}_{abc} \otw
B^{ci}=0,\eqno(9.35)$$
where we have used the fact that $(\psi^{-1})_{ij}$ is symmetric in $i$ and $j$
while $\utw\n{}_{abc}$ is anti-symmetric in $b$ and $c$. Similarly,
substituting (9.34) into the scalar constraint (9.27), we get
$$\eqalign{\e^{ijk}\otw E_i^a\otw E_j^b F_{abk}&=\e^{ijk} \otw E_i^a\otw E_j^b
\utw\n{}_{abc}\otw B_k^c\cr
&=-i\e^{ijk}\otw E_i^a\otw E_j^b\utw\n{}_{abc}\psi_k^l\otw E_l^c\cr
&=-i q\e^{ijk}\e_{ijl}\psi_k^l\cr
&=-2i q\psi_k^k\cr
&=0,\cr}\eqno(9.36)$$
where we have used the fact that $\psi_{ij}$ is trace-free. Thus, the four
diffeomorphism constraints (9.27)-(9.28) are automatically satisfied for $\otw
E_i^a$ having the form given by (9.34). That this is the most general solution
follows if $\otw B_i^a$ is non-degenerate.  Then for a given $A_a^i$, $\otw
E_i^a$ will have 5 degrees of freedom (per space point) corresponding to the 5
degrees of freedom of the symmetric trace-free tensor $\psi_{ij}$.

What remains to be solved is the Gauss constraint (9.29), which in terms of
$\otw B^{ai}$ and $(\psi^{-1})_{ij}$ can be written as
$$\eqalign{0&=\cD_a\otw E_i^a\cr
&=i\cD_a\big((\psi^{-1})_{ij}\otw B^{aj}\big)\cr
&=i\otw B^{aj}\cD_a(\psi^{-1})_{ij}.\cr}\eqno(9.37)$$
To obtain the last line of (9.37), we used the Bianchi identity $\cD_a\otw
B^{aj}={1\over 2}\otw\n^{abc}\cD_{[a} F_{bc]}^j=0$.
\vskip 1cm

\noindent{\bf 10. Discussion}
\vskip .5cm

Let me begin by briefly summarizing the main results reviewed in this paper.

\begin{enumerate}
\item The standard \EH theory is a geometrodynamical theory of gravity in
which a spacetime metric is the fundamental field variable. The phase space
variables consist of a positive-definite metric $q_{ab}$ and its canonically
conjugate momentum $\otw p{}^{ab}$. These variables are subject to a set of 1st
class constraints, which are non-polynomial in $q_{ab}$ and which have a \Pb
algebra involving structure functions. This theory is valid in $n+1$
dimensions.
\item The 2+1 \P theory is a connection dynamical theory defined for any Lie
group $G$. The fundamental field variables consist of a $\cL_G$-valued
connection 1-form and a $\cL_G^*$-valued covector field. The phase space is
coordinatized by a connection 1-form $A_a^I$ and its canonically conjugate
momentum (or ``electric field") $\otw E_I^a$.  These are fields defined on a
2-manifold $\S$, and they are subject to a set of 1st class constraints.  The
constraints are polynomial in $(A_a^I,\otw E_I^a)$ and provide a representation
of the Lie algebra of the inhomogeneous Lie group associated with $G$. One
recovers 2+1 gravity by taking $G=SO(2,1)$.
\item \CS theory is a connection dynamical theory defined for any Lie group
that admits an invariant, non-degenerate bilinear form. In 2+1 dimensions, the
fundamental field variable is a Lie algebra-valued connection 1-form, and the
phase space variables are $A_a^i$---the pull-back of the field variable to the
2-dimensional hypersurface $\S$. There are 1st class constraints, which are
polynomial in $A_a^i$ and provide a representation of the defining Lie algebra.
\CS theory is related to 2+1 \P theories as follows: (i) 2+1 \P theory based on
any Lie group $G$ is equivalent to \CS theory based on the inhomogeneous Lie
group $IG$ associated with $G$; and (ii) the reduced phase space of \CS theory
based on a Lie group $G$ is the same as the reduced configuration space of the
2+1 \P theory based on the same $G$.  As a special case of (i), \ 2+1 gravity
is equivalent to \CS theory  based on the 2+1 dimensional Poincar\'e group
$ISO(2,1)$.
\item  One can couple matter to 2+1 gravity via the 2+1 \P action.  2+1 \P
theory coupled to a cosmological constant $\L$ is defined for any Lie group $G$
that admits an invariant, totally antisymmetric tensor $\e^{IJK}$. This theory
is equivalent to 2+1 dimensional \CS theory based on the $\L$-deformation of
$G$. As a special case, 2+1 gravity coupled to a \cc is equivalent to \CS
theory based on $SO(3,1)$ or $SO(2,2)$ (depending on the sign of $\L$).  \ 2+1
\P theory can also be coupled to matter fields with local degrees of freedom
provided $G=SO(2,1)$. The constraints remain polynomial in the canonically
conjugate variables and form a 1st class set.  However, due to the presence of
structure functions, they no longer form a Lie algebra.
\item The 3+1 \P theory is a geometrodynamical theory of 3+1 gravity in which a
co-tetrad and a Lorentz connection 1-form are the fundamental field variables.
Due to the presence of 2nd class constraints, the Hamiltonian formulation of
this theory reduces to that of the standard \EH theory in 3+1 dimensions.
Unlike the 2+1 \P theory, the 3+1 \P theory does not provide a connection
dynamical theory of 3+1 gravity.
\item The \sd theory is a connection dynamical theory of complex 3+1 gravity in
which a complex co-tetrad and a \sd connection 1-form are the fundamental field
variables.  The phase space variables consist of an $C\cL_{SO(3)}$-valued
connection 1-form $A_a^i$ and its canonically conjugate momentum (or ``electric
field") $\otw E_i^a$, both defined on a 3-manifold $\S$. These variables are
subject to a set of 1st class constraints, which are polynomial in $(A_a^i,\otw
E_i^a)$ but which have a \Pb algebra involving structure functions. In a
solution, $\otw E_i^a$ is a (densitized) spatial triad.  Since none of the
equations involve the inverse of $\otw E_i^a$, the \sd theory makes sense even
if $\otw E_i^a$ is non-invertible.  Thus, the \sd theory provides an extension
of complex \gr that is valid even when the induced spatial metric $\otw{\otw
q}{}^{ab} \ (=\otw E_i^a \otw E{}^{bi})$ is degenerate. One must impose reality
conditions to recover real general relativity.
\item One can couple matter to complex 3+1 gravity via the \sd action. The
constraints remain polynomial in the canonically conjugate variables and form a
1st class set. Since none of the equations involves the inverse of $\otw
E_i^a$, the \sd theory coupled to matter provides an extension of complex \gr
coupled to matter that includes degenerate spatial metrics. Reality conditions
must be imposed to recover the real theory.
\item The pure spin-connection formulation of \gr is a connection dynamical
theory of complex 3+1 gravity in which a $C\cL_{SO(3)}$-valued connection
1-form and a scalar density of weight $-1$ are the fundamental field variables.
The Hamiltonian formulation of this theory is equivalent to that of the \sd
theory provided the \sd part of the Weyl tensor is non-degenerate. When the
``magnetic" field associated with the connection 1-form $A_a^i$ is
non-degenerate, one can write down the most general solution to the four
diffeomorphism constraints. Reality conditions must be imposed to recover the
real theory.
\end{enumerate}

So what can we conclude from all these results? Is gravity a theory of a metric
or a connection? In other words, is gravity a theory of geometry, where the
fundamental variable is a \st metric which specifies distances between nearby
events, or is it a theory of curvature, where the fundamental variable is a
connection 1-form which tells us how to parallel propagate vectors around
closed loops? The answer: Either. As far as the classical equations of motion
are concerned, both a metric and a connection describe gravity equally well in
2+1 and 3+1 dimensions.  Neither metric nor connection is preferred.

As we have shown in this review and have summarized above, 2+1 and 3+1 gravity
admit formulations in terms of metrics and connections. But despite the
apparent differences (i.e., the different actions and field variables; the
different Hamiltonian formulations and canonically conjugate momenta; and the
possiblity of extending the theories to include arbitrary gauge groups and
solutions with degenerate spatial metrics), we have seen that the classical
equations of motion for all these formulations are the same. For instance, we
saw that the 2+1 \P theory reproduces vacuum 2+1 gravity when we choose
$G=SO(2,1)$ and solve the \eom for the connection. Similarly, we saw that \CS
theory reproduces 2+1 gravity coupled to a \cc $\L \ (>0)$ when we choose the
gauge group to be $SO(2,2)$. At the level of field equations, all the theories
are mathematically equivalent. The difference between the theories is, instead,
one of emphasis.

Now such a small change may not seem, at first, to be worth all the effort.
Recall that the shift in emphasis from metric to connection came only after we
successively analyzed the Einstein-Hilbert, Palatini, and Chern-Simons theories
in 2+1 dimensions, and the Einstein-Hilbert, Palatini, self-dual, and pure
spin-connection theories in 3+1 dimesions. This analysis required a fair amount
of work and, as we argued in the previous paragraph, did not lead to anything
particularly new at the classical level modulo, of course, the extensions of
the theories to include arbitrary gauge groups and solutions with degenerate
spatial metrics. But as soon as we turn to quantum theory and consider the
recent results that have been obtained there, the question as to whether the
shift in emphasis from metric to connection was worth the effort has a simple
affirmative answer.  Yes!  Indeed, almost all of the recent advances in quantum
general relativity can be traced back to this change of emphasis. As mentioned
in the introduction, Witten \cite{Witten} used the equivalence of the 2+1  \P
theory based on $SO(2,1)$ with \CS theory based on $ISO(2,1)$ to quantize 2+1
gravity. Others (e.g., Carlip \cite{Carlip1,Carlip2} and Anderson
\cite{Anderson}) are now using Witten's quantization to analyze the problem of
time in the 2+1 theory. In 3+1 dimensions, Jacobson, Rovelli, and Smolin
\cite{JSnp,RS} took advantage of the simplicity of the constraint equations in
the \sd formulation of 3+1 gravity to solve the quantum constraints
exactly---something that nobody could accomplish for the quantum version of the
scalar constraint in the traditional metric variables.  And the list goes on.
(See, e.g., \cite{npcg,poona} and \cite{abhay,lee,carlo,kodama} for more
details.) Where this list will end, and whether or not the change in emphasis
from metric to connection will lead to a mathematically consistent and
physically reasonable quantum theory of 3+1 gravity, remains to be seen.
\vspace{1cm}

\noindent ACKNOWLEDGEMENTS
\vspace{.3cm}

I would like to thank Abhay Ashtekar, Joseph Samuel, Charles Torre, and
Ranjeet Tate for many helpful discussions. This work was supported in part by
NSF grants PHY90-16733 and PHY91-12240, and by research funds provided by
Syracuse University and by the University of Maryland at College Park.

\end{document}